\documentclass[longauth]{aa}

\usepackage{amsmath}
\usepackage{amssymb}
\usepackage{url}
\usepackage{graphicx}
\usepackage{times}
\usepackage[utf8]{inputenc}
\usepackage{xspace}
\usepackage{color}
\usepackage{xcolor}
\usepackage{multirow}
\usepackage{upgreek}
\usepackage{multicol}
\usepackage{float}
\usepackage{supertabular} 
\usepackage{booktabs}
\usepackage{threeparttable}
\usepackage{longtable}
\usepackage[colorlinks=true,linkcolor=blue,citecolor=blue,urlcolor=blue]{hyperref}
\usepackage{comment}
\usepackage{enumitem}
\usepackage{adjustbox}

\bibpunct{(}{)}{;}{a}{}{,}

\usepackage{etoolbox}


\def\apjl{ApJL}
\def\apjs{ApJSS}

\def\aap{A\&A}

\newcommand{\tardis}{$\textsc{tardis}$}
\newcommand{\msun}{\mbox{M$_{\odot}$}}

\newcommand{\degree}{\mbox{$^\circ$}}

\newcommand{\kms}{\mbox{\,$\rm{km}\,s^{-1}$}}

\newcommand{\HeI}{He\,{\sc i}}

\newcommand{\CaII}{Ca\,{\sc ii}}

\newcommand{\I}{\,{\sc i}}

\newcommand{\ips}{\ensuremath{i_{\rm P1}}}
\newcommand{\zps}{\ensuremath{z_{\rm P1}}}

\newcommand{\grizy}{\ensuremath{grizy_{\rm P1}}}

\newcommand{\galex}{\textit{GALEX}}

\newcommand{\wise}{\textit{WISE}}
\newcommand{\mhtwo}{\mbox{$M_{\rm H_2}$}}

\newcommand{\gwtrig}{S191213g\xspace}
\newcommand{\GWDist}{$201\pm81$\,Mpc} 
\def\wxt{SN\,2019wxt}
\newcommand{\wxtDisc}{58833.32}           
\newcommand{\hostz}{0.035785}    
\newcommand{\hostDL}{$154\pm11$}

\newcommand{\Ar}{0.129}

\newcommand{\revtwo}[1]{}
\newcommand{\revthree}[1]{}

\DeclareRobustCommand{\VAN}[3]{#2}
\let\VANthebibliography\thebibliography
\def\thebibliography{\DeclareRobustCommand{\VAN}[3]{##3}\VANthebibliography}

\begin{document}

\title{Panning for gold, but finding helium: Discovery of the ultra-stripped supernova \wxt\ from gravitational-wave follow-up observations}

\titlerunning{\wxt\ -- an ultra-stripped supernova}

\author{
I. Agudo\inst{1}\and
L. Amati\inst{2}\and
T. An\inst{3,4}\and
F.~E. Bauer\inst{5,6,7}\and
S. Benetti\inst{8}\and
M.~G. Bernardini\inst{9}\and
R. Beswick\inst{10}\and
K. Bhirombhakdi\inst{11}\and
T. de Boer\inst{12}\and
M. Branchesi\inst{13,14}
% ...
}


\institute{} 

\authorrunning{The ENGRAVE Collaboration}

\date{\today}

\abstract{We present the results from  multi-wavelength observations of a transient discovered during an intensive follow-up campaign of S191213g, a gravitational wave (GW) event reported by the LIGO-Virgo Collaboration as a possible binary neutron star merger in a low latency search. This search yielded \wxt, a young transient in a galaxy whose sky position (in the 80\% GW contour) and distance ($\sim$150\,Mpc) were plausibly compatible with the localisation uncertainty of the GW event. Initially, the transient's tightly constrained age, its relatively faint peak magnitude ($M_i \sim -16.7$\,mag), and the $r-$band decline rate of $\sim 1$\,mag per 5\,days appeared suggestive of a compact binary merger. However, \wxt\ spectroscopically resembled a type Ib supernova, and analysis of the optical-near-infrared evolution rapidly led to the conclusion that while it could not be associated with S191213g, it nevertheless represented an extreme outcome of stellar evolution. By modelling the light curve, we estimated an ejecta mass of only $\sim 0.1\,M_\odot$, with $^{56}$Ni comprising $\sim 20\%$ of this. We were broadly able to reproduce its spectral evolution with a composition dominated by helium and oxygen, with trace amounts of calcium. We considered various progenitor channels that could give rise to the observed properties of \wxt\, and concluded that an ultra-stripped origin in a binary system is the most likely explanation. Disentangling genuine electromagnetic counterparts to GW events from transients such as \wxt\ soon after discovery is challenging: in a bid to characterise this level of contamination, we estimated the rate of events with a volumetric rate density comparable to that of \wxt\ and found that around one such event per week can occur within the typical GW localisation area of O4 alerts out to a luminosity distance of 500\,Mpc, beyond which it would become fainter than the typical depth of current electromagnetic follow-up campaigns.
}


\maketitle

\section{Introduction}

The first detection of astrophysical gravitational waves (GWs) in 2015 \citep{GW150914discovery} opened up a new window on the transient sky, and has since led to concerted efforts to locate their electromagnetic counterparts \citep[e.g.][]{GW150914JointFollowupPaper,GW150914DECam,GW150914Swift,GW150914Fermi,GW150914PanSTARRS,GW150914JGEM,GW150914Lipunov}. 
Despite this effort, to date only a single GW source has a confirmed counterpart at optical wavelengths, AT2017gfo from the neutron star merger that produced GW170817 \citep{2017ApJ...848L..12A} and GRB170817A \citep{FERMI17gfo, Integral17gfo}. The discovery of the event at optical wavelengths
\citep{2017Natur.551...64A,2017Sci...358.1556C,2017ApJ...850L...1L,2017ApJ...848L..16S,2017ApJ...848L..27T,2017ApJ...848L..24V} and the rapid follow-up from the UV to near-infrared 
\citep{Andreoni2017,Chornock2017,2017ApJ...848L..17C,Drout2017,Evans2017,Kilpatrick2017,Levan17,McCully2017,Nicholl2017,Pian2017,Shappee2017,2017Sci...358.1559K,2017Natur.551...75S,Troja2017,Utsumi2017} produced spectacular coverage of this fast declining
and unprecedented transient. The emission and spectra were shown to be compatible with the thermal emission of few tenths of a solar mass, expanding with velocity $0.2\,c$ and heated by the radioactive decay of heavy elements.
Much has been learned from this source, both pertaining to its nature as well as fundamental physics \citep[e.g.][]{2017PhRvL.119y1301B,2017ApJ...850L..34B,2017Natur.551...85A,2017ApJ...851L..21V,2018MNRAS.481.3423W,2018MNRAS.480.3871C,2018ApJ...852L..29R,2018ApJ...856L..18M,2019Sci...363..968G}. However, subsequent searches during the most recent third observing run (O3) of the gravitational wave interferometers LIGO, VIRGO and KAGRA did not yield further high-significance detections of electromagnetic counterparts (e.g. \citealp{Anand21,Antier20,deJaeger22,Gompertz20,Kasliwal20,Paterson21}; \citealp[also see the review of O3 follow-up in][]{Coughlin20nat}).

The challenge in finding a counterpart to GW events originates from a combination
of factors. Perhaps most prominently, GW events have relatively poor sky localisation. 
Even those that were the best localised in O3 had positional uncertainty regions extending over tens of square degrees (at the 90\% confidence level), and for many events the sky localisation regions were as large as thousands of square degrees. In addition to this, the cosmic rate of GW events involving at least one neutron star appears relatively low, leading to most discoveries in O3 being well beyond a distance of 100\,Mpc \citep{2019PhRvX...9c1040A,2021arXiv211103606T}. Such distances
require surveys down to an appreciable depth and over a large sky area in order to detect faint kilonovae.
Inevitably, this leads to (re-)discovering large numbers of optical transients unrelated to the GW trigger. A clear example is the case of GW190814 where
over 75 unique transients were found within the 20 deg$^2$ error box for the GW trigger \citep{Ackley2020,2019ApJ...884L..55G,2020ApJ...890..131A,2020MNRAS.492.5916W,2020ApJ...895...96V,2020MNRAS.499.3868T,2021arXiv210606897K,2021A&A...649A..72D,Oates21}. These unrelated events included supernovae (SNe), active galactic nucleus (AGN) flares and variability, cataclysmic variables (CVs), foreground stellar flares, as well as moving objects.

The difficulties in searching for counterparts are further compounded because their expected
properties place them not just amongst the least luminous transients, 
but also among the most rapidly evolving \citep[e.g.][]{Kasen2017,Metzger17}. Hence, rapid response and high cadence deep observations are required over
a wide field.
The frequency of binary neutron star (BNS) merger events within $\sim~40$\,Mpc (i.e.\ similar to GW170817) is estimated to be one in ten years \citep{2021ApJ...913L...7A}, so future searches must be optimised to match the cadence, luminosity, distance, and colours of the expected sources.

Coupled with this is the requirement to understand the transient population to a sufficient degree so as to be able to select and prioritise the most promising candidate counterparts identified within a
GW skymap. In practice, this means characterising the faint and fast transients that are associated with binary mergers and the GW signal, as well as those that are not. Examples of the need to understand the unrelated transient population have already been seen in other GW counterpart searches: AT2019ebq \citep{Smith19} was initially proposed as a possible counterpart to the GW S190425z based on an early spectrum \citep{Jonker19}, but it was subsequently shown to be a Type Ib SN \citep{Jencson19}.

However, even when these searches do not result in detections of EM counterparts to GW triggers, they can still yield valuable insights. For instance, faint and fast transients are often found in regions of parameter space similar to kilonovae; they have been hitherto difficult to discover, but they may represent extremes of stellar evolution and death. 
Thus, searching for GW-EM counterparts therefore also offers the opportunity to significantly improve our understanding of the faint transient sky. 

A particularly important and interesting group of such transients are the so-called ultra-stripped SNe \citep{Tauris13}. These are believed to arise from a particular phase of binary star evolution leading to the formation of a double neutron star system. Specifically, following the formation of a tight X-ray binary system containing a NS and a He star, further mass transfer on to the NS can occur by Roche Lobe overflow following core He exhaustion. This extreme stripping of the He star can result in Fe core collapse of a core that is barely above the Chandrasekhar mass. The resulting explosion produces no more than a few tenths of a \msun\ of ejecta, and synthesising no more than a few hundredths of a \msun\ of $^{56}$Ni. The resulting transient is therefore faint, and evolves rapidly.

Here, we present observations of one such event, \wxt, identified as a faint, and rapidly evolving transient inside  the error localisation of a possible binary neutron star merger (Sect. \ref{sect:discovery}). These data were taken by the ENGRAVE collaboration\footnote{\url{engrave-eso.org}}, a large pan-European project which is using European Southern Observatory facilities to identify and study the electromagnetic counterparts of gravitational waves. We augment the ENGRAVE data with supporting observations from a number of other collaborations and facilities.\footnote{Our followup data are available for download from the ENGRAVE webpages (\url{engrave-eso.org/data}) and through WISeREP (\url{wiserep.org}; \citealp{2012PASP..124..668Y}).}

The detection time and early light curve evolution of \wxt\ are broadly consistent with those expected for kilonovae. However, as we show in the following sections, our multi-wavelength analysis (Sect. \ref{sect:obs_properties}) demonstrates that it is unrelated to the GW trigger. Following modelling of its photometric lightcurve, spectral energy distribution (SED) and spectra (Sect. \ref{sect:modeling}) and analysis of its environment (Sect. \ref{sect:env}), we consider possible origins of \wxt\ (Sect. \ref{sect:nature}). We also estimate the rate of \wxt-like events  and discuss their presence as a contaminant for future GW counterpart searches (Sect. \ref{sect:contaminant}). 

We note that \cite{Shivkumar22} have also recently reported their observations and analysis of \wxt. In the following we compare our results to theirs, highlighting some important difference in our findings and interpretation.

\section{Source discovery}
\label{sect:discovery}

\subsection{GW discovery and EM counterpart search}
\label{sec:S191213_discovery}

\begin{figure*}[h]
\centering
\adjincludegraphics[width=\textwidth]{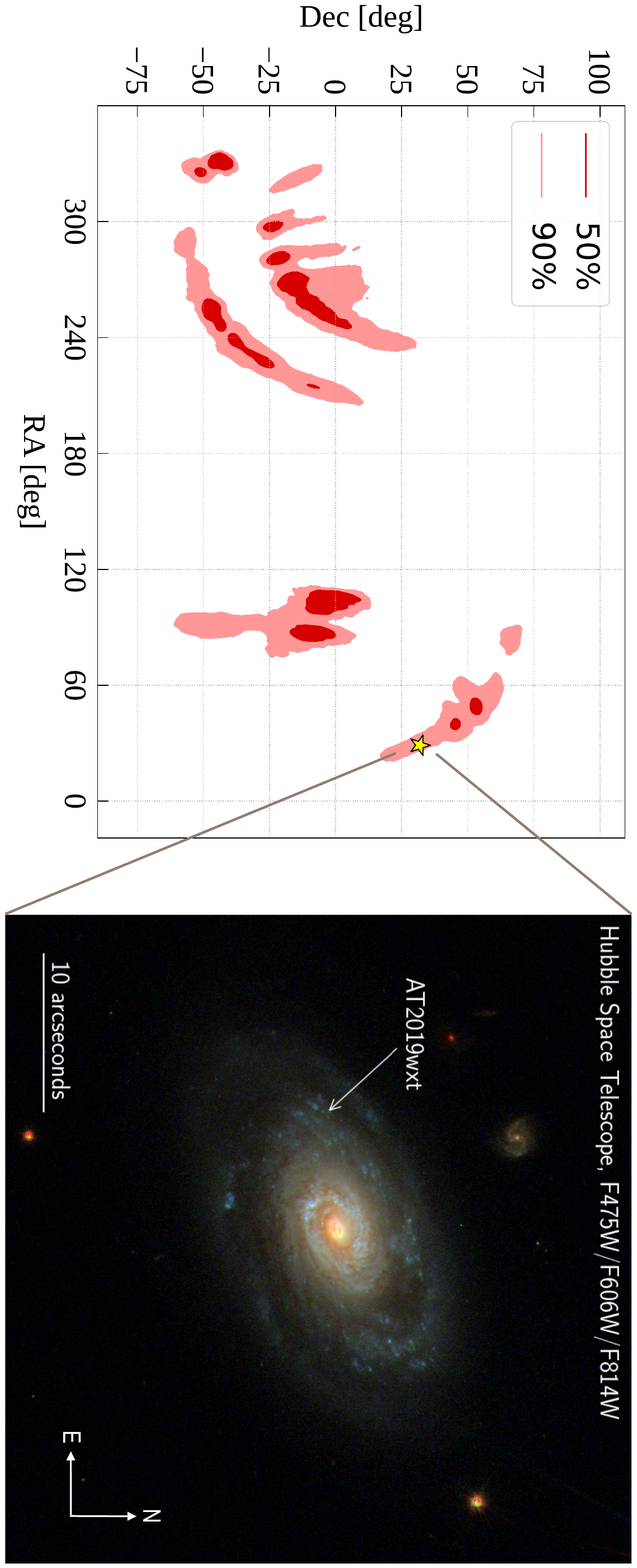}
 \caption{\wxt\ localisation. Left-hand panel: Cartesian projection of the sky localisation error box of S191213g (50\% and 90\% localisation uncertainty contours -- red and pink filled contours, respectively) compared to the position of \wxt\ (yellow star). The transient is located within the 90\% localisation uncertainty contour. The GW localisation is based on the real-time parameter estimation reported in \cite{LVC2019GCN_update}. Right-hand panel: Colour composite HST image for \wxt\  and its host galaxy KUG 0152+311.  }
 \label{fig:sky_localisation}
\end{figure*}

On 13 Dec 2019, the LIGO Scientific Collaboration and the Virgo Collaboration (LVC hereafter) issued a public alert to announce trigger S191213g, a candidate GW signal from a binary neutron star merger \citep{LVC2019GCN_identification}. According to the low-latency classification of the signal \citep{Messick2017}, the probability that the event was due to a BNS merger was  estimated as $p_\mathrm{BNS}\sim 0.77$, with the remaining $0.23$ being attributed to a possible terrestrial origin. Despite the three detectors being online and taking data, the localisation uncertainty was very large (90\% credible area 4480 deg$^2$; distance $201 \pm 81$\,Mpc; \citealp{LVC2019GCN_identification,LVC2019GCN_update}), as shown in Fig.~\ref{fig:sky_localisation}. Despite the low GW signal significance and large sky area, searches for electromagnetic counterparts were carried out across the optical, X-ray and gamma-ray regions \citep{Coughlin20}.

A number of gamma-ray telescopes were actively observing a significant fraction of the localisation region at the time of S191213g. Konus-\textit{Wind} \citep{Ridnaia19} and \textit{Fermi}-GBM \citep{Wilson-Hodge19} were in fact sensitive to the entire region, but reported no detections. Non-detections were also reported from \textit{Fermi}-LAT (80\% instantaneous coverage; \citealp{Cutini19}); \textit{INTEGRAL} SPI-ACS (however, the orientation of the spacecraft led to low sensitivity, \citealp{Diego19}); {\it Swift}-BAT  (80\% instantaneous coverage; \citealp{Barthelmy19}); and \textit{AGILE}-MCAL \citep{Verrecchia19}. \textit{AGILE}-GRID \citep{Casentini19} and CALET \citep{Marrocchesi19} also did not report a detection. 
In soft gamma-rays / hard X-rays, \textit{Insight}-HXMT/HE \citep{Xiao19} and \textit{AstroSat} CZTI \citep{Shenoy19} were both observing around 80\% of the localisation region at the time of the merger, while in soft X-rays \textit{MAXI}/GSC covered 92\% of the region around an hour after the GW trigger \citep{Sugita19}. None of these satellites detected a significant new source.
No temporally and spatially coincident neutrinos were found by the ICECUBE, ANTARES or Pierre Auger detectors \citep{Icecube19,Ageron19,Alvarez-Muniz19}.

Optical surveys were more successful in finding possible counterparts to S191213g. While the galaxy-targeted J-GEM and GRANDMA searches did not find any candidates \citep{Tanaka19,Ducoin19}, wide-field imaging surveys reported a number of transients. Pan-STARRS found a single candidate (\wxt; \citealp{McBrien2019GCN}) which is the subject of this paper (and discussed in more detail in Sect.~\ref{sect:19wxt_discovery}), while the MASTER survey also found a single transient \citep{Lipunov19a, Lipunov19b}, which was subsequently classified as a dwarf nova \citep{Denisenko19}. The Zwicky Transient Facility (ZTF) reported 19 candidate counterparts over two nights following the discovery of \wxt\ \citep{Andreoni19,Stein19}. These candidates were found within the 29\% of the localisation region that was accessible to and observed by ZTF. The first tranche of ZTF candidates \citep{Andreoni19} were all eliminated as possible counterparts through follow-up spectroscopy \citep{Brennan19,Castro-Tirado19,Elias-Rosa19,Perley19}. Out of the candidates from the second night, AT2019wrt and AT2019wrr were flagged as particularly interesting by ZTF as they had constraining non-detections immediately prior to S191213g. AT2019wrt was subsequently found to have a photometric evolution inconsistent with a GRB afterglow or kilonova \citep{Xu19}, while AT2019wrr was spectroscopically classified as a Type Ia SN \citep{Kasliwal20}. The remainder of the candidates from \cite{Stein19} were similarly discounted by \cite{Kasliwal20} from either their photometric evolution, associated with a stellar counterpart, or on the basis of their spectra. Finally, the GOTO prototype \citep{Steeghs22} covered 1557.5 sq. deg. encompassing 34.1\% of the 2D probability for S191213g. Conditions were variable leading to a median survey depth of 18 mag. Following the methodology of \cite{Gompertz20}, this implied a limited search horizon and no viable counterpart candidates were identified.

\subsection{Discovery of \wxt}
\label{sect:19wxt_discovery}

The Pan-STARRS telescopes are used for following up GW sources when the skymap area is less than about 1000\,deg$^2$ and the source has a high probability of being real and containing a neutron star \citep[e.g.][]{GW150914PanSTARRS,Ackley2020}. \gwtrig\ did not meet the Pan-STARRS trigger criteria, and so normal survey operations, primarily for near-earth object detection were in place at the time of the GW detection and over the following few days. We processed these data and searched for transients of interest withor without GW and high-energy counterparts \citep{2019TNSAN..48....1S}. On 18 Dec 2019, Pan-STARRS 1 (PS1) reported the discovery of a potential optical counterpart during these normal survey operations, PS19hgw \citep{McBrien2019GCN}, located at \mbox{$\mathrm{RA(J2000)} = 28.92473^{\circ} (01^h55^m41^s.94)$}, \mbox{$\mathrm{Dec(J2000)} = +31.41791^{\circ} (+31^\circ25'04\farcs4$)}. It is clearly associated to the host galaxy KUG 0152+311 at $z= \hostz$, corresponding to $d_\mathrm{L}=$\hostDL\,Mpc (NASA Extragalactic Database, NED) assuming \cite{Planck2016} cosmological parameter (flat $\Lambda$CDM, $H_0=67.8$\kms\,Mpc$^{-1}$, $\Omega_{\rm M}=0.31$) and correcting to the reference frame of the Cosmic Microwave Background. The position (marked with a yellow star in Fig.~\ref{fig:sky_localisation}, left-hand panel) was compatible with the localisation uncertainty region of the GW trigger. The object was later assigned the IAU identifier \wxt\ \citep{McLaughlin2019TNS}.  We adopt the foreground extinction of $A_r= $~\Ar\,mag, and the equivalent values in other filters from the \cite{2011ApJ...737..103S} map as reported in the NED. In Sect. \ref{sect:local_env} we use the NaD lines to estimate \revthree{that} the host galaxy reddening for \wxt\ \revthree{is} low, at $E(B-V)\sim0.1$~mag, however as this value is quite uncertain (and relatively small) we do not consider this in our analysis.

The association of \wxt\ with a host galaxy at a distance consistent with that of S191213g, its relatively faint absolute magnitude ($M_i = -16.7$\,mag), and tight constraints on the explosion epoch (non-detections 0.2 mag fainter in $i$-band on the preceding night; and between 1.4 and 2.4 mag fainter 3 to 5 days prior in $z$-band) led many groups to prioritise it for spectroscopic classification. \cite{Dutta2019GCN} first reported the spectrum of \wxt\ to be blue and featureless, using the Indian Astronomical Observatory 2.0\,m telescope + Hanle Faint Object Spectrograph Camera. \cite{Dutta2019GCN} obtained their spectrum two hours after the discovery of \wxt\ was publicly announced, and reported it in a Global Coordinates Network (GCN) circular less than two hours later. Shortly thereafter, other groups also reported \wxt\ to appear blue and spectroscopically featureless \citep{Izzo2019GCN,Srivastav2019GCN} using the Alhambra faint object spectrograph and camera (ALFOSC) on the Nordic Optical Telescope (NOT) and the spectrograph for the rapid acquisition of transients (SPRAT) on the Liverpool Telescope (LT) respectively. These spectra are discussed in Section \ref{specevol}.

A few hours later, \cite{MullerBravo2019GCN} reported on behalf of the ePESSTO+ collaboration that they detected a possible broad feature around $5400 - 6200$\,\AA, and this was subsequently confirmed by \cite{Vogl2019GCN} in a higher signal-to-noise ratio (S/N) Very Large Telescope (VLT) spectrum taken with the focal reducer/low dispersion spectrograph 2 (FORS2). \cite{Vogl2019GCN} suggested that the broad features were due to He, and made the first tentative spectroscopic classification of \wxt\ as a SN~Ib or IIb. The same broad He lines were also seen and reported by \cite{Vallely2019GCN} in their Large Binocular Telescope (LBT) multi-object double spectrograph (MODS) data; and by \cite{Becerra-Gonzalez2019GCN} using the Gran Telescopio Canarias (GTC) equipped with the optical system for imaging and low-intermediate-resolution integrated spectroscopy (OSIRIS).

We note that \cite{Antier20} also considered \wxt\ in their compilation paper for O3 events. In the offline search in \cite{2021arXiv211103606T} S191213g was not identified as a significant candidate and therefore the preliminary results (on e.g.\ sky localisation) were not updated.

\section{Observational properties of \wxt}
\label{sect:obs_properties}

\subsection{Discovery and photometric evolution}
\label{sect:phot}

\begin{figure}[h]
\centering
\includegraphics[width=\columnwidth]{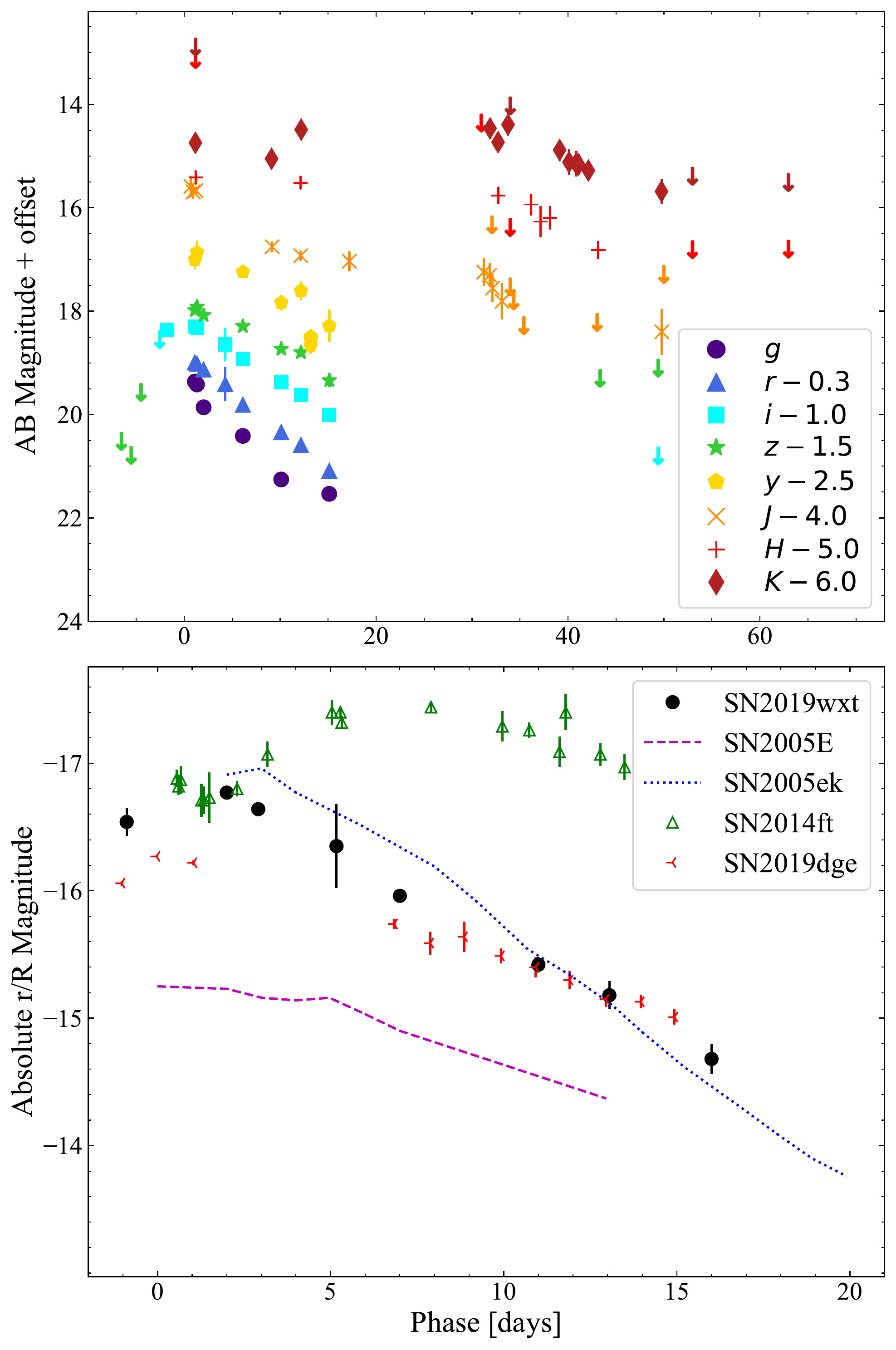}
\caption{Light curves. \textbf{Top panel:} $grizyJHK$ light curves of \wxt.
\textbf{Lower panel:} Comparison of absolute $r$ and $R$-band light curves with those of
other faint and rapidly evolving transients.
For both panels, the phase is given with respect to the epoch of $i$-band maximum, i.e.\ MJD 58835.1.
}
\label{fig:lc}
\end{figure}

The Pan-STARRS1 telescope had been observing the position of \wxt\ in the week leading up to the discovery with a shallow upper limit in \ips\ one day before discovery and deeper
\zps\ limits 3 to 5 days before discovery. The combination of these non-detections
and the discovery on MJD = \wxtDisc\ at 3.3\,days after \gwtrig\ indicated a young transient
in a galaxy with a redshift consistent with the GW luminosity distance (\GWDist). The plausible
4D spatial and temporal coincidence prompted extensive photometric and spectroscopic 
follow-up observations (details of the data reduction are given in the Appendix), which further 
indicated an interesting $r$-band decline of 1\,mag over 5 days (see Table\,\ref{tab:optical}).  The observed multi-band light curves are  shown in Fig. \,\ref{fig:lc}, where the phase is with respect to the epoch of $i$-band maximum on MJD~=~58835.1 (2019 Dec 18 02:24 UTC).

The host redshift of $z =$~\hostz\ corresponds to \mbox{$d_\mathrm{L} = 154 \pm 11$\,Mpc} (with cosmological parameters as defined in Sect.~\ref{sect:19wxt_discovery}) implying an absolute magnitude of \wxt\ of $M_i = -16.7$~mag. This is somewhat brighter than the kilonova AT2017gfo at peak ($M_i= -15.5$~mag). 

The non-detections of \wxt\ (and as shown in Sect. \ref{specevol}, the appearance of the early spectra) are consistent with \revthree{a young transient that appeared a few days before} the discovery. We see a slight rise over the first two epochs in $i$-band, and a decline in all other filters. We see no sign of the shock-cooling emission that has been seen in some other ultra-stripped events \citep{De14ft}. The ultra-stripped SN2014ft displayed shock cooling emission for around 1.5 days, however, this was only visible in the bluer filters -- and in fact there is no sign of shock cooling emission for SN2014ft in $i$-band. Unfortunately, the first detection of \wxt\ is in $i$-band, while the pre-discovery limits are in $z$-band. These data are hence not sensitive to any shock cooling emission similar to that seen in SN2014ft.

In contrast to our findings, \cite{Shivkumar22} reported the detection of a shock-cooling tail for \wxt. \citeauthor{Shivkumar22} suggest that after the initial detection of \wxt\ by PanSTARRS at $i$=19.36 on MJD 58833.3, it subsequently faded by $\sim$0.7~mag to $i$=20.00 on MJD 58835.0. One day later, on MJD 58836.0 \wxt\ had apparently brightened slightly to $i$=19.74. The two photometric points on which this is contingent are both from the 2~m telescope at the Wendelstein Observatory, and were originally reported in a GCN by \cite{Hopp2020GCN}, before being combined with other measurements from the literature by \citeauthor{Shivkumar22}. However, we have $i$-band photometry from PanSTARRS contemporaneous to the Wendelstein measurement on MJD 58835.0 that is 0.5 mag brighter, and as noted previously we find the reported shock cooling tail to be inconsistent with our lightcurve.

\begin{figure*}[h]
\centering
\includegraphics[width=\textwidth]{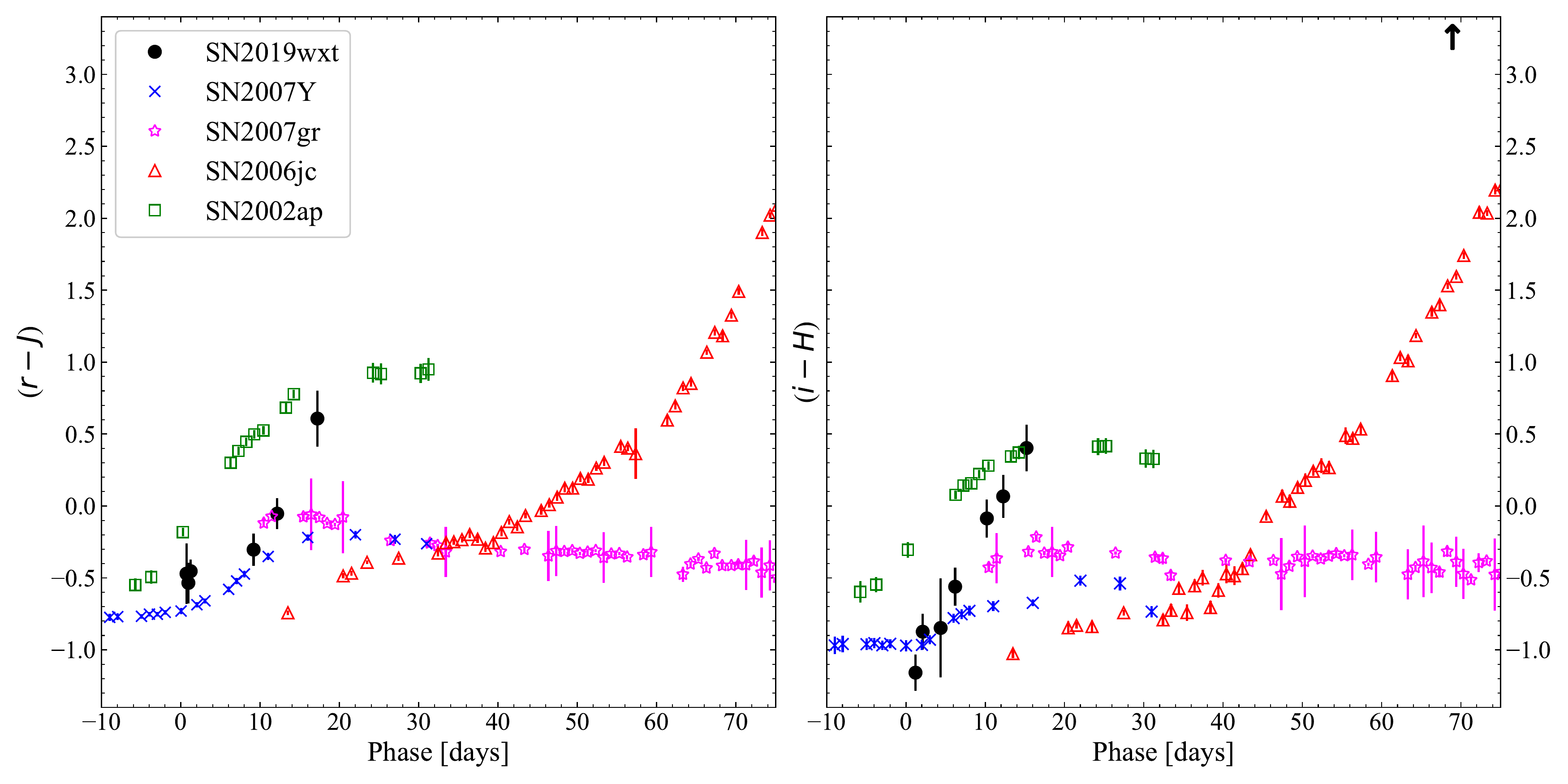}
\caption{($r-J$) and ($i-H$) colour evolution of \wxt\, compared to the colour evolution of the well-observed Type Ibc events SN 2007Y \citep{2009ApJ...696..713S}, SN 2007gr \citep{2014ApJS..213...19B}, SN 2006jc \citep{2014ApJS..213...19B} and SN 2002ap \citep{2003ApJ...592..467Y}. All the magnitudes are in the AB system. }
\label{fig:colour}
\end{figure*}

A noteworthy aspect of the photometric evolution of \wxt\ is the rapid shift to redder colours.
This is illustrated in Fig. \ref{fig:colour}, where we show the optical - near-infrared (NIR) colour evolution of \wxt\ compared to a set of other stripped envelope SNe. At +16\,d, \wxt\ has similar colours to the Type Ic SN~2002ap, but this change in colour occurred very rapidly. The $i-H$, colour changed by nearly two mag in only two weeks. This dramatic change appears to have continued over the next two months, and by the time of our HST observations $i-H > 3.0$\,mag.

After about two weeks from maximum light, \wxt\ is no longer detected at optical wavelengths. However, detections in the J, H, and K-bands show that this trend in NIR evolution persists (Table \ref{tab:IR}).

\subsection{Spectroscopic evolution}
\label{specevol}

The earliest spectroscopic observations of {\wxt} \citep{Dutta2019GCN,Izzo2019GCN,Srivastav2019GCN,MullerBravo2019GCN} showed a blue, featureless continuum. Subsequent spectroscopy \citep{Vogl2019GCN,Vallely2019GCN,Becerra-Gonzalez2019GCN,Valeev2019GCN} revealed the presence of broad emission lines consistent with expansion velocities $7000 - 10000$\kms\ (purportedly H), eventually leading to the classification of the transient as a SN\,IIb, also based on the similarity with the spectrum of SN\,2011fu \citep{Kumar2013SN2011fu}.

\begin{figure*}[h]
 \includegraphics[width=\textwidth]{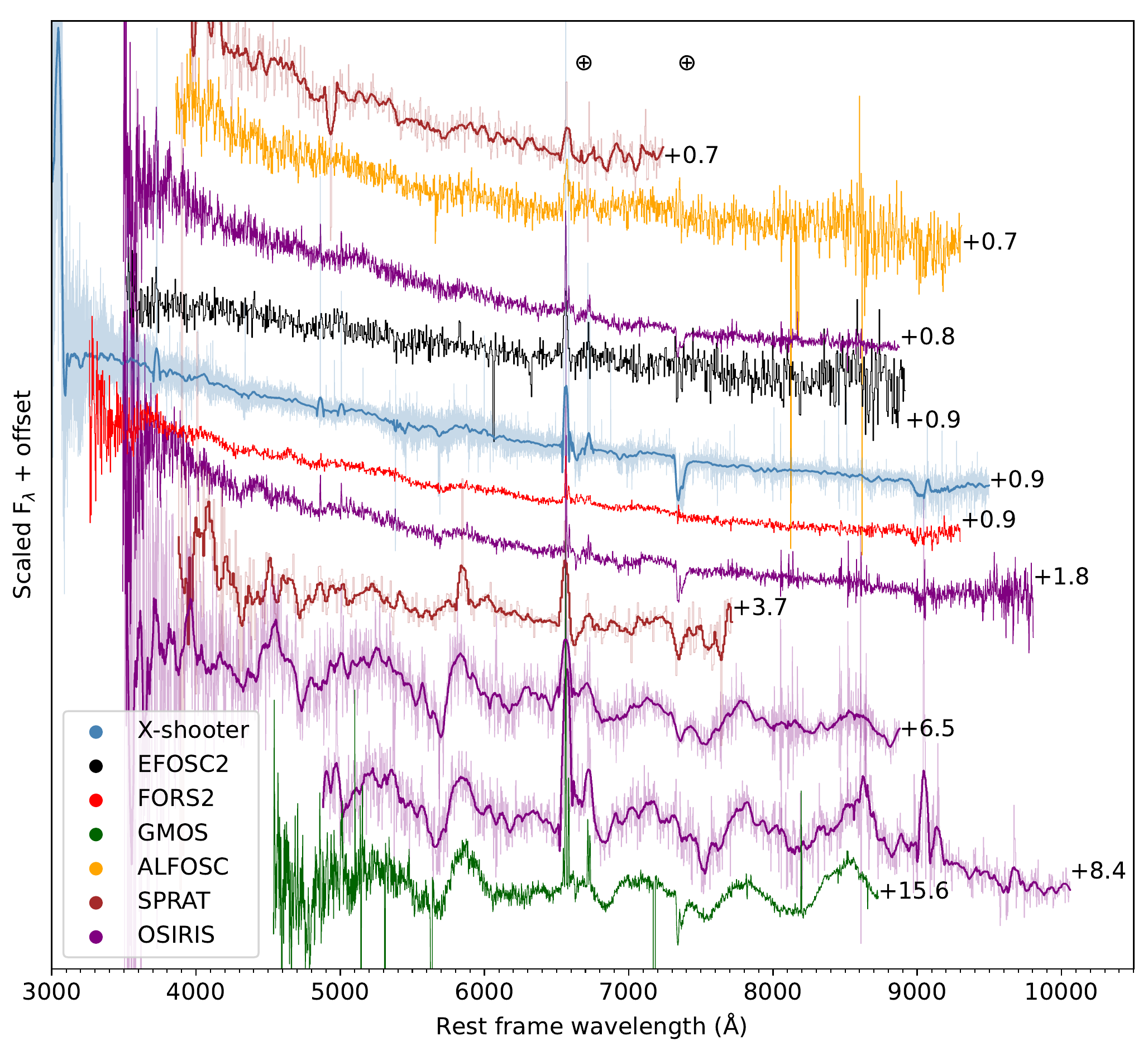}
\caption{Sequence of optical spectra obtained of \wxt. The phase (in rest frame days relative to our adopted $i$-band maximum MJD 58835.1) is listed beside each spectrum. Telluric absorptions are indicated with a $\oplus$ symbol. In the case of the X-shooter, SPRAT and OSIRIS data, we plot a smoothed version of each spectrum, with the unsmoothed spectrum shown underneath in a lighter colour. Smoothed spectra have had a Savitzky-Golay filter applied, with window length of 50 \AA\ for the X-Shooter data, and 100 \AA\ for the SPRAT and OSIRIS spectra.}
 \label{fig:spectra}
\end{figure*}

This evolution is clear from our sequence of spectra (Fig.~\ref{fig:spectra}), where the first eight spectra (covering phases from +0.7 to +3.7~d) are at first glance featureless. At +6.5~d, broad SN-like features have emerged, and revisiting the earlier spectra we can see that the same broad features, albeit very weak, were in fact present in the higher S/N spectra from OSIRIS at +0.8 and +1.8 days, and from FORS2 at +0.9 days.

Turning to the +6.5 day spectrum, we see clear broad, high velocity lines typical of SNe. The strongest feature is consistent with 
He~{\sc i}~$\lambda$5876 with a broad P-Cygni profile with a minimum at a velocity of 11,000\,\kms. Aside from the narrow emission (which we attribute to the host galaxy), we see no signs of strong H$\alpha$ in \wxt, leading us to formally classify this as a Type Ib SN.  
\footnote{At least some of the class of ultra-stripped SNe have been suggested to contain small masses of H, for example the Type IIb SN~2019ehk (\citealp{De21}, although see \citealp{Yao20} and \citealp{Jacobson20} who disagree on this point). Moreover, theoretical modelling suggests that as little as 0.001 \msun\ of H can produce a Type IIb spectrum \cite{Dessart11}. We test for the presence of H through spectral modelling in Sect. \ref{sec:TARDIS}.}
Our final spectrum at +15.6~d more clearly reveals the He lines, now including He~{\sc i} $\lambda$7065, as well as the O~{\sc i} recombination line at $\lambda$7774 and the Ca NIR triplet.

\begin{figure*}[h]
 \includegraphics[width=\textwidth]{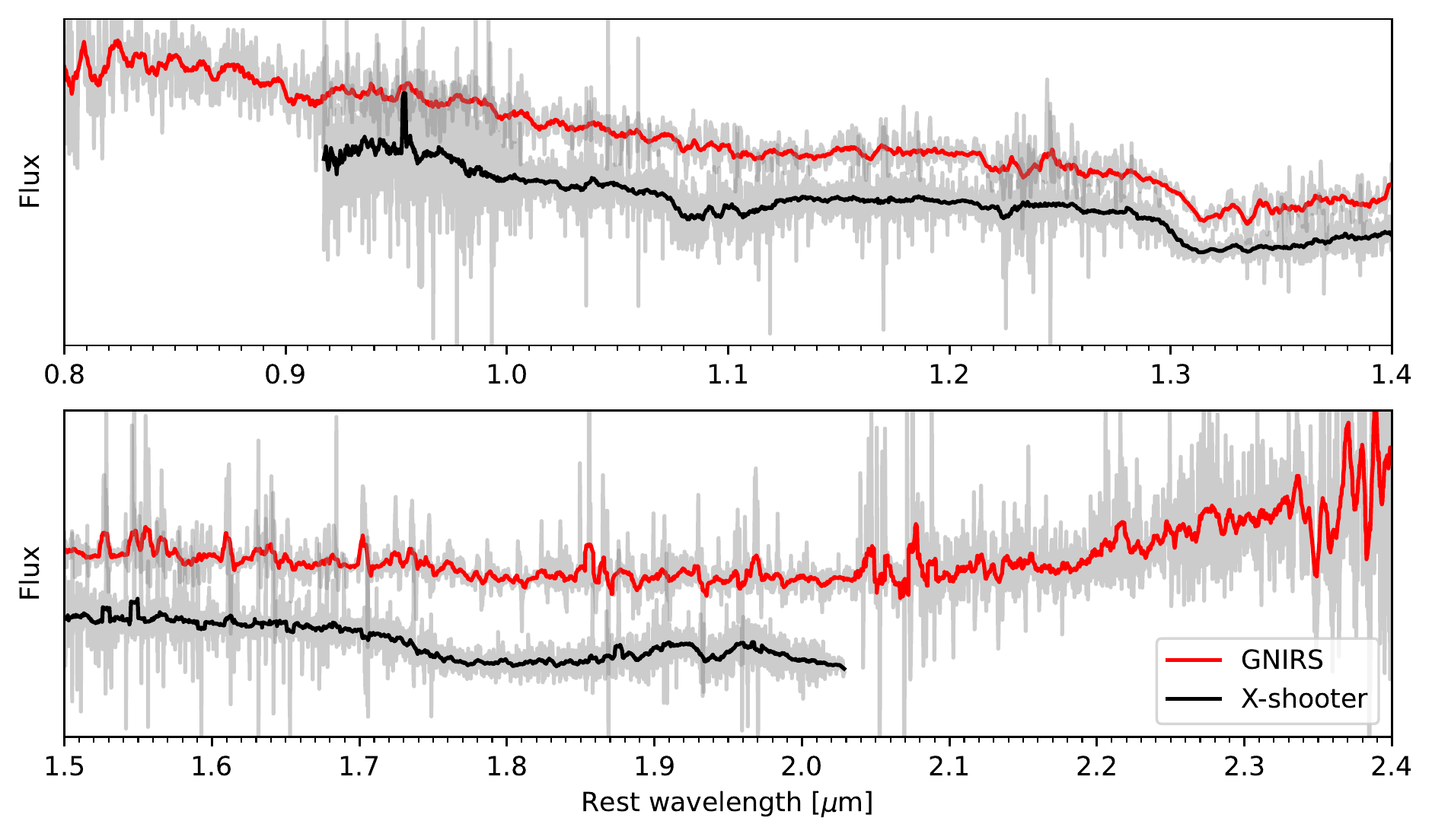}
 \caption{NIR spectra for \wxt\ taken on 19 Dec 2019. Grey lines show original spectra, colour lines have been smoothed with a Savitzky-Golay filter with window length 50\AA. We do not include the GTC/EMIR spectrum which has low S/N.}
 \label{fig:nir_spec}
\end{figure*}

Unfortunately our NIR spectra (Table \ref{tab:spec}) were all taken on the same night close to maximum light. We show the higher S/N GNIRS and X-shooter spectra in Fig. \ref{fig:nir_spec}; even after smoothing and rebinning, no features are evident (aside from telluric absorption). 

\begin{table}
    \centering
    \renewcommand{\arraystretch}{1.3}
    \caption{
    Parameters derived from fitting a blackbody spectrum to the \wxt\ photometry.  Errors are 90\% credible statistical errors. 
    }
    \begin{tabular}{ccccc}
    \toprule
    Phase & $L$ & $T_\mathrm{eff}$ & $R_\mathrm{ph}$ & Method$^{a}$ \\
    $\mathrm{[d]}$ & $\mathrm{[10^{41}erg/s]}$ & $\mathrm{[10^3 K]}$ & $\mathrm{[10^{14}cm]}$ \\
    \midrule
    0.6-0.8 & ${29.5}_{-1.7}^{+2.4}$ & ${11.0}_{-0.5}^{+0.4}$ & ${5.4}\pm{0.4}$ & MBS \\
    1.11 & ${19.7}^{+1.3}_{-1.2}$ & $9.03 \pm 0.2$ & $6.5 \pm 0.2$ & SB \\
    1.1-1.3 & ${19.7}\pm{1.0}$ & ${9.4}\pm{0.3}$ & ${6.0}_{-0.2}^{+0.3}$ & MBS \\
    2.02 & ${12.6}^{+1.6}_{-1.4}$ & $7.6 \pm 0.5$ & $7.1 \pm 0.6$ & SB \\
    4.28 & ${9.6}^{+0.6}_{-0.5}$ & $6.7 \pm 0.1$ & $8.1 \pm 0.3$ & SB\\
    6.11 & ${7.1}^{+0.5}_{-0.4}$ & $6.1 \pm 0.2$ & $8.7 \pm 0.6$ & SB\\
    6.0-6.2 & ${7.4}_{-0.3}^{+0.2}$ & ${6.3}\pm{0.3}$ & ${8.0}_{-0.6}^{+0.5}$ & MBS\\
    9.1-11.1 & ${4.4}\pm{0.1}$ & ${5.4}\pm{0.2}$ & ${8.4}_{-0.6}^{+0.5}$ & MBS\\
    10.11 & $4.1 \pm 0.2$ & $5.3 \pm 0.2$ & $8.7 \pm 0.4$ & SB\\
    12.16 & $3.6 \pm 0.2$ & $5.3 \pm 0.2$ & $8.1 \pm 0.7$ & SB\\
    15.11 & $2.5 \pm 0.2$ & $5.9 \pm 0.4$ & $5.4 \pm 0.6$ & SB\\
    15.1-16.1 & ${2.6}_{-0.15}^{+0.3}$ & ${6.1}_{-0.5}^{+0.6}$ & ${5.1}\pm 0.8$ & MBS \\
    25.0-35.0 & ${0.8}_{-0.1}^{+0.2}$ & ${2.2}_{-0.4}^{+0.3}$ & ${22.3}_{-9.7}^{+8.0}$ & MBS\\
    35.0-45.0 & ${0.72}_{-0.06}^{+0.05}$ & ${1.9}\pm{0.1}$ & ${27.1}_{-4.9}^{+3.9}$ & MBS\\
    45.0-65.0 & ${0.29}_{-0.09}^{+0.28}$ & ${1.5}\pm{0.2}$ & ${29.7}_{-15.8}^{+14.0}$ & MBS\\

    \bottomrule
    \end{tabular}\\
    \footnotesize{$^a$Methods: SB = \texttt{SuperBol}; MBS = MCMC on binned SED.}
    \label{tab:SED_fitting_results}
    \renewcommand{\arraystretch}{1.}
\end{table}

\section{Modelling the light curves and spectra of \wxt}
\label{sect:modeling}

\subsection{Bolometric light curves and blackbody fits}
\label{sect:arnett}

In order to get deeper insights on the intrinsic nature of \wxt, we constructed bolometric and quasi-bolometric light curves with two different methods.

First, we obtained quasi-bolometric fluxes from the multi-band photometry of \wxt, integrated within the wavelength intervals corresponding to the filter response curves, using the \texttt{SuperBol} code \citep{2018RNAAS...2d.230N}. We used \texttt{SuperBol} to also perform a full blackbody integration from a fit to the SED, in order to account for the contribution of missing passbands.

\begin{figure}[h]
\centering
\includegraphics[width=\columnwidth]{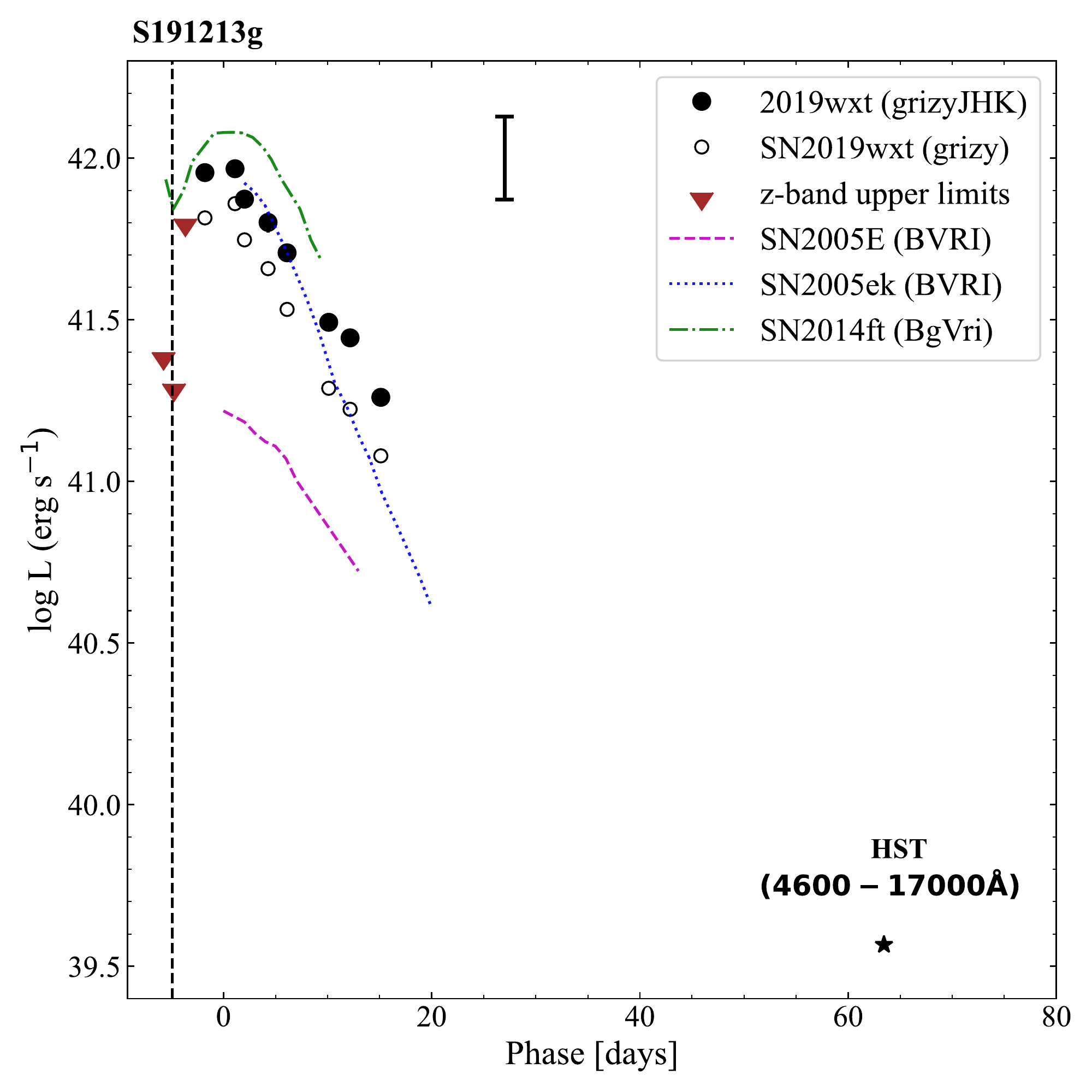}
\caption{Quasi-bolometric $grizyJHK$ light curve of \wxt, together with our late time {\it HST} measurement. Also shown are the quasi-bolometric light curves of the fast declining ultra-stripped SNe 2005E \citep{perets_05E}, 2005ek \citep{Drout05ek} and 2014ft \citep{De14ft} for comparison.
The dashed vertical line indicates the time at which the GW event occurred. The errorbar represents the systematic error that stems from the uncertainty on the distance modulus. \revtwo{The z-band upper limits were converted to bolometric luminosity assuming a blackbody spectrum with effective temperature $T=1.1\times 10^4\,\mathrm{K}$, similar to that at the peak of the light curve, see Table \ref{tab:SED_fitting_results}.}}

\label{fig:bolometric}
\end{figure}

The quasi-bolometric light curve of \wxt, integrated within the wavelength limits defined by our $grizyJHK$ photometry, is shown in Fig.~\ref{fig:bolometric}. Also shown for comparison are the quasi-bolometric light curves of the Ca-strong SN\,2005E \citep{perets_05E}, and ultra-stripped core-collapse candidates SN\,2005ek \citep{Drout05ek} and SN 2014ft \citep{De14ft}. We used \texttt{SuperBol} to compute the bolometric light curves of the comparison objects for consistency.

\begin{figure*}[h]
    \centering
    \includegraphics[width=\textwidth]{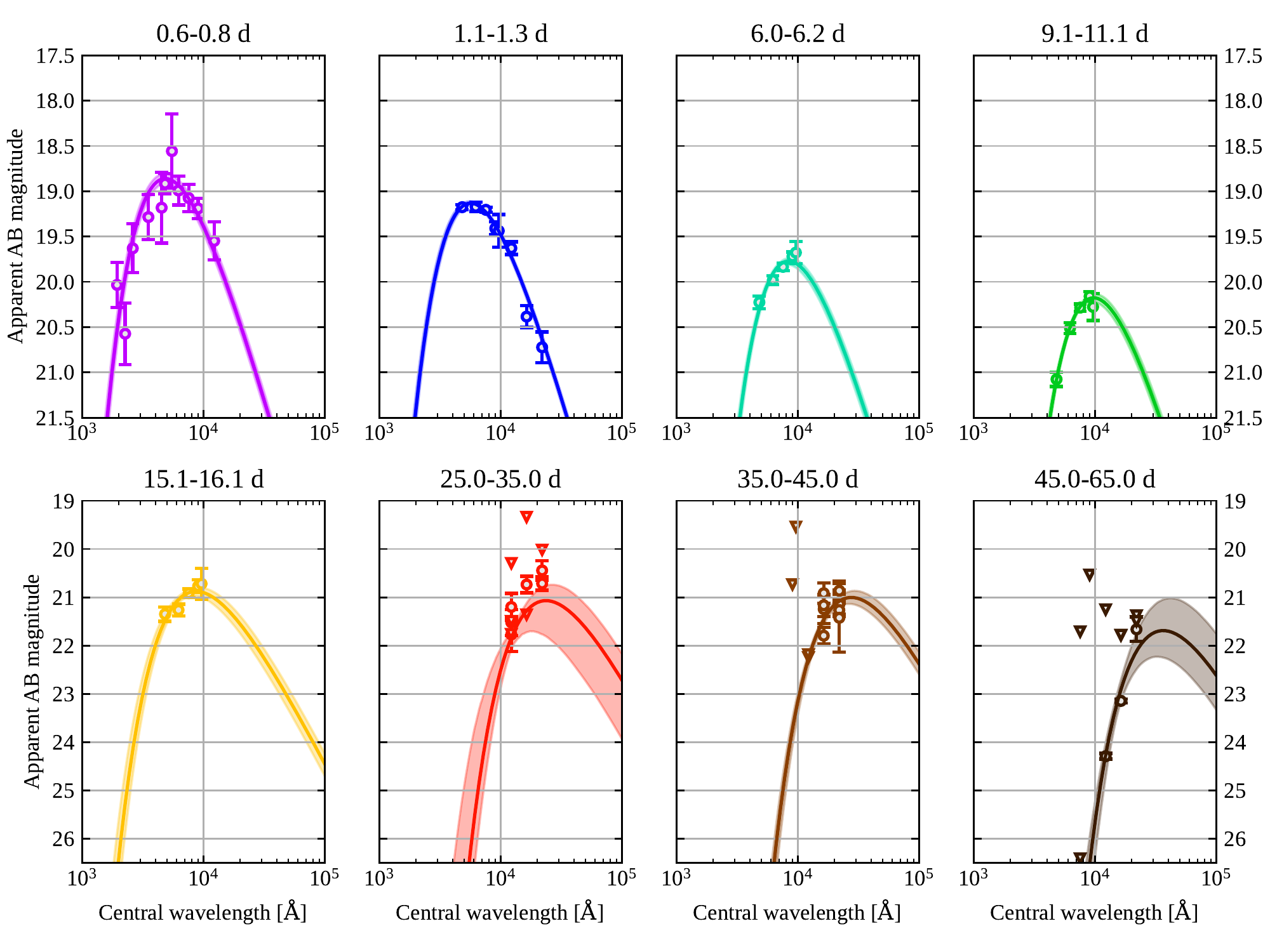}
    \caption{Spectral energy distributions fitted with a blackbody model. 
    Each panel shows a SED of SN2019wxt (circles represent detections with one-sigma error bars; triangles represent upper limits) constructed by considering photometric measurements binned within a time window (annotated above each panel, in days post $i$-band maximum). The formally best-fitting blackbody is shown by a solid line, while the filled regions span the 16th to the 84th percentile (equivalent to one-sigma uncertainties) of the model magnitudes corresponding to the posterior samples at each wavelength. The fits are performed adopting a log-uniform prior on the BB luminosity in the $10^{35}-10^{43}$ erg/s range and a uniform prior on the temperature in the $1000-20000$ K range.}
    \label{fig:SEDs_BB}
\end{figure*}

Since \texttt{SuperBol} relies on polynomial interpolation of the photometric evolution, the reliability of its results at epochs with sparse wavelength coverage can be uncertain. For that reason, as a cross-check and as a way to extend the bolometric light curves to later times, we also performed Bayesian blackbody parameter estimations on SEDs constructed by collecting the photometric measurements in time bins. This was done by sampling the model posterior probability through the affine-invariant Markov Chain Monte Carlo (MCMC) sampler \texttt{emcee} \citep{Foreman-Mackey2013}. The model employed was a simple blackbody with luminosity $L$ and effective temperature $T_\mathrm{eff}$ emitting at the distance and redshift of the source, and the likelihood for the observed extinction-corrected magnitudes was assumed Gaussian. Upper limits were conservatively treated by adding a one-sided Gaussian penalty with 0.1 mag standard deviation to the likelihood, and a systematic relative error contribution parameter $f_\mathrm{sys}$ was introduced such that the effective error on each observed magnitude $m_i$ was defined as $\sigma_{i,\mathrm{eff}} = \sqrt{\sigma_i^2 + f_\mathrm{sys}^2 m_{i,\mathrm{mod}}^2}$, where $\sigma_i$ is the magnitude error from the observation and $m_{i,\mathrm{mod}}$ is the model magnitude at the corresponding time and frequency. The posterior probability was defined as the product of the likelihood times a log-uniform prior on $L$ in the range $10^{35}-10^{43}\,\mathrm{erg/s}$, a uniform prior on $T_\mathrm{eff}$ in the range $1000-20000\,\mathrm{K}$ and a log-uniform prior on $f_\mathrm{sys}$ in the range $10^{-10}-1$. The resulting posterior probability density was then marginalised over the $f_\mathrm{sys}$ nuisance parameter. Figure \ref{fig:SEDs_BB} shows the projections of the resulting posterior probability density in the data space for this part of the SED modelling.

Blackbody parameter estimates obtained by the two methods are summarised in Table~\ref{tab:SED_fitting_results}. The temporal evolution of these parameters is shown in Fig.~\ref{fig:L,Teff,Rph}, along with the corresponding parameters estimated from the spectroscopic modelling with \tardis\ (Sect.~\ref{sec:TARDIS}). Comparing to \cite{Shivkumar22}, we find a similar evolution of the luminosity, temperature and radius of \wxt\, aside from their putative early shock cooling tail.

\subsection{Modelling the photometric evolution}

\begin{figure}[h!]
\includegraphics[width=\columnwidth]{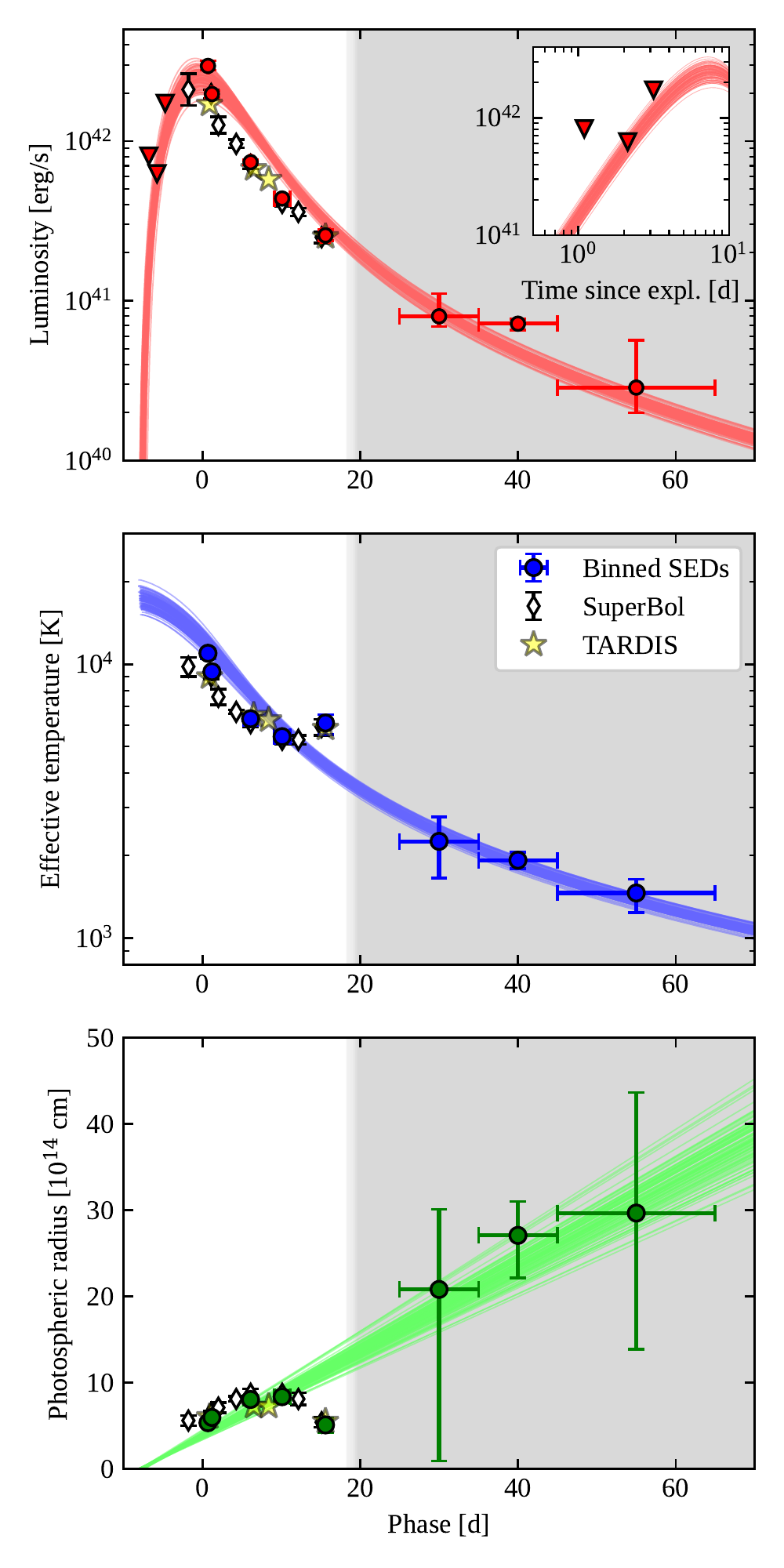} 
\caption{Evolution of photospheric quantities from blackbody fits to photometric data. The figure shows the luminosity (top panel), effective temperature (middle panel) and photospheric radius (bottom panel) derived from fitting a blackbody spectrum to our photometric data at various epochs (circles with error bars show results from MCMC fitting of binned SEDs; white diamonds show the \texttt{SuperBol} bolometric fit results; star symbols show the results from our \tardis\ model -- sec.~\ref{sec:TARDIS}). Thin lines are 100 posterior samples from our simple SN model (Appendix \ref{sect:simple_SN_model}) fitted to the photometric dataset. The grey shaded area corresponds to times when the ejecta are formally optically thin (nebular phase). The inset in the top panel shows the model samples plotted against the time since explosion, with a logarithmic x-axis to better display the agreement with the upper limits. The latter are derived from our $z$-band upper limits assuming the blackbody spectrum corresponding to the best-fit model.
}
\label{fig:L,Teff,Rph}
\end{figure}

The blackbody parameters obtained as described in the preceding sub-section appear to evolve smoothly in time. Encouraged by this, we fitted the simple SN model described in Appendix \ref{sect:simple_SN_model} to the entire photometric dataset. The model consists of an ejecta shell of mass $M_\mathrm{ej}$ and UVOIR grey opacity $\kappa$ expanding at a constant speed $v_\mathrm{ej}$ and heated by gamma-rays emitted by a central radioactive source of mass $M_\mathrm{Ni}$, initially composed entirely of $^{56}$Ni. The ejecta luminosity is computed in the diffusion approximation and its photosphere is assumed to simply track the expansion, $R_\mathrm{ph}=v_\mathrm{ej}t$, where $t$ is the time since the explosion, which was assumed to take place at a phase $t_0$ from our reference time MJD 58835.1. The resulting model has five free parameters ($M_\mathrm{ej}$, $\kappa$, $v_\mathrm{ej}$, $M_\mathrm{Ni}$, $t_0$) plus the $f_\mathrm{sys}$ parameter described in the previous sub-section, which is included as it avoids datapoints with very small formal uncertainties dominating the likelihood. We sampled the posterior probability on this parameter space with the same MCMC approach as described in the preceding section, adopting log-uniform priors on all parameters except for $t_0$, for which we used a uniform prior. The result is shown in the corner plot in Figure \ref{fig:SNfit_corner} in Appendix \ref{sect:simple_SN_model} and summarised in Table \ref{tab:SNfit_result_table}. \cite{Shivkumar22} also find \revthree{an} ejecta and $^{56}$Ni mass consistent with these results.

The model reproduces correctly the main trends, but slightly overestimates the luminosity and temperature at 3-15 d, and deviates significantly in temperature with respect to the SED at $\sim 15-16\,\mathrm{d}$. We interpret this deviation as due to the simplistic treatment of the photosphere evolution in our model, especially in the transition between the photospheric phase and the nebular phase (grey shaded region in Figure \ref{fig:L,Teff,Rph}). Moreover, the model suggests that the effective temperature drops below 2000~K at late times, which is lower than that typically seen in stripped envelope SNe. In the next sub-section, we explore the possibility that the emission in the nebular phase is instead due to dust.
 
\begin{table}
\caption{SN model fitting results. Parameter values represent maximum \textit{a posteriori} estimates, with errors encompassing 90\% credible ranges of the marginalised posteriors.}
    \centering
    \begin{tabular}{lll}
        \toprule
        Parameter & Value & Prior$^a$   \\
        \midrule
        $M_\mathrm{ej}/\mathrm{M_\odot}$ & $0.09\pm0.04$ & l.u. (0.01, 1)\\
        $M_\mathrm{Ni}/\mathrm{M_\odot}$ & $(1.69\pm0.37)\times 10^{-2}$ & l.u. ($10^{-3}$, 1)\\
        $\kappa/\mathrm{cm^2\,g^{-1}}$ & $0.14\pm0.06$ & l.u. (0.05, 0.3)\\
        $v_\mathrm{ej}/10^4\,\mathrm{km\,s^{-1}}$ & $0.58\pm0.07$ & l.u. (0.03, 3)\\
        $t_\mathrm{0}/\mathrm{d}$ & $-7.9\pm0.28$ & u. (-10, 0)\\
        \bottomrule
    \end{tabular}\\
    \flushleft\footnotesize{$^a$ l.u. = log-uniform; u. = uniform. The numbers in parentheses bracket the prior support.}
    \label{tab:SNfit_result_table}
\end{table}

\subsection{SED modelling with blackbody + dust}
\label{sect:dust}

Motivated by the NIR evolution, we explored whether some fraction of this emission can be attributed to pre-existing or newly forming dust grains. 
To do so, we carried out a two-component fit to the $grizyJHK$-band photometry (Fig.~\ref{fig:lc}, Tables~\ref{tab:IR} and \ref{tab:optical})  using a combination of a black-body function and a modified black-body function \citep{1983QJRAS..24..267H,2017ApJ...849L..19G}. 
This allowed us to simultaneously fit for the parameters of a blackbody representing the supernova, $T_\mathrm{SN}$ and $L_\mathrm{SN}$, and a cooler dust component with temperature $T_{\mathrm{d}}$ and mass $M_{\mathrm{d}}$. In analogy to the formalism described in \cite{2017ApJ...849L..19G}, we assumed that the dust mass absorption coefficient, $\kappa_{\mathrm{abs}}(\nu, a)$ (in units of [cm$^{2}$ g$^{-1}$]) can be approximated as a $\lambda^{-x}$ power law, with $x$ as the power-law slope, within the NIR wavelength range 0.9--2.5 $\mu$m covered by the $zyJHK$ bands. We assumed a power law slope $x = 1.2$, mimicking large grains (but we obtain similar result with $x=1.5$, appropriate for smaller grains) and we adopted $\kappa_{\mathrm{abs}}(\lambda = 1 \mu$m) = 1.0 $\times$ 10$^{4}$ cm$^{2}$ g$^{-1}$, which is appropriate for carbonaceous dust \citep{1991ApJ...377..526R}. Such a simple model is well justified based on the limited data available.

Since the SED data do not show clear signs of two emission components (see Figure \ref{fig:SEDs_BB}), in order to obtain sensible results from the fit we had to impose priors based on expectations for the dust and SN components. In particular, we imposed $100\leq T_\mathrm{d}/\mathrm{K} \leq 2500$, while $4000\leq T_\mathrm{SN}/\mathrm{K} \leq 20000$, therefore distinguishing dust and SN based on their plausible temperatures. 

The result, shown in Fig.~\ref{fig:SEDs_BB+dust} and summarised in Table \ref{tab:BB+Dust_fitting_result_table}, shows that the latest three SEDs at $t\geq 25\,\mathrm{d}$ (Tab. \ref{tab:SED_fitting_results}) can be attributed to a small amount ($M_\mathrm{d}\sim 10^{-5}\,\mathrm{M_\odot}$) of relatively hot ($T_\mathrm{d}\sim 1500\,\mathrm{K}$) dust. Prompted by this, we repeated the simple SN model fit taking the photometric data at $t>20\,\mathrm{d}$ as upper limits. The result is consistent within the uncertainties with the fit performed on all photometric data (Table \ref{tab:SNfit_result_table}, see also Appendix \ref{sec:BB+dust_sed_modelling}) and it therefore does not change our interpretation of the nature of the transient. 

We note that dust has been found in some other stripped envelope SNe with small ejecta masses, such as the Type Ibn SN~2006jc, where \cite{Mattila08} found evidence for both newly formed and pre-existing dust. 
SN ejecta at 6,000 \kms\ will reach a radius of $2\times10^{15}$~cm after 40 days, which is comparable to the blackbody radius of \wxt\ at this phase. So, if dust is the cause of the IR emission in \wxt\ at late times, then it has likely formed in the ejecta. Further investigation of the late-time evolution of ultra-stripped SN in the IR will help settle this question.

\subsection{\tardis\ spectral modelling} \label{sec:TARDIS}
\par
To model the photospheric phase spectral evolution, we used \tardis\ \citep{tardis, tardis2}, a one-dimensional Monte Carlo radiative transfer code capable of rapidly generating synthetic supernova spectra. The underlying methodology assumes a spherically symmetric explosion and approximates the inner region of optically thick SN ejecta material as a single-temperature blackbody. The outer region of optically thin material is divided into shells, and $r$-packets (analogous to bundles of photons) are launched from the boundary between the optically thick and optically thin regions. These $r$-packets are assigned properties based on the model properties at this boundary, and are free to propagate through the optically thin shells and interact with the material within. The escaping packets are then used to compute a synthetic spectrum, based on how they last interacted with the ejecta material.

\par
Using \tardis, we were able to produce a sequence of self-consistent models for a subset of the observed spectra, in order to constrain ejecta properties. Specifically, we focused our modelling efforts on the +0.9\,d X-shooter, +0.9\,d FORS2, +6.5 \& +8.4\,d OSIRIS, and +15.6\,d GMOS spectra, as these spanned most of the photospheric phase that we have data.
We flux-calibrated these spectra \citep[using the \textsc{sms} code; see][]{Inserra2018}, corrected for extinction, and shifted to rest-frame for our modelling. The input parameters for our sequence of models are included in Table~\ref{tab:TARDIS_model_parameters}. We used an exponential density profile to model the ejecta, which has the general form:
\begin{equation}
    \rho \left( v, t_{\rm exp} \right) = \rho_0 \left( \frac{t_0}{t_{\rm exp}} \right)^3 \textrm{exp} \left( -\frac{v}{v_0} \right),
    \label{eqn:TARDIS density profile}
\end{equation}
for ${v_{\rm min}} \leq v \leq {v_{\rm max}}$, where $\rho_0$, $t_0$, $v_0$ and $v_{\rm max}$ are constants.
The values for these constants were chosen empirically to best match the observed spectra. We obtain good agreement with $\rho_0 = 2.5 \times 10^{-12}$\,g\,cm$^{-3}$, $t_0 = 2$~days, $v_0 = 6000$\kms, and $v_{\rm max} = 18000$\kms.
$t_{\rm exp}$ is the time since explosion for each epoch we model, and we find good agreement to the data invoking an explosion epoch 3.2 days before $i$-band maximum.
We used the \tardis\ \texttt{nebular} treatment for ionisation, and \texttt{dilute-LTE} for excitation. In order to correctly reproduce the observed He features in the spectra, we adopted the He NLTE treatment as outlined by \citet{tardis_he}. By considering non-thermal excitation processes, we were able to more accurately predict the strengths of the He features produced by the SN ejecta. We note that this He NLTE treatment is a simple, empirically derived approximation. For our modelling, we applied minor corrections to the relative populations for some of the levels. We did this in order to produce feature strengths that were more in line with the observations. This was to demonstrate that He is capable of reproducing the features in the observed spectra. We do not place any emphasis on our modification of the He NLTE treatment here, beyond the fact that this was purely an empirical exercise, to allow the models to better replicate the observed He features.

\begin{table}
    \centering
    \caption{
        Input parameters used to generate the various \tardis\ models presented in this work.
    }
    \begin{threeparttable}
        \centering
        \begin{tabular}{lcccc}
            \toprule
                &\multicolumn{4}{c}{Phase (days)} \\
            \cmidrule{2-5}
                &+0.9     &+6.5      &+8.4   &+15.6     \\
            \midrule                                              
            $t_{\rm exp}$\,(days)                                 &4.1      &9.7      &11.6     &18.8   \\
            $L_{\rm ph}$\,(10$^{41}$\,erg\,s$^{-1}$)              &17.0     &6.71     &5.75     &2.52   \\
            $v_{\rm min}$\,(km\,s$^{-1}$)                         &17000    &9300     &7700     &3500   \\
            $T_{\rm ph}$\,($10^3$\,K)                             &8.99     &6.51     &6.26     &5.84   \\
            $M_{\textsc{tardis}}$\,($10^{-3}\,\msun$)\tnote{a}    &1.35     &13.1     &15.4     &19.6   \\
            \bottomrule
        \end{tabular}
        \begin{tablenotes}
            \small
            \item[a] $M_{\textsc{tardis}}$ is a derived property of our models, but is included here for reference. It represents the mass bound by the \tardis\ computational domain, and so it represents a \textit{lower limit} for a model's total ejecta mass. 
        \end{tablenotes}
    \end{threeparttable}
    \label{tab:TARDIS_model_parameters}
\end{table}

\par
\begin{figure}[h!]
    \centering
    \includegraphics[width=\linewidth]{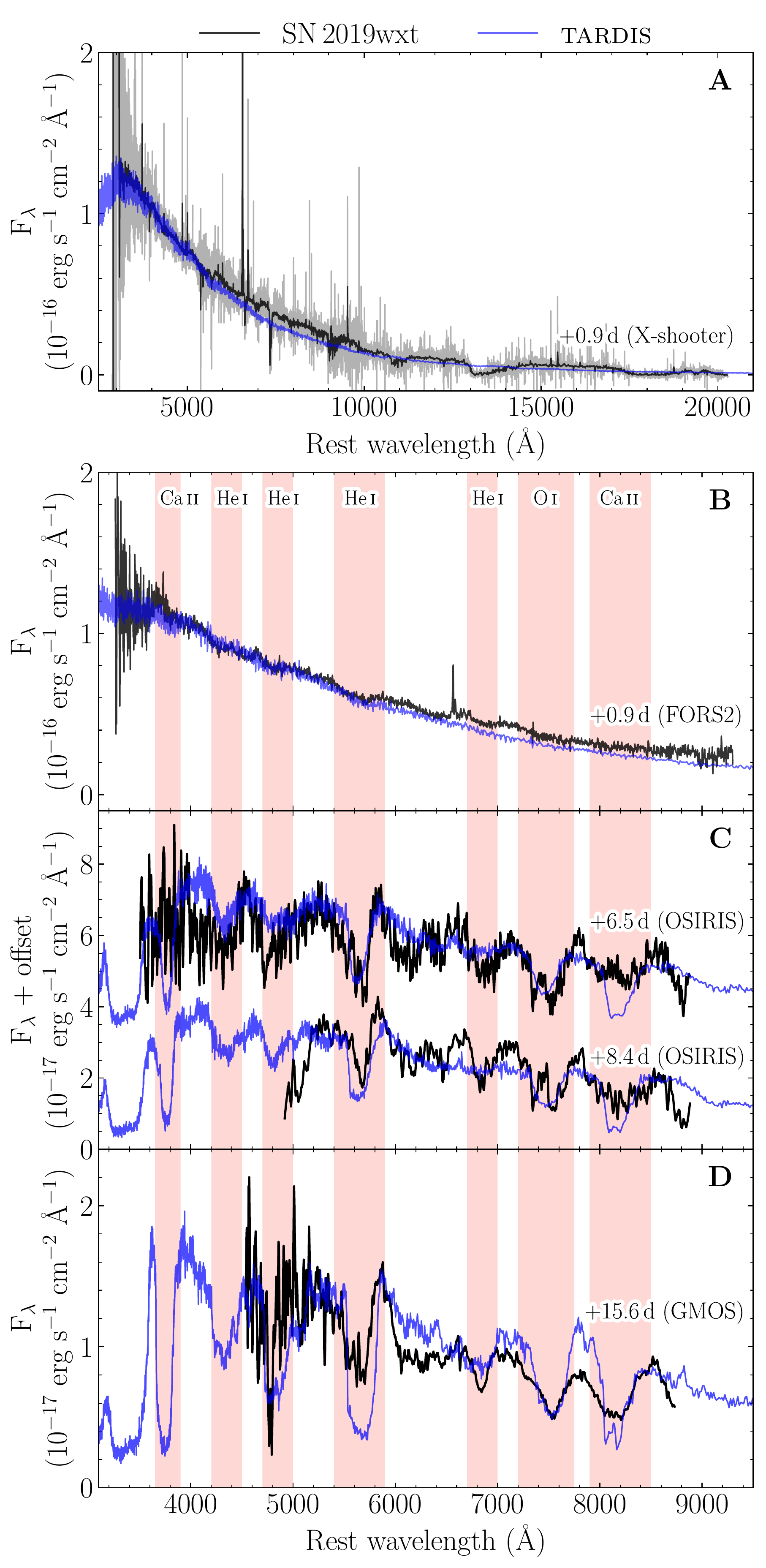}
    \caption{
        Comparisons between our model spectra (blue) and observations (black).
        The observed spectra have been rebinned by a factor of 10 \citep[using \textsc{spectres},][]{spectres}.
        \textit{Panel A:} Comparison of the +0.9\,d X-shooter spectrum with our best-fitting \tardis\ model.
        \textit{Panel B:} Comparison of the early FORS2 spectrum (+0.9\,d), and our best-fitting \tardis\ model.
        \textit{Panel C:} Comparison of the two later OSIRIS spectra (+6.5 and +8.4\,d) with their corresponding \tardis\ models. The +6.5\,d spectrum and model have been offset by \mbox{$3 \times 10^{-17}$\,erg\,s$^{-1}$\,cm$^{-2}$\,\AA$^{-1}$}, for clarity.
        \textit{Panel D:} Comparison of the late-time GMOS spectrum (+15.6\,d), and its corresponding \tardis\ model.
        The vertical shaded bands in panels B, C and D correspond to regions of absorption in the observed spectra. The species dominating these same absorption features in our best-fitting \tardis\ models have added to the top of each band in panel B.
    }
    \label{fig:TARDIS_model_spectral_sequence}
\end{figure}

The sequence of model spectra that best match the observations are presented in Fig.~\ref{fig:TARDIS_model_spectral_sequence}. Here we can see that across all epochs, the \tardis\ model spectra reproduce most of the observed features, with an ejecta composition dominated by helium and oxygen, and trace amounts of calcium at high velocities. We observe features at $\sim 4300$, 4800, 5600, 6800, 7500 and 8200\,\AA.
We attribute the features at $\sim 4300$, 4800, 5600 and 6800\,\AA\ to the \HeI\ 4471.5, 5015.7, 5875.6, and 7065.2\,\AA\ lines (in air).
We reproduce the feature at $\sim$ 7500\,\AA\ with a blend of the O\,$\textsc{i}$ 7771.9, 7774.2, and 7775.4\,\AA\ lines (in air).
Finally, we attribute the feature at $\sim 8200$\,\AA\ to the commonly observed \CaII\ NIR triplet.
Our models also include a small amount of $^{56}$Ni (and its decay products) to generally improve the SED beyond a few days.
We summarise our compositions in Table~\ref{tab:TARDIS composition}.
Our model compositions are quite simple -- we do not require the presence of many elements to reproduce the prominent observed features in the spectra of \wxt. Nonetheless, the presence of some commonly identified elements in SN spectra have been explored, and we also present the upper limits we obtain for these species in Table~\ref{tab:TARDIS composition}.

\begin{table}
    \centering
    \caption{
        Mass fractions of the different elements included in our sequence of best-fitting \tardis\ models.
        The mass fractions included apply only to the velocity ranges quoted (outside these ranges the mass fractions are set to zero, with any deviation from a mass fraction of unity compensated for by slightly increasing or decreasing the He mass fraction).    }
    \begin{threeparttable}
    \renewcommand{\TPTminimum}{\linewidth}
        \centering
        \makebox[\linewidth]{
        \begin{tabular}{lcc}
            \toprule
            Element     &Mass fraction  &Velocity range (km\,s$^{-1}$)     \\
            \midrule
            He                      &0.69       &$3500 - 18000$     \\
            O                       &0.30       &$3500 - 18000$     \\
            Ca                      &$10^{-4}$  &$12000 - 18000$    \\
            $^{56}$Ni\tnote{a, b}   &0.01       &$3500 - 17000$     \\
            \addlinespace[0.5em]
            H\tnote{c}  &$\lesssim 0.04$  &$3500 - 18000$   \\
            C           &$< 0.05$         &$3500 - 18000$   \\
            Na          &$< 0.05$         &$3500 - 18000$   \\
            Mg          &$< 0.10$         &$3500 - 18000$   \\
            Si          &$< 0.05$         &$3500 - 18000$   \\
            S\tnote{d}  &$\lesssim 0.5$   &$3500 - 18000$   \\
            \bottomrule
        \end{tabular}}
        \begin{tablenotes}
            \small
            \item[a] Our models included this initial mass fraction of $^{56}$Ni (at \mbox{$t = 0$ days}). Its mass fraction was updated at subsequent epochs, accounting for the effects of radioactive decay.
            \item[b] Our early model (at +0.9 days) cannot accommodate any significant quantity of $^{56}$Ni, and so this outermost region of the ejecta \mbox{($v_{\rm ej} = 17000 - 18000$\kms)} is free of IGEs across our sequence of spectra.
            \item[c] This is the mass fraction of H needed to reproduce the emission at $\sim 6600$\,\AA, although we expect this to be somewhat uncertain, due to the limitations of our LTE approximations.
            \item[d] Our S mass fraction remained reasonably unconstrained across our sequence of models, as evidenced by the fact we can accommodate an unphysically large mass fraction.
        \end{tablenotes}
    \end{threeparttable}
    \label{tab:TARDIS composition}
\end{table}

\par
We are able to reproduce the observed spectra well at all epochs, with a composition dominated by helium and oxygen. We also require some trace quantity of calcium to reproduce the absorption feature at $\sim 8200$\,\AA, which we attribute to the \CaII\ NIR triplet. This calcium is concentrated at high velocities in our models ($v \geq 12000$\kms), as too much of it negatively impacts our fit to the data when extended to lower ejecta velocities.
We have for consistency in our modelling efforts maintained a consistent abundance profile across all epochs (with the exception of $^{56}$Ni and $^{56}$Co decay). However, our +6.5\,d model over-produces the \CaII\ triplet, and as a result has a strong feature at $\sim 3800$\,\AA, which corresponds to the \CaII\ H\&K lines. Although this wavelength region of the +6.5\,d OSIRIS spectrum is extremely noisy, we do not see any evidence for such a strong feature, which would suggest that the early epoch spectra formed in less Ca-rich ejecta.

The spectra exhibit an emission-like feature at $\sim 6600$\,\AA, which our \tardis\ models do not reproduce. One potential line identification for the production of this feature is the He\I\ 6678\,\AA\ line (in air). Despite including a large mass fraction of He, and producing multiple strong features from He\I, we are unable to produce any strong feature from this particular line. Therefore,
we consider the possibility that this feature may be the result of H$\alpha$ emission instead.
To test this identification, we added a small amount of H to our \tardis\ models.

We find that we can produce a prominent P-Cygni feature, with the emission component broadly resembling the position and width of the emission feature at $\sim 6600$\,\AA. However, we see no evidence for the associated strong absorption component, indicating that this feature is in net emission (something that we cannot reproduce with our \tardis\ models). These simple models indicate that, for the ejecta velocities, temperatures and densities invoked for our sequence of models, we can expect to see H features (if H is present). We deduce that a mass fraction of $\sim 4$~per cent is needed to produce the emission observed, although we stress that, due to the limitations of assuming LTE level populations, this inferred mass fraction is somewhat uncertain.

We also include a $^{56}$Ni mass fraction of 0.01 in our model ejecta below $17000$\kms\ (at $t = 0$ days), which decays $\left( ^{56} \rm{Ni} \rightarrow\ \! ^{56} \rm{Co} \rightarrow\ \! ^{56} \rm{Fe} \right)$, resulting in a small amount of Ni, Co, and Fe at the epochs we generate models. We find that this small amount of iron-group element (IGE) material improves overall agreement with the SED, but too much ($\gtrsim 0.01$) negatively impacts the model spectra. This is in seeming contradiction with our invoked $^{56}$Ni mass from the bolometric lightcurve modelling presented in Sect.~\ref{sect:arnett}. There, we showed that we can fit the evolution with an ejecta mass, $M_{\rm ej} \sim 0.1$\,\msun, approximately 20 per cent of which is $^{56}$Ni. We ran a set of \tardis\ models including an additional 10 per cent of material, which we appropriated to $^{56}$Ni (and its subsequent decay products) to test the effect this amount of IGEs would have on our models. We found that the models could not accommodate this amount of IGEs, suggesting that this heavy material remains beneath the inner boundary of our \tardis\ models. We note that \tardis\ assumes an inner boundary to its models, beneath which we cannot infer any ejecta properties. As such, our modelling efforts can only constrain the properties in the line-forming region of the ejecta. The mass enclosed in the \tardis\ line-forming region for the latest epoch $M_{\textsc{tardis}}(t = 18.8\,\rm{d}) \simeq 0.02$\,\msun, which lies comfortably below the mass invoked from the lightcurve modelling (0.1\,\msun), and so it does not seem unreasonable to imagine most of the heavy IGE material residing beneath this optically thick boundary.

The composition and ejecta velocities deduced from our \tardis\ modelling are consistent with that of a SN Ib, indicating that this event is unrelated to the GW trigger (see also Sect. \ref{sect:wxt_as_kn}).

\section{The environment of \wxt}
\label{sect:env}

\subsection{Local host properties}
\label{sect:local_env}

The equivalent width of the narrow interstellar NaD absorption often seen in spectra has been long used to estimate the extinction towards supernovae \citep[e.g.][]{Turatto03}. Using our highest resolution X-Shooter spectra, we measure the equivalent width of the D1 and D2 lines at the redshift of \wxt\ to be $0.27 \pm 0.01$ and $0.45 \pm 0.02$\,\AA, respectively. Applying the calibration of \cite{Poznanski12}, this implies a host galaxy reddening of either $E(B-V)_{D1}=0.08^{+0.03}_{-0.06}$ or $E(B-V)_{D2}=0.12^{+0.05}_{-0.04}$ towards \wxt. If we applied this reddening correction to \wxt\, then the peak of the lightcurve would be 0.25 mag brighter in $r$-band. However, in light of the possible presence of circumstellar dust (which will affect the relation between extinction and equivalent width), we opt not to apply any correction for host galaxy reddening in this paper.

\begin{figure}[h!]
    \centering
    \includegraphics[width=\linewidth]{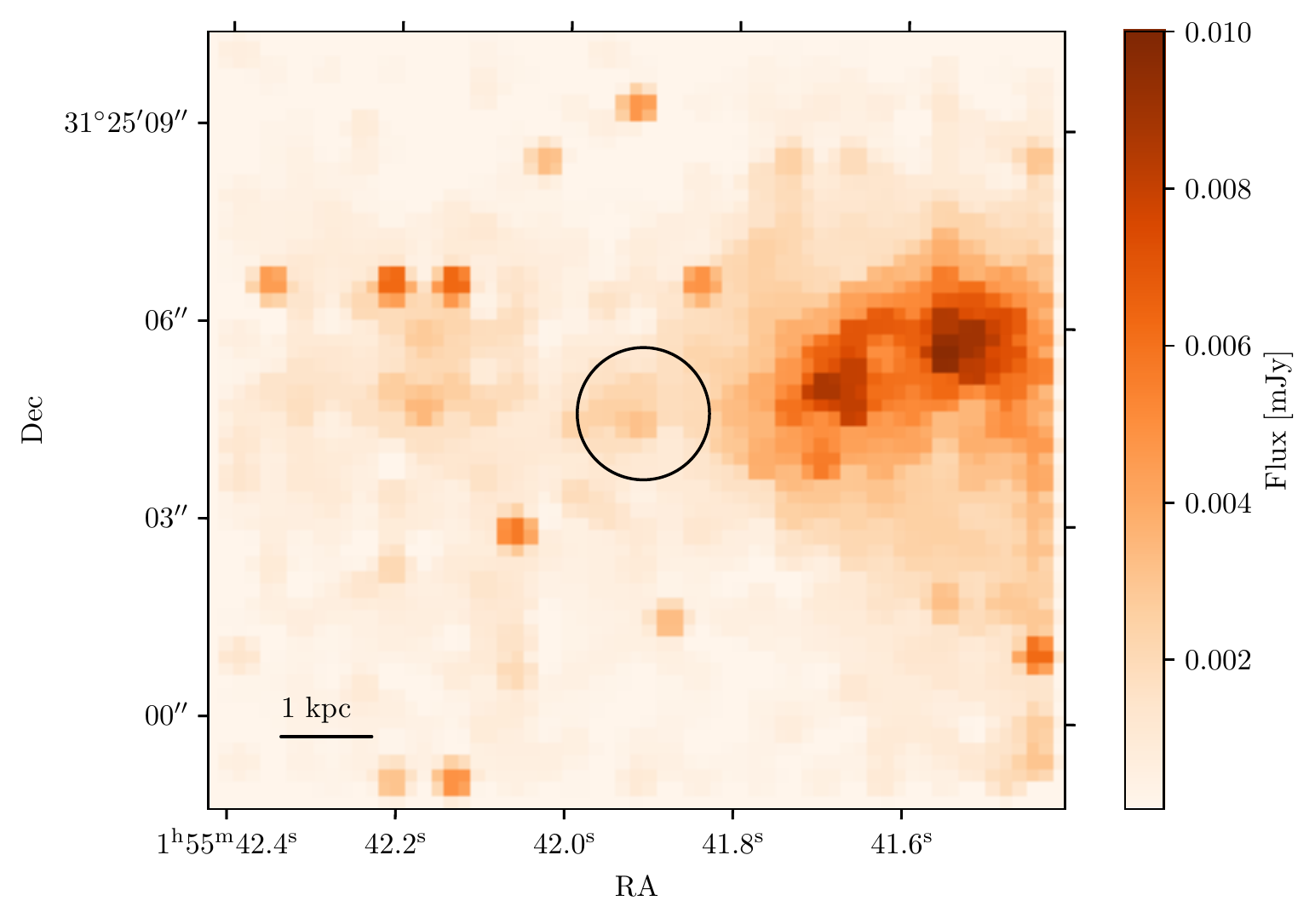}
    \includegraphics[width=\linewidth]{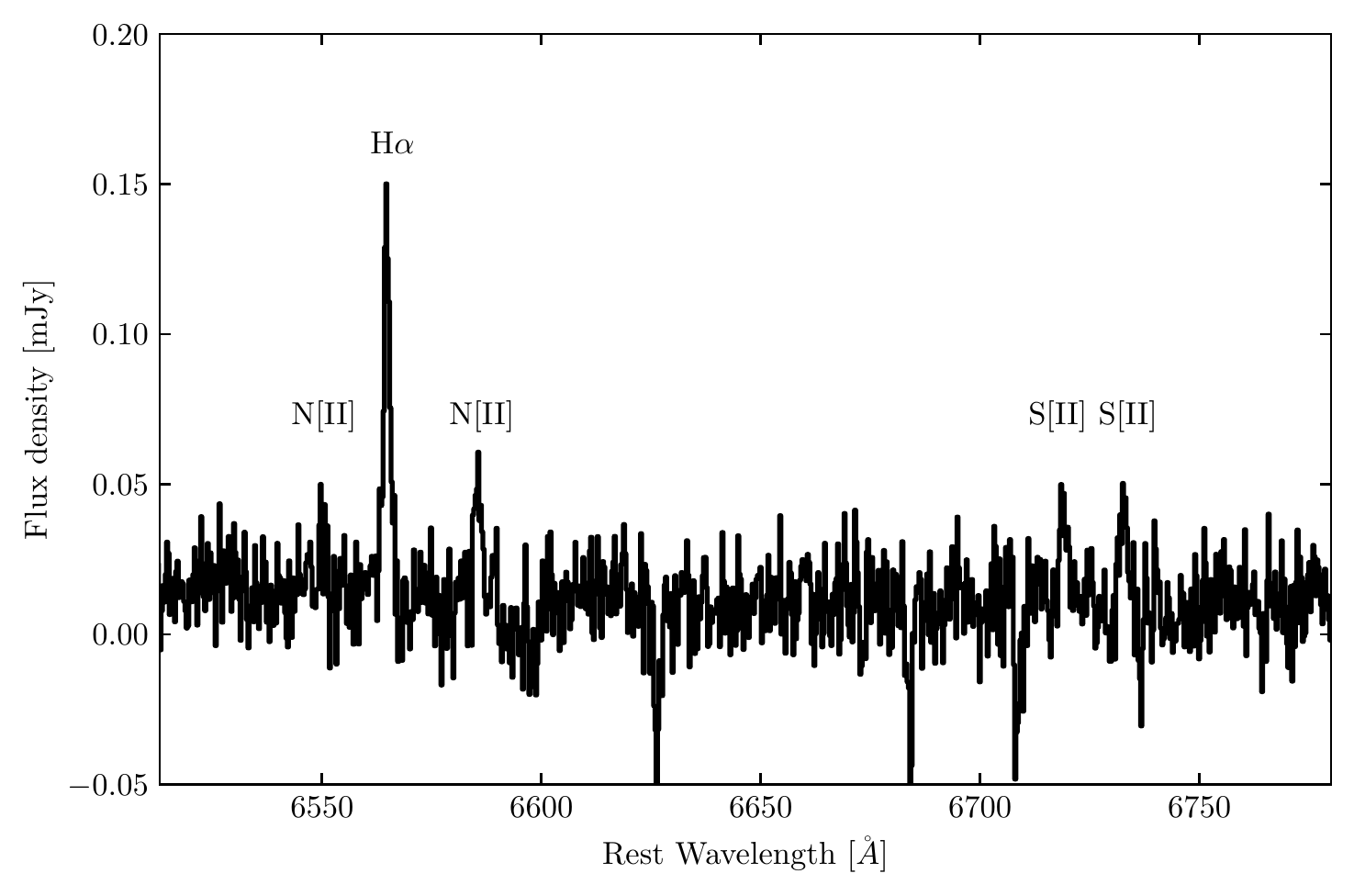}
    \caption{Integral field spectroscopy of the host galaxy. Upper panel: H$\alpha$ map constructed from 2D MEGARA+LR-R spectrum. The aperture used to extract the host galaxy spectrum is indicated with a black circle. Lower panel: MEGARA LR-R spectrum at the position of \wxt\ showing detected host galaxy lines.}
    \label{fig:megara}
\end{figure}

MEGARA IFU spectra taken on 28 Jan 2020, at +40~d were examined to determine the local metallicity at the location of \wxt. The reduction of these data is discussed in Appendix~\ref{sect:IFU}. We extracted a one-dimensional spectrum from both the LR-B and LR-R datacubes using a 5 pixel (1\arcsec, corresponding to 0.73~kpc at the distance of KUG 0152+311) radius aperture centred on the position of \wxt (Fig. \ref{fig:megara}). Unfortunately no continuum or emission line flux was seen in the extracted LR-B spectrum (which covers a rest frame wavelength range 4200 -- 5050 \AA). However, we see a number of emission lines in the LR-R spectrum, including H$\alpha$, 
[N~{\sc ii}] 
$\lambda\lambda$6548,6583, and [S~{\sc ii}] $\lambda\lambda$6716,6730 (Fig. \ref{fig:megara}).

Unfortunately, most metallicity indices rely on H$\beta$ or [O~{\sc ii}] and [O~{\sc iii}] lines that lie in the blue. We hence use the N2 metallicity indicator from \cite{Pettini04}. We measure the flux in H$\alpha$ and [N~{\sc ii}] $\lambda6583$ and from this calculate the N2 index to be $-0.37 \pm 0.03$. Using the calibration in \citeauthor{Pettini04}, we determine the metallicity at the location of \wxt\ to be 12 + log(O/H) = $8.7 \pm 0.2$ dex (the 1$\sigma$ uncertainty here is dominated by the intrinsic scatter in the N2-metallicity index). The more recent calibrations of \cite{Mar13} and \cite{Curti20} give consistent values of 8.6$\pm$0.2 and 8.7$\pm$0.1~dex respectively.
These values for the metallicity are approximately solar, although we must caution that this is the average metallicity measured over a large physical region in the host galaxy, using a diagnostic with relatively large scatter.

Using the same extracted 1D spectrum, we measure the H$\alpha$ line luminosity within  1\arcsec\ of \wxt. From this, we use the calibration in \cite{Kennicutt98} to estimate the star formation rate in this region to be $(4.4\pm0.3)\times10^{-4}$ \msun~yr$^{-1}$. As the MEGARA field of view only covers part of the host galaxy, we cannot estimate the global star formation rate for this galaxy.

\begin{figure*}[h!]
    \centering
    \includegraphics[width=\linewidth]{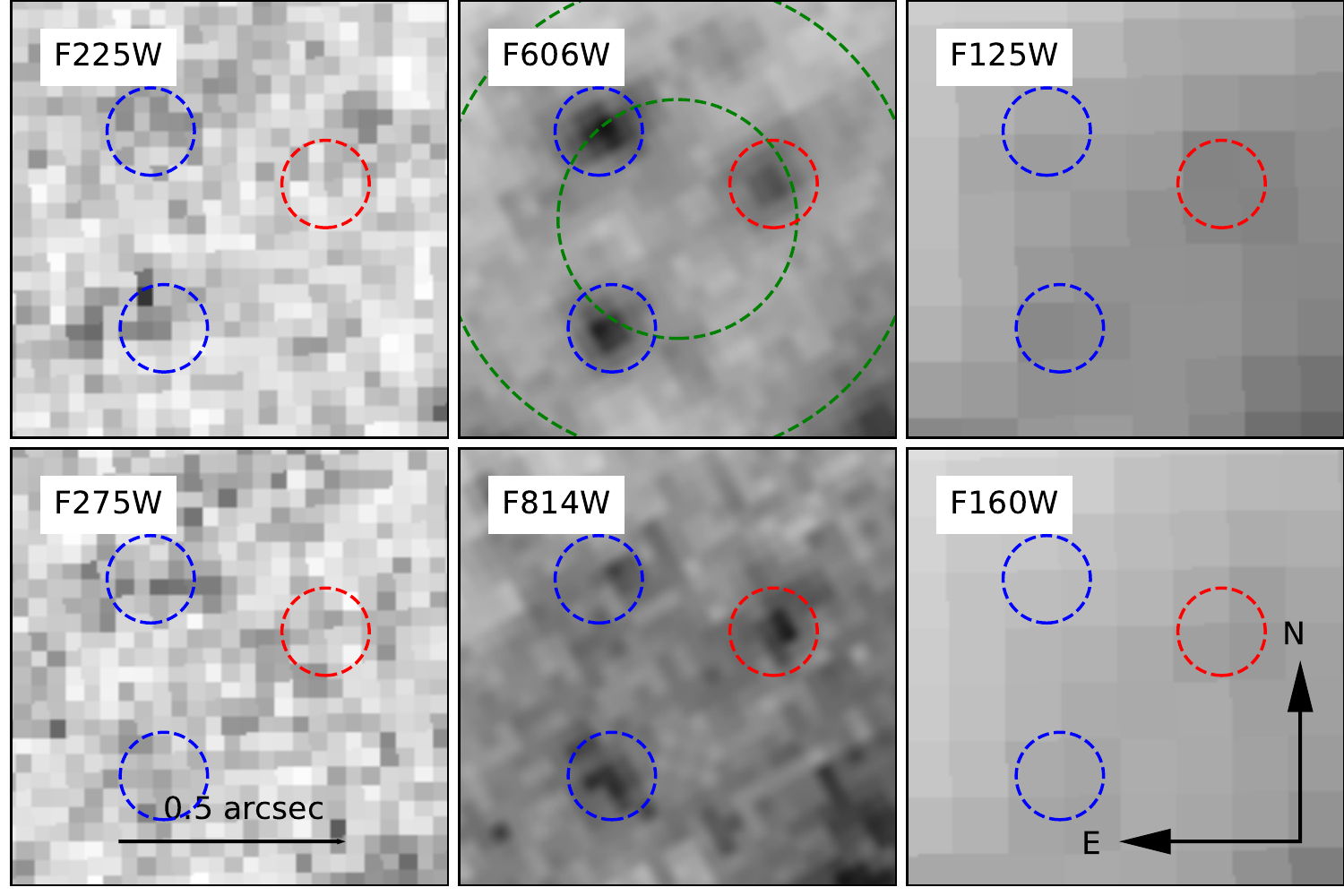}
    \caption{
    $1 \arcsec \times$  1\arcsec\ cutouts from the late time (Jan 2021) HST imaging of the site of \wxt. Each panel is centred on the position of \wxt, filters are indicated in each panel. In the F606W cut-out, the dashed green circles mark a radius of 200 and 400 parsec from \wxt. The sources discussed in the text are also indicated; the redder source with a red circle, and the two blue sources with blue circles.
    }
    \label{fig:HST_cutouts}
\end{figure*}

We also examined the late time HST images covering the site of \wxt\ (Fig. \ref{fig:HST_cutouts}). In order to accurately locate the position of \wxt\ on these images we first measured the pixel coordinates on the F125W image from Feb 2020. We then used around 20 sources common to this image and each of the F606W, F814W, F125W and F160W images from 2021 to derive a geometric transformation between the two frames. The rms uncertainty in the transformation ranged between 0.14 to 0.18 pixels for the IR filters, and between 0.34 and 0.40 pixels for the UVIS filters. This corresponds to an uncertainty of a few tens of mas in position.

The location of \wxt\ lies approximately equidistant between three extended sources, seen most clearly in the F606W filter (Fig. \ref{fig:HST_cutouts}). One of these sources (to the N-W of \wxt) is relatively red, being brighter in F814W and also showing some flux in the F125W band. On the other hand, the two sources to the East of \wxt\ are blue, with some faint UV emission in F225W and F275W filters that is suggestive of a young stellar population. However, as \wxt\ is at least $\sim$200 pc from each of this regions, we cannot securely associate it with any of them. Assuming a modest velocity of 20 \kms, the progenitor of \wxt\ could have plausibly traveled the 200 pc distance to any of these sources in $\sim$10 Myr. We note that the three sources have a magnitude in the F606W band of 25~mag or fainter, implying an absolute magnitude of $\lesssim -10$. Such a magnitude is consistent with that expected for a stellar cluster.
However, as we cannot distinguish which (if any of these sources) \wxt\ is associated with, we opt not to analyse these further.

\subsection{Global host properties}

We retrieved science-ready coadded images from the \textit{Galaxy Evolution Explorer} (\galex) general release 6/7 \citep{Martin2005a}, the Sloan Digital Sky Survey data release 9 (SDSS DR 9; \citealt{Ahn2012a}), the Panoramic Survey Telescope and Rapid Response System (Pan-STARRS, PS1) DR1 \citep{Chambers2016}, the Two Micron All Sky Survey \citep[2MASS;][]{Skrutskie2006a}, and preprocessed \wise\ images \citep{Wright2010a} from the unWISE archive \citep{Lang2014a}\footnote{\href{http://unwise.me}{http://unwise.me}}. The unWISE images are based on the public \wise\ data and include images from the ongoing NEOWISE-Reactivation mission R3 \citep{mainzer2014, Meisner2017a}. We measured the brightness of the host using \textsc{lambdar}\footnote{\href{https://github.com/AngusWright/LAMBDAR}{https://github.com/AngusWright/LAMBDAR}} \citep[\textsc{lambda adaptive multi-band deblending algorithm in r};][]{Wright2016a} and the methods described in \citet{Schulze2021a}. Table \ref{tab:hostphot} shows the measurements in the different bands. We also used the data from the  Atacama Large Millimeter Array centred at the frequency of the CO(1-0) line. The data cube has a resolution of 1\farcs66 $\times$~0\farcs9. The channel width is 16\,MHz, corresponding to 44\,{\kms}.

We modelled the UV to mid-IR spectral energy distribution with the software package \textsc{prospector} version 0.3 \citep{Leja2017a}. \textsc{prospector} uses the \textsc{flexible stellar population synthesis} (\textsc{fsps}) code \citep{Conroy2009a} to generate the underlying physical model and \textsc{python-fsps} \citep{ForemanMackey2014a} to interface with \textsc{fsps} in \textsc{python}. The \textsc{fsps} code also accounts for the contribution from the diffuse gas (e.g.\ \ion{H}{ii} regions) based on the \textsc{cloudy} models from \citet{Byler2017a}. Furthermore, we assumed a Chabrier initial mass function \citep{Chabrier2003a} and approximated the star formation history (SFH) by a linearly increasing SFH at early times followed by an exponential decline at late times [functional form $t \times \exp\left(-t/\tau\right)$, where $t$ is the age of the SFH episode and $\tau$ is the $e$-folding timescale]. The model was attenuated with the \citet{Calzetti2000a} model.

\begin{figure}[h!]
    \centering
    \includegraphics[width=1\columnwidth]{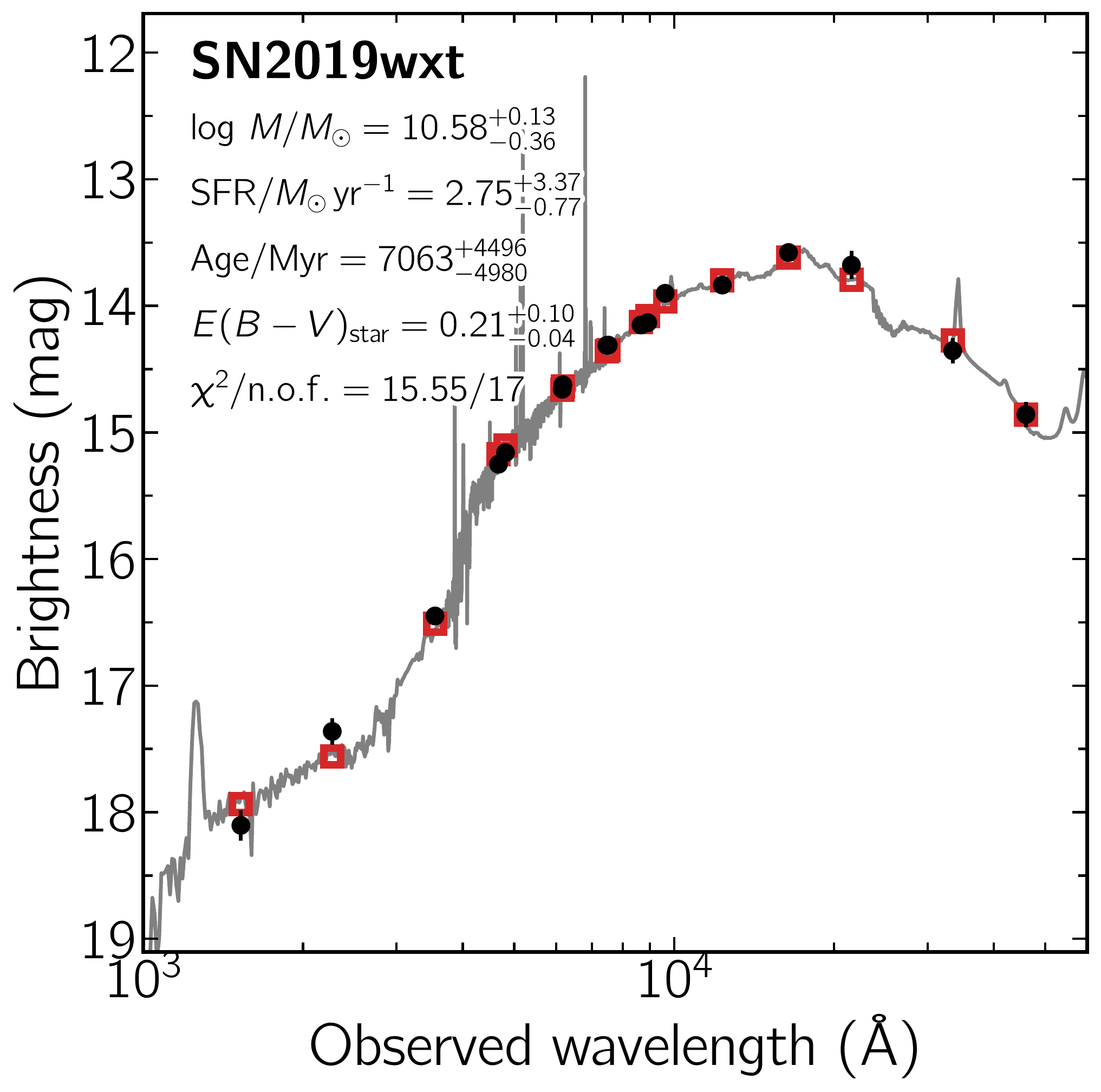}
    \caption{Spectral energy distribution of the host galaxy of \wxt\ from 1000 to 60,000~\AA\ (black data points). The solid line displays the best-fitting model of the SED. The red squares represent the model-predicted magnitudes. The fitting parameters are shown in the upper-left corner. The abbreviation "n.o.f." stands for numbers of filters.}
    \label{fig:gal_sed}
\end{figure}

Figure \ref{fig:gal_sed} shows the observed SED and its best fit. The  median values of the marginalised posterior probability functions and their 1$\sigma$ confidence intervals are shown in the same plot. The galaxy SED is adequately described by a moderately attenuated ($E(B-V)\sim0.2$~mag) massive ($\sim4\times10^{10}~M_\odot$) star-forming ($\sim3~M_\odot\,{\rm yr}^{-1}$) galaxy dominated by an old stellar population ($\sim7$~Gyr).
Comparing to the global host properties of a sample of Type Ibc SN hosts in \cite{Galbany14}, we find that the mass and star formation rate we derive for \wxt\ are within 1$\sigma$ of the mean; while the age is $\sim 2\sigma$ older than the mean.
In light of the possible presence of H in the spectra of \wxt, we also compared to the host galaxies of 61 SNe IIb from the Palomar Transient Factory \citep{Schulze2021a}, again finding the properties of \wxt\ to be fairly typical.

\begin{figure}[h!]
    \centering
    \includegraphics[width=1\columnwidth]{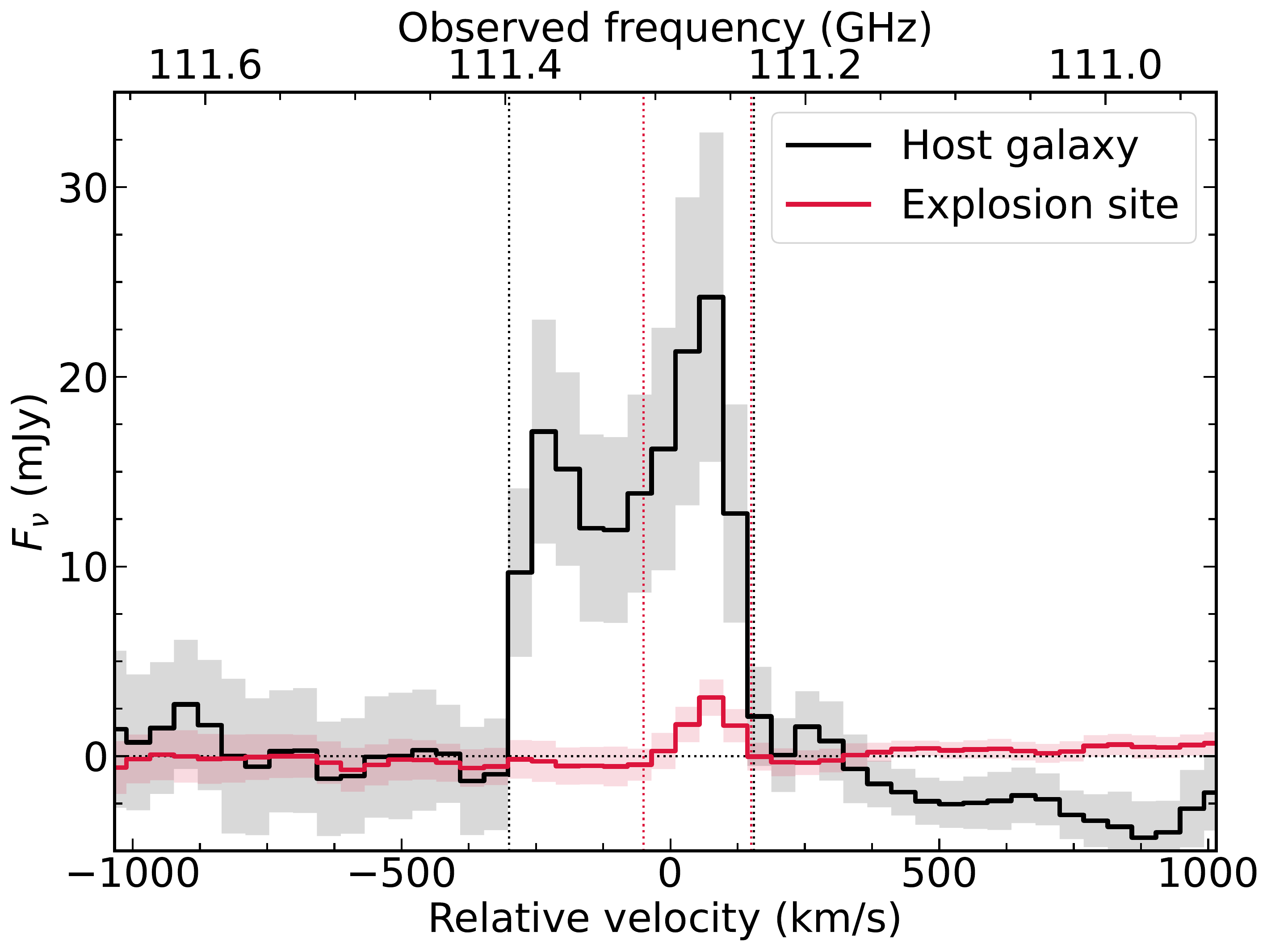}
    \caption{CO spectra extracted over the entire SN host galaxy (black) and of the explosion site (red). The vertical dotted lines denote the spectral region from which the fluxes were extracted.
    }
    \label{fig:cospec}
\end{figure}

Figure~\ref{fig:cospec} shows the CO spectra extracted for the entire detected emission of the host in the aperture of $10\arcsec$ and at the SN site in the aperture of $2\arcsec$. Table~\ref{tab:mhtwo} presents the derived redshift of the CO line, the line flux, luminosity and the corresponding molecular gas mass, assuming  the Galactic CO-to-$H_2$ conversion factor $\alpha_{\rm CO}=5\,M_\odot\, (\mbox{K km s}^{-1} \mbox{ pc}^2)^{-1}$.

Using the SFR-{\mhtwo} relation of \citet[][eq.~1]{Michalowski2018} for star-forming galaxies, the SFR of the host galaxy (Fig.~\ref{fig:gal_sed}) implies the expected molecular gas mass of $\log(\mhtwo/\msun)=9.49^{+0.36}_{-0.15}$. This is consistent with our ALMA measurements of $\log(\mhtwo/\msun)=9.282 \pm 0.053$, showing that the host has normal molecular gas properties. We also detect CO emission at the SN explosion site. Applying the same method as for the host galaxy, we derive a $\log(\mhtwo/\msun)=7.9\pm0.1$.

We inspected the NRAO VLA Sky Survey \citep[][45\arcsec\ angular resolution]{Condon1998} and the recently released Apertif imaging survey \citep[][$12\arcsec \times18\arcsec$ angular resolution at the location of the host]{Adams2022}. Both surveys show significant continuum emission at 1.4 GHz associated with KUG 0152+311, with a flux density of about 3 mJy. This corresponds to a monochromatic radio luminosity of $8.9\times 10^{21}$~W\,Hz$^{-1}$ and a radio-derived star formation rate of $\sim 4.3$~M$_\odot$\,yr$^{-1}$ \citep{Greiner2016}, in remarkably good agreement with the rate estimated from the optical SED fitting.
At higher angular resolution, the radio emission associated to star formation is resolved out and a much more compact (subarcsecond), yet fainter (sub-mJy), source is detected with e-MERLIN and ALMA (see  Appendix \ref{sect:radio}). This source is most likely a weak AGN (with\ $\nu L_\nu\sim 1.7\times10^{37}$~erg\,s$^{-1}$, from the 5~GHz e-MERLIN observations) at the centre of the host galaxy.

\begin{table}
\caption{Host photometry}\label{tab:hostphot}
\tiny
\centering
\begin{tabular}{cccc}
\toprule
Instrument/Filter & Brightness & Instrument/Filter & Brightness\\
 & (mag) & & (mag) \\
\midrule
\galex/$FUV$	&$ 18.50 \pm 0.07 $ & PS/$i$    &$ 14.41 \pm 0.01 $\\
\galex/$NUV$	&$ 17.81 \pm 0.04 $ & PS/$z$    &$ 14.23 \pm 0.01 $\\
SDSS/$u$	&$ 16.69 \pm 0.03 $ & PS/$y$    &$ 13.97 \pm 0.03 $\\
SDSS/$g$	&$ 15.44 \pm 0.01 $ & 2MASS/$J$	&$ 13.88 \pm 0.05 $\\
SDSS/$r$	&$ 14.79 \pm 0.01 $ & 2MASS/$H$	&$ 13.61 \pm 0.05 $\\
SDSS/$i$	&$ 14.42 \pm 0.01 $ & 2MASS/$K$	&$ 13.70 \pm 0.05 $\\
SDSS/$z$	&$ 14.21 \pm 0.04 $ & \wise/W1   &$ 14.36 \pm 0.02 $\\
PS/$g$	    &$ 15.34 \pm 0.02 $ & \wise/W2   &$ 14.86 \pm 0.02 $\\
PS/$r$	    &$ 14.76 \pm 0.01 $\\
\bottomrule
\end{tabular}
\tablefoot{All magnitudes are reported in the AB system and are not corrected for reddening.}
\end{table}

\begin{table*}
\small
\centering
\caption{Molecular gas properties.\label{tab:mhtwo}}
\begin{tabular}{lcccc}
\toprule
Region & Redshift & $F_{\rm int}$   &  $\log\, L({\rm CO})$          & $\log\,M\left({\rm H}_2\right)$ \\
    & &   (Jy km s$^{-1}$) & ($\mbox{K km s}^{-1} \mbox{ pc}^2$) & ($M_\odot$) \\
(1)    & (2)     & (3)  & (4)           & (5)   \\
\midrule
host & $0.035579 \pm        0.000056$ & $6.54 \pm 0.85$ & $8.583 \pm 0.053$ & $9.282 \pm 0.053$  \\
SN & $0.036026 \pm        0.000067$ & $0.27 \pm 0.08$ & $7.205 \pm 0.113$ & $7.904 \pm 0.113$  \\
\bottomrule
\end{tabular}
\tablefoot{
(1) Region (the entire host or the SN site). (2) Redshift determined from the emission-weighted frequency of the CO line. (3) Integrated flux within the dotted lines of the top panel in Fig.~\ref{fig:cospec}. (4) CO line luminosity using equation 3 in \citet{Solomon1997a}. (5) Molecular hydrogen mass  using the Galactic CO-to-$H_2$ conversion factor $\alpha_{\rm CO}=5\,M_\odot\, (\mbox{K km s}^{-1} \mbox{ pc}^2)^{-1}$.
}
\end{table*}

\section{The nature of \wxt}
\label{sect:nature}

\subsection{\wxt\ as a kilonova}
\label{sect:wxt_as_kn}
Although an association with S191213g is unlikely as previously mentioned, for completeness we considered whether the properties of \wxt\ are at all compatible with a merging neutron star system.

First, \wxt\ appears to be too luminous for a kilonova powered by the decay of unstable heavy isotopes synthesised by the r-process. Assuming an ejecta heating rate per unit mass $\dot\varepsilon(t) \sim \dot\varepsilon_0 (t/t_0)^{-1.3}$ with $\dot\varepsilon_0=1.1\times 10^{10}\,\mathrm{erg\,s^{-1}\,g^{-1}}$ for $t_0=1\,\mathrm{day}$ \citep{Korobkin2012}, a peak luminosity of $L_\mathrm{pk}\sim 3\times 10^{42}\,\mathrm{erg\,s^{-1}}$ at a time $t_\mathrm{pk}\sim 5\,\mathrm{days}$ would require an implausible kilonova ejecta mass of $M_\mathrm{ej}\sim L_\mathrm{pk}/\dot\varepsilon(t_\mathrm{pk})\sim 1\,\mathrm{M_\odot} (t_\mathrm{pk}/5\,\mathrm{d})^{1.3}$. One would have to invoke an additional powering mechanism for kilonova ejecta to reach the observed luminosity of \wxt, such as magnetar spin down from a massive neutron star remnant or accretion onto the central black hole.

Second, the luminosity and colour evolution of \wxt\ is slower than plausible kilonova light curves. To demonstrate this, we compared the \wxt\ photometric data with synthetic kilonova light curves computed using the multi-component model of \cite{2021MNRAS.505.3016N}, which is tailored on BNS mergers and includes dynamical ejecta produced during the merger, winds from the accretion disk of the merger remnant and possibly cooling emission from the cocoon of a putative relativistic jet. The main model parameters are the progenitor chirp mass, and mass ratio (or equivalently binary masses $m_1$ and $m_2$), the maximum mass $M_\mathrm{TOV}$ that can be supported by the neutron star matter equation of state (EoS), the neutron star radius $R_\mathrm{NS}$ determining the compactness, and the fraction of the remnant disk ejected in winds. The dynamical ejecta and disk masses are estimate using scaling relations from \cite{2017CQGra..34j5014D} and \citet{2019MNRAS.489L..91C}. While the grey opacity of the wind is determined by the lifetime of the merger remnant (with longer-lived remnants leading to more neutrino irradiation, increasing the electron fraction and hence lowering the opacity), the opacities of the equatorial and polar dynamical ejecta components are free parameters, and the relative emission seen from each component depends on the angle of the orbital axis relative to the observer (see \citealt{2021MNRAS.505.3016N} for a complete description). Here we fixed $M_\mathrm{TOV}=2.2\,\mathrm{M_\odot}$ and $R_\mathrm{NS}=11\,\mathrm{km}$ and simulated $\sim17,000$ synthetic lightcurves over a broad parameter space, with chirp masses covering $0.7-2.0\,\mathrm{M_\odot}$, mass ratios between $0.5-1.0$, viewing angles of $0^{\circ},30^{\circ},60^{\circ},75^{\circ}$ and $90^{\circ}$, disk efficiencies ranging $10\%-40\%$ and opacities of $0.5, 1.0$\,cm$^{2}$g$^{-1}$ and $10.0, 25.0$\,cm$^{2}$g$^{-1}$ for the blue and red ejecta components respectively. For a subset of models with a mass ratio of $1$, disk efficiency of $20\%$ and default opacities, cocoon models were produced with opening angles of $10^{\circ},20^{\circ}$ and $30^{\circ}$.

\begin{figure*}[h!]
\includegraphics[width=\textwidth]{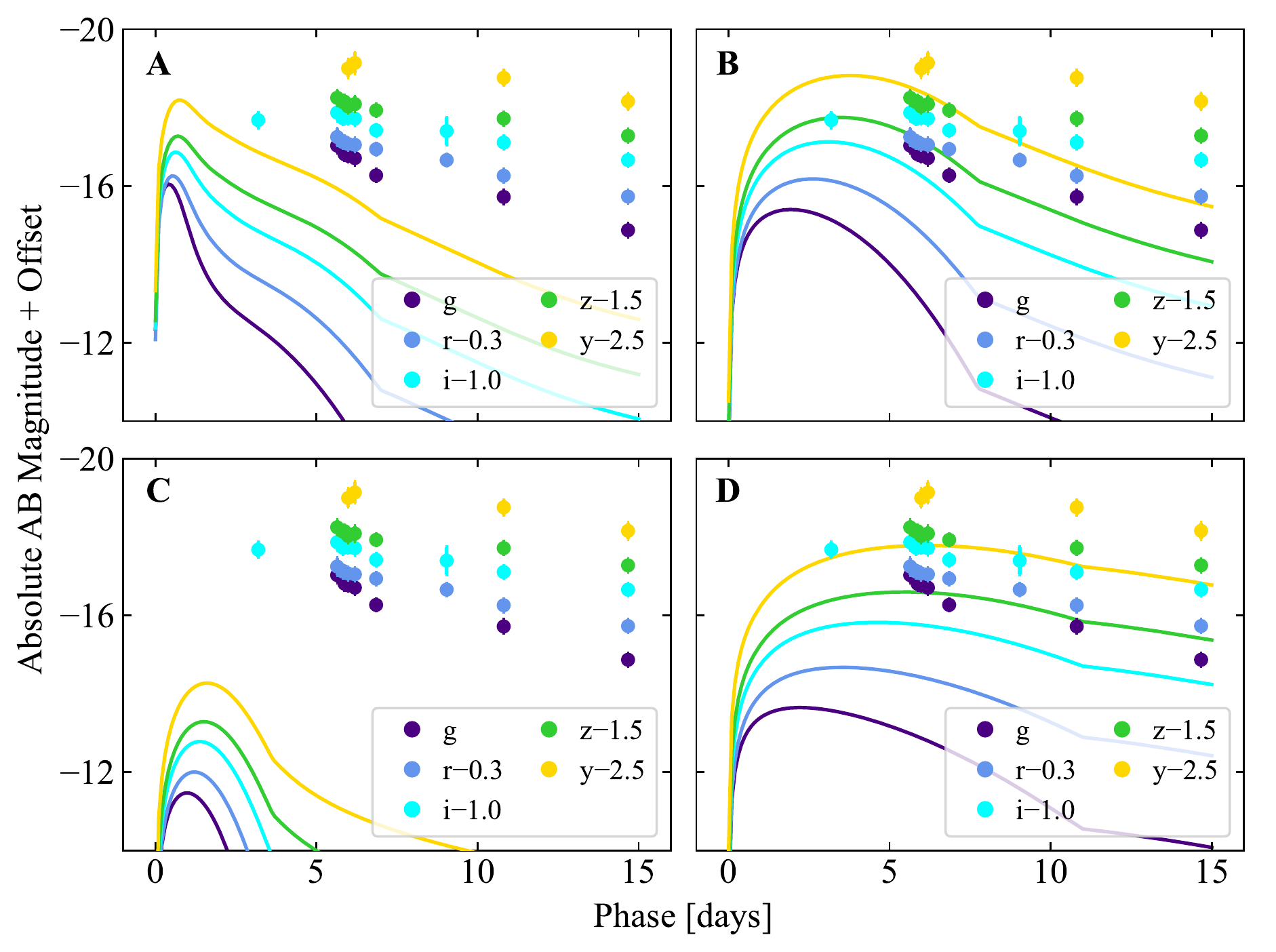}
\caption{Comparison of kilonova models with \wxt\ data. Each panel shows our $grizy$ data of \wxt\ (offset by constant magnitude values -- reported in the legend -- for presentation purposes) together with a kilonova model from \cite{2021MNRAS.505.3016N}. {\bf A:} `bright and blue' case. {\bf B:}  `bright and red'. {\bf C:} `highest chirp mass'. {\bf D:} `highest ejecta mass'. The parameters of each model are reported in the text. The phase is with respect to the time of \gwtrig. }
\label{fig:knmodels}
\end{figure*}

The light curves shown in Fig.\,\ref{fig:knmodels} represent four extreme cases: model A is a `bright blue' case with $m_1=m_2=1.67\,\mathrm{M_\odot}$ (chirp mass of $1.45\,\mathrm{M_\odot}$) and an ejecta that is largely influenced by the blue component with cocoon cooling emission. This model had the brightest emission in the $g$-band out of all the models and, given the blue colour of \wxt\ at peak, it was chosen for comparison. We see that while there are similarities surrounding peak brightness, the rapid decline and reddening of the model is very unlike \wxt\ . Model B is a `bright red' case with asymmetric masses $m_1=1.34\,\mathrm{M_\odot}$ and $m_2=0.81\,\mathrm{M_\odot}$ with a total ejecta mass of $0.096\,\mathrm{M_\odot}$, that is largely influenced by the red component and produces the brightest emission in the y-band. The $i$, $z$ and $y$ bands of this model are close to observed data at early times, however, \wxt\ is still too blue and does not decline as fast as the model at later times. In an attempt to match both the longevity and brightness of \wxt, we made further comparisons to models that had very large masses. Model C is a case with the largest chirp mass of $1.9\,\mathrm{M_\odot}$, and largest binary system mass with $m_1=m_2=2.18\,\mathrm{M_\odot}$. Model D is a case that yielded the largest ejecta mass of $0.13\,\mathrm{M_\odot}$, with very asymmetric masses $m_1=1.97\,\mathrm{M_\odot}$ and $m_2=0.99\,\mathrm{M_\odot}$. We found that model D produced the best match to \wxt\ in terms of longevity, but not in brightness or colour evolution. Model C produced a very faint light curve, due to most of the neutron star mass becoming bound in the remnant with very little being ejected. We note that it would be possible to match the observed brightness in model D by increasing the total mass of the binary system, but it would require a primary with an unrealistically large mass. While we do not simulate here kilonovae from neutron star - black hole (NS-BH) mergers, we argue that these would also fail at reproducing the observed properties of \wxt, given the extreme requirements in terms of ejecta mass and the blue colour at peak.

While BNS and NS-BH mergers are the expected sites of heavy element production via the r-process, they might also produce lighter elements. \cite{Perego2022} calculated light element yields for BNS mergers, and found that He could be present with a number abundance of between $5\times10^{-5} - 10^{-3}$. In contrast, we find a lower limit to the He content of the ejecta that is at least an order of magnitude larger
(Table \ref{tab:TARDIS composition}).

Taken together, the photometric and spectroscopic properties of \wxt\ allow us to rule out a kilonova origin with high confidence.

\subsection{\wxt\ as a peculiar thermonuclear explosion}

Several distinct scenarios involving the disruption of CO white dwarf have been put forward to
explain faint and fast evolving transients. We consider some of these below in the context of \wxt.

Thermonuclear explosions may occur in systems consisting of a CO white dwarf accreting He from a companion star. For certain combinations of binary parameters and accretion rates, the surface He layer may detonate, resulting in an explosion, often dubbed .Ia SN \citep{bildsten07}. Numerous models and predictions are available, and while the detailed physical treatment differs, the consensus is that such explosions should produce faint ($-18 \lesssim M_V/\mathrm{mag} \lesssim -15$) and rapidly evolving transients. Although the decline rate of \wxt\ ($\sim 0.14$ mag/day, $r$-band) can plausibly be matched by some of these models, the spectral features are at odds with model predictions. Detonation of a He shell should result in heavily line blanketed spectra dominated by features due to Ca II and Ti II, and lacking intermediate mass elements \citep{shen10}. While we do detect features due to Ca, and perhaps also He, the overall shape and evolution of the spectra do not provide a convincing match to these models.

The detonation of a white dwarf may also occur via extreme tidal forces due to a black hole \citep{rosswog09}, or in a nuclear dominated accretion flow \citep{metzger12}. This may result from
a chance encounter with a black hole in a dense cluster environment, or via three-body interaction \citep{sell15}. The resulting transient is expected to be faint and rapidly evolving, with peak luminosities and ejecta velocities that are broadly consistent with the observations of \wxt, but the lack of intermediate mass elements in the spectra of \wxt\ is a concern. Also, some fraction of the shredded WD material should fall back onto the black hole generating high-energy photons, but the lack of x-ray detections of \wxt\ over a time span of $\sim$5 months (\S\ref{sec:chandra}) provides another argument against this scenario.

Calcium-strong transients are defined by their strong Ca emission at late times but their early time spectra and light curve properties are diverse, with some suggested to be from a thermonuclear white dwarf origin and some associated with massive stars \citep[see][for a discussion]{de2020}. Typically their spectra at maximum light can be split into those that show He (Ib-like) and those that do not (Ia-like) but there is ambiguity; SN\,2005E  \citep{perets_05E} is most likely associated with an old stellar population given its remote offset galaxy location but showed He in its spectra and so would be classified as Ib-like based on its peak spectra. The origin of the thermonuclear class of Ca-strong transients is uncertain, with \cite{perets_05E} suggesting the detonation of He-shell on the surface of the white dwarf as a likely explanation, although this has not been proven. Alternate models have been suggested, such as the disruption of a CO white dwarf by a hybrid HeCO white dwarf \citep{zenati2022arXiv} or the tidal disruption of a white dwarf by a intermediate-mass hole but the predicted X-ray signature was not detected  \citep{sell15, sell2018}. Based on the presence of He in the spectrum of \wxt\ and its association with a massive star-forming host, we conclude that \wxt\ is not associated with any of these  thermonuclear scenarios, although \revthree{the} Ca-strong scenario can not be conclusively ruled out.

\subsection{\wxt\ as a peculiar CCSN}

After discounting the possibilities of \wxt\ being a thermonuclear SN or a genuine GW counterpart, we consider the possibility that it is a peculiar core-collapse supernova (CCSN).

Multiple lines of evidence now point towards relatively low mass ($10 - 15$\,\msun) progenitors in binary systems giving rise to the majority of Type Ibc SNe \citep[e.g.][]{Yoon10,Eldridge13}. For these supernovae, a binary companion strips the progenitor of its H (and in some cases He) envelope. However, even after binary stripping the pre-explosion mass is still typically a few \msun\ \citep[e.g.][]{Vartanyan21}, while ejecta masses in Type Ibc SNe are generally in the range of $0.5 - 4$\,\msun\ \citep{Lyman16,Barbarino21}.
However, it is possible in some cases for binary evolution to result in a pre-explosion progenitor mass of only $\sim 1.5$\,\msun, which will undergo Fe core-collapse but produce only a few 0.1\,\msun\ of ejecta \citep{Tauris13}. Such supernovae are often referred to as ultra-stripped SNe (USSNe), and can occur in a close binary containing a He star and a NS. If the He star expands at the end of core He-burning, then so-called Case BB Roche-Lobe overflow can occur onto the NS. This process can produce an almost bare C/O core that is slightly above the Chandrasekhar mass, and that will hence explode as an Fe core-collapse SN.

A number of very rapidly evolving H-deficient SNe have been discovered during optical transient surveys, with absolute magnitudes ranging from $-16$ to $-$19 in the \textit(r)-band (e.g. SN\,2002bj, \citealp{Poznanski10,Perets11}; SN\,2005E \citealp{perets_05E} SN\,2005ek, \citealp{Drout05ek}; SN\,2010X, \citealp{Kasliwal10X}; SN\,2014ft, \citealp{De14ft}; SN\,2018kzr \citealp{18kzr_mcbrien}; SN\,2019bkc, \citealp{19bkc_chen,19bkc_prentice}; SN\,2019dge, \citealp{Yao20}). We compare the bolometric lightcurves of a subset of these SNe to \wxt\ in Fig. \ref{fig:bolometric}, and find good matches in both timescale and luminosity (especially for SN~2005ek; \citealp{Drout05ek}). Most of these events (SNe\,2005ek, 2010X, 2014ft, 2018kzr, 2019bkc and potentially 2002bj\footnote{See \citep{Kasliwal10X} for a discussion of He versus Al line identifications.}) do not show spectra consistent with He-rich ejecta material. SN 2014ft does show early He emission features most likely resulting from He-rich CSM and an early flux excess in its light curve \citep{De14ft} but no He in its underlying ejecta spectra. Of these fast-evolving transients listed above, only SNe\,2005E and 2019dge show signatures of He absorption in their spectra and both are on the fainter end of distribution of absolute peak magnitudes at $-$15.5 and $-$16.3 mag, respectively. Similarly to SN 2014ft, SN 2019dge shows signatures of interaction with He-rich CSM at early times and both have been suggested to result from the explosions of ultra-stripped stars \citep{Tauris13,De14ft, Yao20}\footnote{We caution however that not all of these events may be CCSNe: although the spectra of SN\,2005E show clear signature of He absorption, it is considered unlikely to be an ultra-stripped SN due to the lack of recent star formation at the SN location in the halo ($\sim11$ kpc from the centre) of its S0/a host galaxy \citep{perets_05E}.}

\begin{figure}[h!]
 \includegraphics[width=\columnwidth]{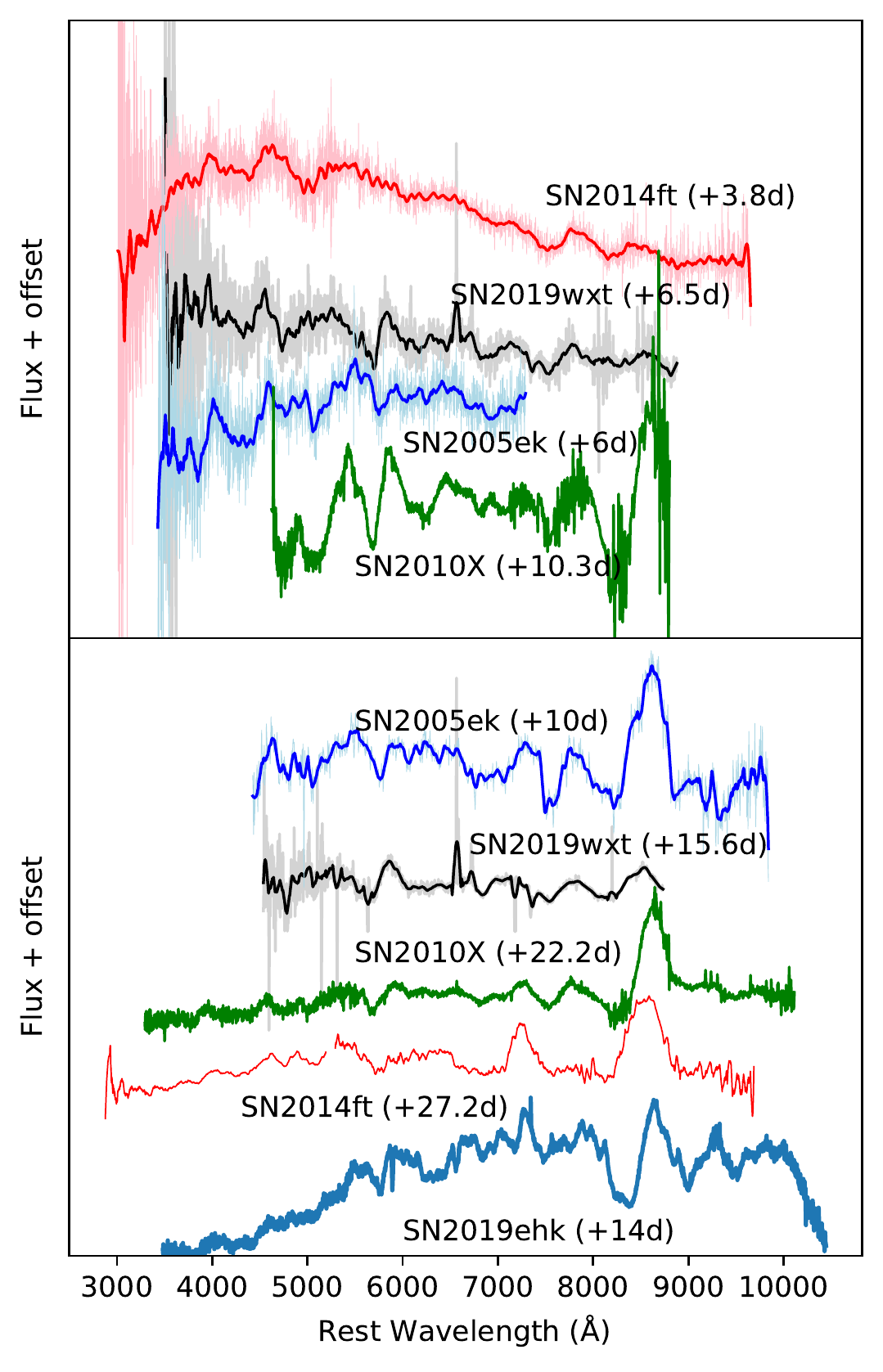}
 \caption{Spectral comparison of \wxt\ at +6.5~d (upper panel) and +15.6~d (lower panel) to a sample of ultra-stripped SNe at similar phases. Low S/N spectra have been smoothed with a Savitzky-Golay filter (with a 100\AA\ window) for presentation purposes. We also show the possible Type IIb ultra-stripped SN 2019ehk in the lower panel. The flat topped line at $\sim 6500$\AA\ that has been suggested to be H$\alpha$ \citep{De21} is either much weaker or absent in \wxt.}
 \label{fig:spec_comp}
\end{figure}

We compare to a set of ultra-stripped SNe in Fig. \ref{fig:spec_comp}, namely SNe 2010X \citep{Kasliwal10X},  2005ek \citep{Drout05ek} and  2014ft \citep{De14ft}. The comparison is made harder by the low S/N, however it is clear that many of the broad features seen in the spectrum of \wxt\ are consistent with those seen in other ultra-stripped SNe. In particular, the strong 
He~{\sc i}~$\lambda$5876
line is seen prominently in both \wxt\ at +6.5~d and in SN\,2010X at +10.3~d, while at later phases (lower panel in Fig. \ref{fig:spec_comp}) we also see good agreement in the red part of the spectrum (albeit with a weaker Ca NIR triplet in \wxt). The presence of He rules out at least some CCSNe scenarios in \cite{Tauris15}.

One puzzle posed by \wxt\ is that a large fraction of the ejecta is Ni. While typical Type Ibc SNe are found to have $f_{Ni}\lesssim0.1$ \revthree{(for example, \cite{2022arXiv220905552R} find a value of $0.042^{+0.031}_{-0.018}$ for Type Ic SNe)}; in the case of \wxt\ we find $f_{Ni}=0.19$. Interestingly, SN 2014ft also showed a surprisingly high Ni fraction of $0.17 - 0.33$ \citep{De14ft}. Turning to theoretical calculations, the 3-D explosion models from \cite{2019MNRAS.484.3307M} do not predict ejected $^{56}$Ni masses, however the total mass of iron-group elements (which must be greater than the $^{56}$Ni mass) for ultra-stripped SNe is 0.01 to 0.04 \msun. Nucleosynthesis calculations for ultra-stripped SNe were also presented by \cite{2017MNRAS.466.2085M}, who suggest that $^{56}$Ni masses of 0.03 \msun are plausible. Alternatively, the luminosity of \wxt\ may be supplemented by a central engine (viz. accretion or spin-down energy from a neutron star, as suggested by \citealt{Sawada22} for SN~2019dge), \revthree{or through shock breakout and cooling \citep[as explored by][for SN2014ft]{2023arXiv230403360K}.}

\section{Ultra stripped SNe as contaminants for KN searches}
\label{sect:contaminant}

\subsection{Volumetric rate estimate}
\label{sec:rate_estimate}

A rough estimate of the local volumetric rate of \wxt-like objects can be obtained exploiting the fact that this event was found in a search for a counterpart to the S191213g GW event: this amounts to one event over the effective time-volume of our search, $R_\mathrm{0,wxt-like}\sim 1/V_\mathrm{eff}T$. Since the transient was discovered by Pan-STARRS approximately $5\,\mathrm{d}$ after the GW public alert, and since the average waiting time for a Poisson process is equal to the mean time separation between events, we can take $T\sim 5\,\mathrm{d}$. We estimate the effective volume as $V_\mathrm{eff}=(\Omega_\mathrm{GW 90\times PS1}/4\pi)V_\mathrm{c,PS1}$, where $\Omega_\mathrm{GW 90\times PS1}$ is the portion of the GW 90\% localisation region that is visible to PS1 (given its declination constraint $\mathrm{dec}>-30\,\mathrm{deg}$), and $V_\mathrm{c,PS1}$ is the comoving volume within the distance out to which PS1 would have been sensitive to a \wxt-like transient.  Considering the peak magnitudes $grizy = (19.1, 19.12, 19.17, 19.26, 19.36)$~mag from Table \ref{tab:optical} and the PS1 limiting magnitudes\footnote{\url{https://panstarrs.stsci.edu/}} $grizy < (22.0, 21.8, 21.5, 20.9, 19.7)$, the source would have been detectable in principle out to $d_\mathrm{L}\sim 490\,\mathrm{Mpc}$ in three bands ($gri$), and out to $d_\mathrm{L}\sim 660\,\mathrm{Mpc}$ in one band ($g$). Taking the smaller limiting distance among the two, we have $V_\mathrm{eff}\sim 3.2\times 10^{-2}\,\mathrm{Gpc^3}$. This yields $R_\mathrm{0,wxt-like}\sim 2.7_{-2.3}^{+6.4}\times 10^{3}\,\mathrm{Gpc^{-3}\,yr^{-1}}$, (median and symmetric 90\% credible interval of the rate posterior assuming a Jeffreys $p(R) \propto 1/\sqrt{R}$ prior on the Poisson process rate).
This credible interval comprises $0.4$\% to $10$\%  of the volumetric rate of core-collapse supernovae $R_\mathrm{CCSN}\sim 9.1\times 10^{4}\,\mathrm{Gpc^{-3}\,yr^{-1}}$ \citep{2021MNRAS.500.5142F}.

\subsection{Volumetric rate limits from simulations}
In order to validate our simple rate estimate from the previous section, we estimated the rate of \wxt-like transients in the Palomar Transient Factory \citep[PTF;][]{PTF_REF, 2009PASP..121.1334R}. PTF was an automated optical sky survey that observed in, predominantly, the Mould R-band between 2009 to 2012. Covering over 8000 deg$^{2}$ with cadences from one to five days, PTF is an excellent archival resource to search for \wxt-like events. Supernova rates in PTF have been extensively studied and the detection efficiencies are well-understood \citep{Frohmaier17}. \citet{2021MNRAS.500.5142F} presented the rates of core-collapse and stripped-envelope supernovae from PTF, allowing us to adopt their method and simulated survey footprint to calculate an intrinsic \wxt-like rate. Firstly, we searched the footprint for candidate supernovae with a SN Ib, Ic, IIb or inconclusive spectroscopic classifications. We also included photometrically-identified candidates with three or more detections on their light curve. We visually inspected the resulting 34 candidate supernovae and found zero events with similar brightness and rapid light curve evolution as \wxt. This is corroborated by \citet{2020ApJ...895L..23C}, who also found no fast-transients in their search of PTF data. We simulated a sample of \wxt-like supernovae in PTF following the methods described in \cite{2021MNRAS.500.5142F}: we assumed a narrow Gaussian spread in brightness  $\mathrm{M}_{r}$ = $-17\pm\, 0.2$ mag and a maximum reliable detection distance of $d_\mathrm{L}\sim 180\,\mathrm{Mpc}$. When compared to zero observed events in the data, the simulations allow us to place a $3\sigma$ upper-limit on the \wxt-like rate of $9\times 10^3\,\mathrm{Gpc^{-3}\,yr^{-1}}$, which is compatible with our estimate in the previous section. This is $\lesssim\,$10$\%$ of the core-collapse SN rate and $\lesssim\,$38$\%$ of the stripped-envelope SN rate from \cite{2021MNRAS.500.5142F}.

\subsection{Comparison with theory and literature}
Our estimated volumetric rate is in agreement with expectations for ultra-stripped supernovae \citep{Tauris13} obtained from population synthesis models.  In particular, using the COMPAS binary population synthesis code \citep{COMPAS}, we find that USSNe comprise between $\approx 1\%$ and $7\%$ of CCSNe, depending on the assumptions made. In particular, the relative rate of USSNe decreases with more stringent assumptions on how close the mass-transferring post He-main sequence donor needs to be in order to fully strip the envelope \citep{Tauris15}, rises with increasing metallicity, and is further sensitive to a number of stellar and binary evolution assumptions such as the typical sizes of natal kicks that may disrupt binaries or the type of accretor that may enable ultra-stripping.  

We also performed a search for USSNe in the Binary Population And Spectral Synthesis simulations \citep{eldridge2017,stanway2018,stevance2020} by selecting hydrogen poor supernovae (as in \citealt{stevance2021}) with ejecta masses $<$0.35 \msun based on the observations of \cite{Yao20}. 
A key take home point from the BPASS search is that USSNe (as defined by their low ejecta mass) are not necessarily associated with the secondary star of the system and can occur at the end of the life of the primary. 
They are however not natively found in our single star models, even at twice solar metallicity, where wind mass-loss is strongest. 
Consequently, although USSNe are not necessarily the second SN in a system they are direct byproducts of binary interactions.
The rate of USSNe in BPASS are about half those seen in COMPAS but in general agreement. 
We find that USSNe account for 0.6 to 3.8\% of CCSNe at SMC metallicity and twice solar metallicity, respectively.

Our estimated rate is also consistent with other observations: for example, \cite{Yao20} conclude that the rate of USSNe similar to SN~2019dge is between 1.4 and $8.2\times 10^{3}\,\mathrm{Gpc^{-3}\,yr^{-1}}$.

\subsection{Expected rate in searches for EM counterparts to GW candidates}
The rapid evolution of SN2019wxt and its initially featureless spectrum made it a relevant contaminant in the search for an EM counterpart to the S191213g GW event candidate, leading to a massive observational effort to characterise it. Here we address the question of how frequently we should expect such type of objects to appear in GW-related searches in the near future. To that purpose, we considered the predicted distribution of 90\% credible binary neutron star merger GW sky localisation areas in O4\footnote{The distribution retains a similar shape in O5 as well, see Figure 2 of \citet{Petrov2022} and Figure 6 of \citet{Abbott2020LRR}.} from \citet{Petrov2022}, and computed the expected weekly number of events with a volumetric rate density equal to that of SN2019wxt (as estimated in sec.~\ref{sec:rate_estimate}) that happen within the extent of such localisation areas and within a luminosity distance $d_\mathrm{L,max}=500$ Mpc (which we take as a representative detectability distance for these kind of events).
\begin{figure}[t]
    \centering
    \includegraphics[width=\columnwidth]{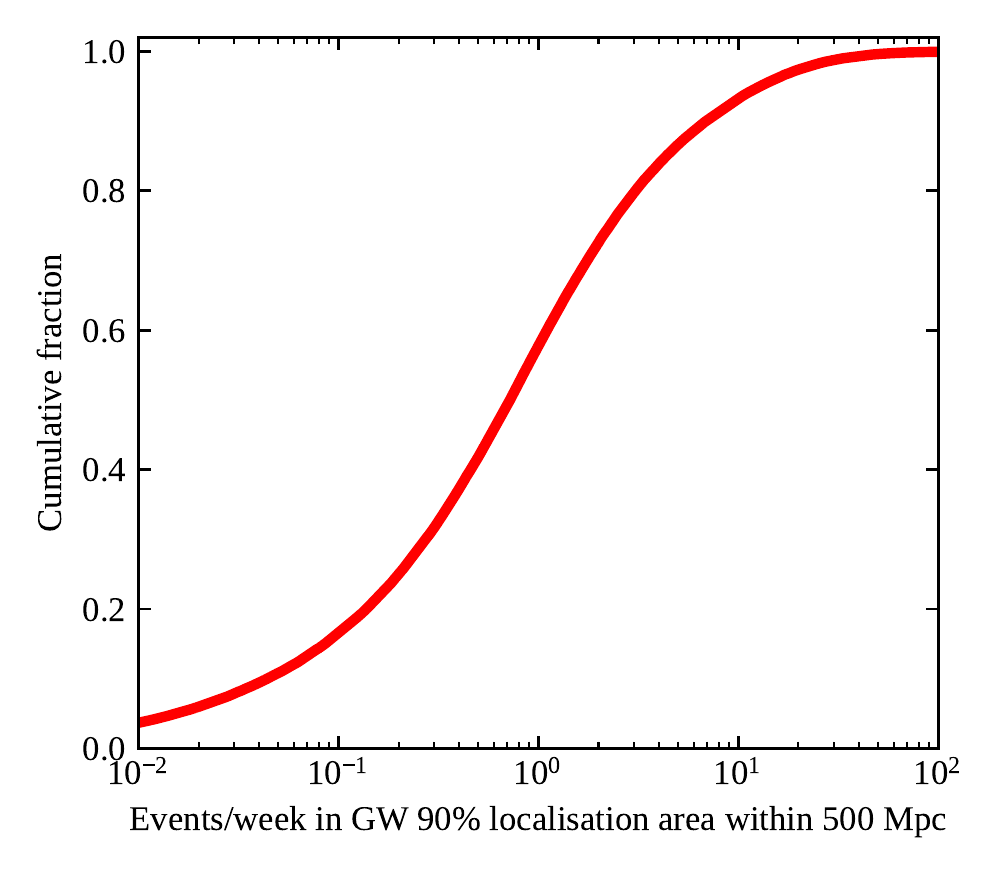}
    \caption{Cumulative distribution of the expected number of SN2019wxt-like events per week within 500 Mpc that will happen within the 90\% sky localisation region of GW candidates in O4.}
    \label{fig:contaminant_rate}
\end{figure}
Figure~\ref{fig:contaminant_rate} shows the resulting cumulative distribution, which shows that we can expect $N \sim 10^{0\pm 1}(d_\mathrm{L,max}/500\,\mathrm{Mpc})^{3}$ such events per week (90\% credible range) to take place within the GW localisation area of O4 alerts and within $d_\mathrm{L,max}$. 

\section{Conclusions}

In this paper, we have presented the results of a comprehensive multi-wavelength observational campaign for \wxt. We have shown that these data are consistent with an USSN (a similar conclusion was reached by \citealt{Shivkumar22}), and conclusively rule out an association between \wxt\ and S191213g. The fast declining lightcurve of \wxt\ suggests a small ejecta mass of $\sim$0.1~\msun, while our spectral modelling implies a photosphere comprised mostly of He and O, together with trace amounts of Ca and Fe-group elements.

While a handful of USSNe have been identified before, to our knowledge none have NIR followup at late phases. These new data allow us to track the temperature evolution of \wxt\ to around 1,500~K by +2 months. This is much lower than is typically seen in stripped envelope SNe, and it is possible that the NIR emission is not coming from the ejecta but rather is re-radiation from $\sim10^{-5}\,$\msun\ of dust. Of course, one must also caution that the ejecta is almost certainly optically thin at this phase, and so the treatment of the SED as a blackbody with a defined photosphere may itself be questionable. Moreover, we note that regardless of the interpretation of the late time SED, our results on ejecta and $^{56}$Ni mass from modelling of the lightcurve are unchanged.

We also note that \wxt\ has a relatively high fraction of $^{56}$Ni compared to the total ejecta (close to 20\%)\footnote{As discussed in Sect. \ref{sect:env}, there may also be additional host galaxy reddening of $E(B-V)\sim0.1$~mag, which we have not accounted for. This would actually make the Ni to ejecta ratio even more extreme, as it would mean the SN is brighter at peak, implying a larger ejected $^{56}$Ni mass, while leaving the ejecta mass unchanged.}. This $^{56}$Ni also cannot be mixed too far into the ejecta, as our spectral modelling requires a low Fe-group element mass above the photosphere. A similarly large $^{56}$Ni to ejecta ratio was also seen in SN 2014ft \citep{De14ft}. In principle, this observation can be used to constrain explosion models for USSNe, and we suggest that computational modelling of this would be useful.

Finally, we return to the question of identifying the counterparts to GW triggers, that was our original motivation for the followup campaign for \wxt. It is clear that this is a challenge - the only case where we have succeeded so far was GW170817, which was unusually nearby and well-localised. Identifying counterparts will remain challenging throughout the O4 observing run of LIGO-Virgo-KAGRA as most GW triggers will likely be found at distances of 100--200 Mpc. Compounding this challenge, we have shown here that we can expect to find unrelated fast declining USSNe similar to \wxt\ in many of the localisation volumes of future GW triggers. Another interesting - albiet unhelpful - conclusion from the analysis of \wxt\ presented here is that a faint, rapidly declining lightcurve with late-time NIR emission is not a unique signature of a KN. As efforts continue to find kilonovae without an associated GW trigger, it will be necessary to either secure spectroscopy or make a convincing association with a GRB to rule out \wxt-like events.

In the case of \wxt\, the sequence of spectra taken over the first two days from discovery were all apparently blue and featureless. However, with the benefit of hindsight one can identify broad features in some of these data that are clearly present in later spectra at +6.5~d. Unsurprisingly, only the early spectra with high S/N allowed for broad lines to be retrospectively identified. It is clear that obtaining further spectra with high S/N for apparently blue featureless targets should be a priority.\\[10pt]

\section*{Acknowledgements}

We thank the referee for their careful reading of the manuscript and helpful suggestions. We also wish to acknowledge the valuable contribution of Alex Kann, who sadly passed away during the preparation of this paper. His contributions to the ENGRAVE collaboration will be missed. The full acknowledgments are available in Appendix \ref{sect:acknowledgments}.

\section*{Author contributions}

IA, TA, RB, SF, SG, MGi, BM, JM, MO, ZP, MP-T and JY were co-investigators (PI: MGi) of the e-MERLIN proposal and contributed to radio data analysis and interpretation.  LA contributed to the discussion and manuscript review. FEB helped with the interpretation and contributed to the manuscript. SB served as on-call team member for ENGRAVE during O3 and provided comments and inputs to the preparation of the manuscript. MGB coordinated the working group that interfaces with external facilities and contributed to operations. KB served in the ENGRAVE/HST team. TDB, CCL, EAM and RW contributed PS1 data and processing. MB contributed to governance as a member of the ENGRAVE Governing Council and provided comments on the manuscript. SJB performed photometry on optical and IR images with \texttt{autophot}. EB served in the Governing Council of ENGRAVE contributing in the governance activities. EC performed data reduction and presentation. KCC and MEH led the PS1, Gemini and UKIRT observing, proposals and data management. SC contributed to the revision of the manuscript and to scientific discussions. TWC coordinated the reduction and analysis of GROND data, compared bolometric light curves and joined the weekly discussion meetings during the preparation of the draft. AC served on the on-call operations team triggering VLT observations. SC contributed to the revision of the final text and to scientific discussions. FD provided comments to the paper during the second circulation. PDA contributed to governance as a Governing Council member, served on the operations team and reduced the TNG data. VD contributed to triggering, worked in the ENGRAVE WG-2 spec and reviewed the paper. AFi reduced the FORS2 and GNIRS spectra and contributed to discussions and editing of the paper. AFl served during his PhD programme for 6 ENGRAVE observing subruns as on-call or writing team member. MFr co-led the writing team, produced Figures \ref{fig:lc}, \ref{fig:colour}, \ref{fig:spectra}, \ref{fig:nir_spec}, \ref{fig:spec_comp} and \ref{fig:chandra}, contributed to spectroscopic and photometric analysis, interpretation and paper writing. MFu run the kilonova models and produced Figure \ref{fig:knmodels}. LG served on the on-call operations team and contributed to the revision of the final text. CG contributed to discussion and modelling of dust. JG-R contributed with reduction of MEGARA data. GG participated to the discussion on event rates. JHG led the spectroscopic modelling with $\textsc{tardis}$, and assisted with writing the manuscript. In addition to analysing and interpreting the e-MERLIN data, MGi inspected the VLASS and Aperitif survey data and used it to estimate the host galaxy SFR. BPG calculated the GOTO statistics used in \S\ref{sec:S191213_discovery}. MGr contributed to observations and data reduction. KEH contributed to NOT and VLT observations and data reduction. JH is a member of the ENGRAVE Governing Council and contributed to discussions on dust.
 YDH performed optical observations with GTC. AI carried out ACAM and LIRIS observations and data reduction. LI contributed to the data analysis of the NOT spectrum and provided comments to the manuscript. ZPJ did observation-related duties and participated in early discussions. PGJ is a Governing Council member and PI of part of the WHT and GTC data, helped in reducing the data and contributed to the analysis. DAK provided comments and proofread the paper. ECK analysed the WISE data and provided comments. RK served as a member of the writing team and provided inputs on draft. GL coordinated the WG-POL and provided comments on the manuscript. AJL chairs the Executive Committee, contributed to data collection and led HST observations. JDL is a member of the ENGRAVE operations and spectroscopy teams. KM coordinated the on-call team, scheduled observations and provided scientific interpretation. IM contributed to astrophysical modelling and interpretation. DMS contributed to EMBOSS/ENGRAVE efforts in WHT and GTC data reduction. SM contributed to ENGRAVE and provided comments on the manuscript. AM is a member of the imaging working group and contributed to observations and data reduction. MJM measured and interpreted the molecular gas properties of the host (Table \ref{tab:mhtwo}) and produced the CO spectra (Fig.~\ref{fig:cospec}). MN carried out light curve modelling and worked in the operations team. ANG has been involved in the GROND observations and data reduction. SRO performed the Swift/UVOT data analysis. FO is a member of the operation team and of the WG-SPEC and contributed to the paper review process. SP  is a member of the spectroscopy team and provided comments on the manuscript. RP contributed to ALMA data reduction and analysis. MAPT contributed to coordinate the EMBOSS WHT observations and part of the EMIR and OSIRIS data analysis. EP is a member of the ENGRAVE Governing Council. GP served on the on-call operations team. JQ-V provided comments and suggestions to the writing team. FR has contributed with comments in the first circulation of the paper draft. ARa is the PI of the GROND ToO time project. SR contributed to revise the manuscript. ARo provided comments to the draft. OSS co-led the writing team, produced Figures \ref{fig:sky_localisation} (together with AJL), \ref{fig:L,Teff,Rph}, \ref{fig:SEDs_BB}, \ref{fig:contaminant_rate}, \ref{fig:SNfit_corner}, \ref{fig:SEDs_BB+dust} and \ref{fig:BB+dust_evolution} (and the corresponding pieces of analysis) and derived the SN2019wxt-like transient volumetric rate estimate in sec \ref{sec:rate_estimate}. SSc reduced the X-shooter data and was involved in the ALMA observation. SJS contributed to the PS1 data, light curve and spectral analysis, text and interpretation. KWS is developer and operator of the QUB Pan-STARRS transient science server. MDF contributed to the KN comparison analysis. JS contributed to the text and discussion. SSr contributed with calibrating the Pan-STARRS and UKIRT photometry and modelling the bolometric light curve. RLCS provided comments on and contributed to editing the manuscript. DS is a member of the ENGRAVE governing council and was involved in discussions around this object from the start. HFS conducted the search within the BPASS fiducial models and contributed text. VT participated to pipelines development, paper layout drafting and is an on-duty operations member. SDV is a member of the ENGRAVE Executive Committee and provided comments to the manuscript. DV contributed to the astrophysical interpretation. DW provided comments on the manuscript. KW served on the on-call operations team. LW was part of the on call operation team. SY served on the on-call operations team. DRY developed and maintains many of the software tools essential to the work of the consortium.

\bibliographystyle{aa} 
\bibliography{references}

{\footnotesize 
\begin{enumerate}[label=$^{\arabic*}$]
\item Instituto de Astrof\'isica de Andaluc\'ia (IAA-CSIC), Glorieta de la Astronom\'ia s/n, 18008 Granada, Spain
\item INAF - Osservatorio di Astrofisica e Scienza dello Spazio di Bologna, via Piero Gobetti 93/3, 40129 Bologna, Italy
\item Shanghai Astronomical Observatory, Chinese Academy of Sciences, Shanghai 200030, China
\item Peng Cheng Laboratory, Shenzhen 518066, China
\item Instituto de Astrof{\'{\i}}sica and Centro de Astroingenier{\'{\i}}a, Facultad de F{\'{i}}sica, Pontificia Uni versidad Cat{\'{o}}lica de Chile, Casilla 306, Santiago 22, Chile
\item Millennium Institute of Astrophysics (MAS), Nuncio Monse{\~{n}}or S{\'{o}}tero Sanz 100, Providencia, Santiago , Chile
\item Space Science Institute, 4750 Walnut Street, Suite 205, Boulder, Colorado 80301, USA
\item INAF - Osservatorio Astronomico di Padova, I-35122 Padova, Italy
\item INAF - Brera Astronomical Observatory, via Bianchi 46, 23807, Merate (LC), Italy
\item Jodrell Bank Centre for Astrophysics, Department of Physics and Astronomy, The University of Manchester, M13 9PL, UK
\item Space Telescope Science Institute, 3700 San Martin Drive, Baltimore, MD 21218, USA
\item Institute for Astronomy, University of Hawaii, 2680 Woodlawn Drive, Honolulu, HI 96822, USA
\item Gran Sasso Science Institute, Viale F. Crispi 7, I-67100, L'Aquila (AQ), Italy
\item INFN - Laboratori Nazionali del Gran Sasso, I-67100, L'Aquila (AQ), Italy
\item School of Physics, O'Brien Centre for Science North, University College Dublin, Belfield, Dublin 4, Ireland
\item INAF - Osservatorio Astronomico d'Abruzzo, Via M. Maggini s.n.c.~, I-64100 Teramo, Italy
\item INAF - Osservatorio Astronomico di Roma, Via di Frascati 33, 00078 Monteporzio Catone (RM), Italy
\item Instituto de Astrof\'{i}sica de Canarias, E-38205 La Laguna, Tenerife, Spain
\item Departamento de Astrof\'\i{}sica, Universidad de La Laguna, E-38206 La Laguna, Tenerife, Spain
\item GRANTECAN, Cuesta de San Jos\`e s/n, E-38712 Bre\~na Baja, La Palma, Spain
\item Universit\'e Paris Cit\'e, CNRS, Astroparticule et Cosmologie, F-75013 Paris, France
\item The Oskar Klein Centre, Department of Astronomy, Stockholm University, AlbaNova, SE-10691 Stockholm, Sweden
\item Max-Planck-Institut f{\"u}r Extraterrestrische Physik, Giessenbachstra\ss e 1, 85748, Garching, Germany
\item INAF - Istituto di Radioastronomia, Bologna, via Gobetti 101 40127 Bologna, Italy
\item ASI Science Data Centre, Via del Politecnico snc, 00133 Rome, Italy
\item European Centre for Theoretical Studies in Nuclear Physics and Related Areas (ECT$\star$), Fondazione Bruno Kessler, Trento, Italy
\item INFN-TIFPA, Trento Institute for Fundamental Physics and Applications, Via Sommarive 14, I-38123 Trento, Italy
\item GSI Helmholtzzentrum f\"ur Schwerionenforschung, Planckstra\ss e 1, 64291 Darmstadt, Germany
\item Konkoly observatory, ELKH Research Centre for Astronomy and Earth Sciences, Konkoly Thege Mikl\'os \'ut 15-17, H-1121 Budapest, Hungary
\item CSFK, MTA Centre of Excellence, Konkoly Thege Mikl\'os \'ut 15-17, H-1121 Budapest, Hungary
\item Institute of Physics, ELTE E\"otv\"os Lor\'and University, P\'azm\'any P\'eter s\'et\'any 1/A, H-1117 Budapest, Hungary
\item School of Physics and Astronomy, University of Southampton, Southampton, SO17 1BJ, UK
\item Astrophysics Research Centre, School of Mathematics and Physics, Queen's University Belfast, BT7 1NN, UK
\item Institute of Space Sciences (ICE, CSIC), Campus UAB, Carrer de Can Magrans, s/n, E-08193 Barcelona, Spain
\item Institut d’Estudis Espacials de Catalunya (IEEC), E-08034 Barcelona, Spain
\item DARK, Niels Bohr Institute, University of Copenhagen, Jagtvej 128, 2200 Copenhagen N, Denmark
\item Universit\'a di Bologna, via Zamboni 33, I-40126 Bologna (BO), Italy
\item Institute for Gravitational Wave Astronomy and School of Physics and Astronomy, University of Birmingham, Edgbaston, Birmingham B15 2TT, UK
\item Astronomical Observatory, University of Warsaw, Al. Ujazdowskie 4, 00-478 Warszawa, Poland
\item Cosmic Dawn Center (DAWN), Jagtvej 128, T\r{a}rn I, 2200 Copenhagen, Denmark
\item Niels Bohr Institute, University of Copenhagen, Lyngbyvej 2, 2100 Copenhagen N, Denmark
\item Centre for Astrophysics and Cosmology, Science Institute, University of Iceland, Dunhagi 5, 107 Reykjav\'ik, Iceland
\item Departamento de F\'isica Te\'orica y del Cosmos, Universidad de Granada, E-18071 Granada, Spain
\item Department of Astrophysics/IMAPP, Radboud University, P.O.~Box 9010, 6500 GL, Nijmegen, The Netherlands
\item SRON, Netherlands Institute for Space Research, Niels Bohrweg 4, 2333~CA, Leiden, The Netherlands
\item Key Laboratory of Dark Matter and Space Astronomy, Purple Mountain Observatory, Chinese Academy of Sciences, Nanjing 210008, China
\item Department of Physics and Astronomy, University of Turku, Vesilinnantie 5, Turku, FI-20014, Finland
\item DTU Space, National Space Institute, Technical University of Denmark, Elektrovej 327, 2800 Kongens Lyngby, Denmark
\item Department of Physics, University of Warwick, Coventry, CV4 7AL, UK
\item School of Physics, Trinity College Dublin, University of Dublin, College Green, Dublin 2, Ireland
\item School of Physics and Astronomy, Monash University, Clayton, Victoria 3800, Australia
\item Joint Institute for VLBI ERIC, Oude Hoogeveensedijk 4, 7991 PD Dwingeloo, The Netherlands
\item School of Sciences, European University Cyprus, Diogenes street, Engomi, 1516 Nicosia, Cyprus
\item Astronomical Observatory Institute, Faculty of Physics, Adam Mickiewicz University, ul.~S{\l}oneczna 36, 60-286 Pozna{\'n}, Poland
\item Th\"uringer Landessternwarte Tautenburg, Sternwarte 5, 07778 Tautenburg, Germany
\item Departamento de Ciencias Fisicas, Universidad Andres Bello, Avda. Republica 252, Santiago, Chile
\item Universit\'{a} degli Studi di Milano-Bicocca, Piazza della Scienza 3, I-20126 Milano (MI), Italy
\item INFN - Sezione di Milano-Bicocca, Piazza della Scienza 3, I-20126 Milano (MI), Italy
\item The Oskar Klein Centre, Department of Physics, Stockholm University, AlbaNova, SE-10691 Stockholm, Sweden
\item School of Physics and Astronomy, University of Leicester, University Road, LE1 7RH, UK
\item The Department of Physics, The University of Auckland, Private Bag 92019, Auckland, New Zealand
\item Special Astrophysical Observatory, Russian Academy of Sciences, Nizhnii Arkhyz, 369167, Russia
\item GEPI, Observatoire de Paris, PSL University, CNRS, 5 Place Jules Janssen, 92190 Meudon, France
\item Goethe University Frankfurt, Max-von-Laue-Strasse 1, Frankfurt am Main 60438, Germany
\item Physics Department, Lancaster University, Lancaster, LA1 4YB, UK
\item Department of Space, Earth and Environment, Chalmers University of Technology, Onsala Space Observatory, SE-439 92 Onsala, Sweden
\item Observatorio Astron\'omico de Quito, Escuela Polit\'ecnica Nacional, 170136, Quito, Ecuador
\end{enumerate}

}

\begin{appendix}

\section{Full acknowledgments}\label{sect:acknowledgments}
IA acknowledges support from the Spanish MCINN through the ``Center of Excellence Severo Ochoa'' award for IAA-CSIC (SEV-2017-0709), and through grants AYA2016-80889-P and PID2019-107847RB-C44. LA acknowledges support from the Italian Ministry of Research through grant PRIN MIUR 2020 – 2020KB33TP METE. FEB acknowledges support from CONICYT Basal AFB-170002 and the Ministry of Economy through grant IC120009 to The Millennium Institute of Astrophysics (MAS). MGB acknowledges support from ASI grant I/004/11/5. MB acknowledges support from MIUR PRIN 2017, grant 20179ZF5KS. SJB thanks the Science Foundation Ireland and the Royal Society (RS-EA/3471). EB acknowledges support from the GRAWITA grant funded by INAF. MDC-G and YDH acknowledge support from the Ram\'on y Cajal Fellowship RYC2019-026465-I (funded by the MCIN/AEI/ 10.13039/501100011033 and the European Social Funding). EC acknowledges support from MIUR PRIN 2017. TWC acknowledges Marie Sklodowska-Curie grant H2020-MSCA-IF-2018-842471. AJCT acknowledges support from the Spanish Ministry project PID2020-118491GB-I00 and Junta de Andalucia grant P20\_010168. PDA acknowledges support from ASI grant I/004/11/5 and from MIUR PRIN 2017, grant 20179ZF5KS. AF acknowledges the support of the ERC under the EU Horizon 2020 research and innovation program (ERC Advanced Grant KILONOVA No. 885281). MFr is supported by a Royal Society - Science Foundation Ireland University Research Fellowship. LG acknowledges RYC2019-027683-I, PID2020-115253GA-I00 \& PIE20215AT016 grants. CG is supported by a VILLUM FONDEN Young Investor Grant (project number 25501). JG-R acknowledges support from Spanish AEI under Severo Ochoa Centres of Excellence Programme 2020-2023 (CEX2019-000920-S), and from ACIISI and ERDF under grant ProID2021010074. GG acknowledges the PRIN MIUR `Figaro' for financial support. MG is supported by EU Horizon 2020 programme under grant No 101004719. KEH acknowledges support by a Project Grant (217690-051) from The Icelandic Research Fund. JH was supported by a VILLUM FONDEN Investigator grant (project number 16599). AI acknowledges the research programme Athena with project number 184.034.002, which is financed by the Dutch Research Council (NWO). LI was supported by research grants from VILLUM FONDEN (proj. 16599, 25501). ZPJ has been supported by NSFC under grant No. 11933010. DAK acknowledges support from Spanish National Research Project RTI2018-098104-J-I00 (GRBPhot). ECK acknowledges support from the G.R.E.A.T.\ research environment funded by the Vetenskapsr\aa det, and from The Wenner-Gren Foundations. GL was supported by a research grant (19054) from VILLUM FONDEN. AJL has received funding from the European Research Council search Council (ERC) via grant number 725246. JDL acknowledges support from a UK Research and Innovation Future Leaders Fellowship (MR/T020784/1). KM EU H2020 ERC grant no. 758638. IM is partially supported by OzGrav (ARC project CE17010000). BM acknowledges support from the Spanish MCINN under grant PID2019-105510GB-C31 and through the Mar\'ia de Maeztu award CEX2019-000918-M. DMS acknowledges support from the ERC under Horizon 2020 programme (No. 715051), as well as the Gobierno de Canarias and ERDF (ProID2020010104). AM acknowledge support from ASI grant I/004/11/3. MJM acknowledges the National Science Centre, Poland grant 2018/30/E/ST9/00208. DMS acknowledges support from the Gobierno de Canarias and ERDF (ProID2021010132); as well as from the Spanish Ministry of Science and Innovation via an Europa Excelencia grant (EUR2021-122010). JM acknowledges support from the Spanish MCINN through the ``Center of Excellence Severo Ochoa'' award to the IAA-CSIC (SEV-2017-0709), from the grant RTI2018-096228-B-C31 (MICIU/FEDER, EU) and the grant IAA4SKA P18-RT-3082 (Reg. Govt. of Andalusia). MN is supported by ERC grant 948381 and by a Turing Fellowship. ANG acknowledges support to TLS. FO acknowledges support from the GRAWITA/PRIN project `The new frontiers of the Multi-Messenger Astrophysics' and from the H2020 grant 871158. MAPT was supported by grants RYC-2015-17854 and AYA2017-83216-P. GP is supported by ANID - Millennium Science Initiative - ICN12\_009. JQV acknowledges support from ANID folio 21180886. AR acknowledges support from Premiale LBT 2013. OSS acknowledges the Italian MUR grant 1.05.06.13 and INAF-Prin 1.05.06.13. RS-R acknowledges support under the CSIC-MURALES project with reference 20215AT009. SS acknowledges support from the G.R.E.A.T.\  research environment, funded by Vetenskapsr\aa det project number 2016-06012. SJS sTFC Grant ST/P000312/1 and ST/N002520/1. RLCS acknowledges funding from STFC. HFS acknowledge the support of the Marsden Fund Council managed through Royal Society Te Aparangi. SDV aknowledges fundings from PNHE of INSU/AA. DV acknowledges the financial support of the German-Israeli Foundation (GIF No. I-1500-303.7/2019). DW is supported by Independent Research Fund Denmark grant DFF-7014-00017. The Cosmic Dawn Center is funded by the Danish National Research Foundation. LW acknowledges support from the Polish NCN DAINA No. 2017/27/L/ST9/03221, EC H2020 OPTICON No. 730890 and ORP No. 101004719. SY has been supported by the Knut and Alice Wallenberg Foundation, and the G.R.E.A.T.\  research environment funded by the Swedish Research Council.

Based on observations collected by the ENGRAVE collaboration at the European Southern Observatory under ESO programmes 1102.D-0353, 0102.D-0348, 0102.D-0350; also on observations collected at the European Southern Observatory under ESO programmes 1103.D-0328 (ePESSTO+) and 1104.A-0380 (by the adH0cc team).
Data for this paper has been obtained under the International Time Programme of the CCI (International Scientific Committee of the Observatorios de Canarias
of the IAC) with the GTC operated on the island of La Palma in the Roque de los Muchachos.

This research made use of \tardis, a community-developed software package for spectral synthesis in supernovae. The development of \tardis\ received support from the Google Summer of Code initiative and from ESA's Summer of Code in Space program. \tardis\ makes extensive use of Astropy and PyNE.
We are grateful for use of the computing resources from the Northern Ireland High Performance Computing (NI-HPC) service funded by EPSRC (EP/T022175).
This research is based on observations made with the NASA/ESA \textit{Hubble} Space Telescope obtained from the Space Telescope Science Institute, which is operated by the Association of Universities for Research in Astronomy, Inc., under NASA contract NAS 5–26555. These observations are associated with program 15980.
GROND observations at La Silla were performed as part of the program 0104.A-9099. Part of the funding for GROND (both hardware as well as personnel) was generously granted from the Leibniz-Prize to Prof. G. Hasinger (DFG grant HA 1850/28-1).
Based (in part) on observations made in the Observatorios de Canarias del IAC with the GTC operated on the Island of La Palma in the Roque de los Muchachos Observatory. This research used telescope time awarded by the CCI International Time Programme ("GTC1-18ITP; Coordinated European follow-up of gravitational wave events").
This work was enabled by observations made from the Gemini North telescope and UKIRT telescopes, located within the Maunakea Science Reserve and adjacent to the summit of Maunakea. We are grateful for the privilege of observing the Universe from a place that is unique in both its astronomical quality and its cultural significance. UKIRT is owned by the University of Hawaii (UH) and operated by the UH Institute for Astronomy. When the data reported here were obtained, the operations were enabled through the cooperation of the East Asian Observatory (EAO). The international Gemini Observatory is a program of NSF’s NOIRLab, which is managed by the Association of Universities for Research in Astronomy (AURA) under a cooperative agreement with the National Science Foundation on behalf of the Gemini Observatory partnership: the National Science Foundation (United States), National Research Council (Canada), Agencia Nacional de Investigaci\'{o}n y Desarrollo (Chile), Ministerio de Ciencia, Tecnolog\'{i}a e Innovaci\'{o}n (Argentina), Minist\'{e}rio da Ci\^{e}ncia, Tecnologia, Inova\c{c}\~{o}es e Comunica\c{c}\~{o}es (Brazil), and Korea Astronomy and Space Science Institute (Republic of Korea).
This paper makes use of the following ALMA data: ADS/JAO.ALMA\#2019.1.01406.T. ALMA is a partnership of ESO (representing its member states), NSF (USA) and NINS (Japan), together with NRC (Canada), MOST and ASIAA (Taiwan), and KASI (Republic of Korea), in cooperation with the Republic of Chile. The Joint ALMA Observatory is operated by ESO, AUI/NRAO and NAOJ

\section{Observational data and reductions}

\subsection{Ground-based imaging}
\label{sect:optical_imaging}

Optical and NIR imaging for \wxt\ was obtained with a number of instruments: the Panoramic Survey Telescope and Rapid Response System \citep[Pan-STARRS1;][]{Chambers2016} telescope  equipped with the Gigapixel Camera 1; the Gamma-Ray Burst Optical Near-IR Detector \citep[GROND;][]{2008PASP..120..405G} mounted on the 2.2-m MPG telescope at ESO's La Silla Observatory; Andalucia Faint Object Spectrograph and Camera (ALFOSC) on the Nordic Optical Telescope (NOT) on La Palma, and the Auxiliary-port CAMera (ACAM) on the \textit{William Herschel} Telescope (WHT) on La Palma \citep{acam}, the Italian 3.6-m Telescopio Nazionale \textit{Galileo} (TNG) telescope, located at the Roque de los Muchachos Observatory on La Palma in the Canary Islands of Spain, and Wide Field Infrared Camera (WFCAM) on the United Kingdom Infra-Red Telescope (UKIRT) in Mauna Kea \citep{2007A&A...467..777C}.

The Pan-STARRS1 system (PS1) comprises a 1.8\,m telescope with a 1.4 Gigapixel camera (GPC1)   
with 0\farcs26 pixels and a field-of-view area of 7.06 sq deg 
\citep{Chambers2016}. It is equipped with a filter system, denoted as \grizy\ as described in \cite{2012ApJ...750...99T}, and the PS1 Science Consortium conducted the 
3$\pi$ Survey of the whole sky north of $\delta = -30\degree$ in these filters.  With 
these images as reference frames, all new images can be 
immediately reduced with the Image Processing Pipeline \citep{magnier2017a,waters2017}, 
including difference imaging. Individual detections from survey operations are 
ingested into the PS1 Transient Server database at Queens University Belfast and assimilated into distinct objects with a time variable history, cross-matched with all catalogued galaxies, AGN, CVs and historical transients \citep{GW150914PanSTARRS} and simultaneously a machine learning
algorithm is applied to image pixel stamps at each transient position \citep{2015MNRAS.449..451W}. 
PS1 works both in general survey mode, currently searching for near-earth objects and 
carrying out a transient survey called the Young Supernova Experiment  \citep[YSE;][]{2021ApJ...908..143J}, or the surveys can be interrupted for specific, targeted photometry of targets-of-opportunity. The
advantage of PS1 in the latter mode is that difference imaging can be immediately applied (since templates exist over 3$\pi$ of the sky) producing reliable photometry.

We obtained a single epoch of observations for \wxt\ on 19 Dec 2019 (+0.91 d) using GROND, which provided multi-band imaging simultaneously with \textit{g', r', i', z', J, H} and $K_{\mathrm{s}}$ bands. The data were reduced using the GROND pipeline \citep{2008ApJ...685..376K} that includes standard procedures including bias and flat-field corrections, stacks images and provides astrometric calibration. 

During the night beginning on 18 Dec 2019 (+0.76 d), we obtained two sets of $griz$ images using the ALFOSC camera at the NOT, separated by a few hours. Standard reduction was applied, subtracting a master bias and correcting with sky flats.
 
We obtained a single epoch of observations in \textit{r, i, z} for \wxt\ on 15 Jan 2020 (+28.73 d) using WHT+ACAM \citep{acam}, with exposure times of $9\times100$, $9\times100$ and $9\times200$ s and 5, 5 and 10\arcsec\ dithering for the respective filters. These data were reduced using standard procedures in {\sc iraf} for bias and flat-field corrections. We used {\sc lacosmic} \citep{vanDokkum2001} for cosmic ray cleaning before aligning and stacking the images within {\sc iraf}.

NIR observations from TNG were carried out using the Near Infrared Camera Spectrometer (NICS) instrument in imaging mode \citep{DAvanzo2019GCN}. A series of images were obtained with the J filter on 18 Dec 2019 starting 19:16:04 UT (+0.70 d). The image reduction was carried out using the {\it jitter} task of the ESO-eclipse package.\footnote{\url{https://www.eso.org/sci/software/eclipse/}}  Astrometry was performed using the 2MASS\footnote{\url{https://irsa.ipac.caltech.edu/Missions/2mass.html}} catalogue.

Photometry in $J$, $H$, $K$ bands was also obtained using the WFCAM, that is equipped with four 2048$\times$2048 HgCdTe detectors, with a 0.2 square degree field of view and a pixel scale of $0\farcs4$. The processed data were downloaded from the Cambridge Astronomy Survey Unit (CASU).

WHT images were obtained using the Long-slit Intermediate Resolution Infrared Spectrograph (LIRIS) on 17, 18 and 19 Jan 2020 (+29.66, +30.73 and +31.58 d) in \textit{H}, \textit{J+K$_\mathrm{s}$} and \textit{H+K$_\mathrm{s}$} bands, respectively.
Data reduction was done using {\sc theli} version 3 \citep{theli1, theli2}, which is a tool for automated reduction of astronomical data, which includes (bright and dim) flat-field corrections, background and collapse corrections (to correct for gradients due to residual reset anomaly), astrometry (to construct the dithering pattern for coaddition), sky-subtraction and co-addition. We manually removed any images where effects from the reset anomaly are still visible before applying the astrometry. 
Likewise, the GTC/EMIR photometric observations were reduced using the {\sc theli} package, albeit with version 2 and not version 3 as the WHT/LIRIS images \citep{theli1, theli2}.

Point-spread function (PSF) fitting photometry was performed on the NOT+ALFOSC, GROND and ACAM images using the {\sc AutoPhOT} code \citep{brennan2022automated}. Photometry was calibrated to catalogued Pan-STARRS sources in the field.

\subsection{{\it Swift}-UVOT}

The Ultraviolet Optical Telescope \citep[UVOT;][]{roming} on-board the {\it Neil Gehrels Swift Observatory} ({\it Swift}) took observations of \wxt\ beginning T0+488.59ks (where T0 is the time of the GW trigger) and detected the source above the host galaxy level in all filters $v$, $b$, $u$, $uvw1$, $uvm2$ and $uvw2$ \citep{Oates2019GCN}. In Jun 2020, after the transient light was no longer detectable, we obtained additional observations of the field of \wxt. Using these template observations we measured the host contribution in the aperture of \wxt\ and used that to obtain host-corrected photometry. 
We downloaded the images from the {\it Swift} data archive.\footnote{\url{https://www.swift.ac.uk/archive/index.php}} The source counts were obtained using a circular region with a 3\arcsec\ radius. In order to be consistent with the UVOT calibration, the count rates were corrected to 5\arcsec\ using the curve of growth contained in the {\it Swift} calibration files.\footnote{\url{https://heasarc.gsfc.nasa.gov/docs/heasarc/caldb/swift/}} Background counts were extracted using a circular aperture of 20\arcsec\ radius from a blank area of sky near to the source position. The count rates were obtained from the image lists using the {\it Swift} tool {\textsc {uvotsource}}. From the template images we measured the host count rate using the same source aperture, corrected the count rate to a 5\arcsec\ radius aperture and subtracted this from the measured source count rate. Finally, the source count rates were converted to magnitudes using the UVOT photometric zero-points \citep{bre11}. The analysis pipeline used UVOT calibration 20170922. Since the UVOT detector is less sensitive in a few small patches\footnote{\url{https://heasarc.gsfc.nasa.gov/docs/heasarc/caldb/swift/docs/\\uvot/uvotcaldb\_sss\_01.pdf}} for which a correction has not yet been determined, we checked to see if \wxt\ falls on these patches in any of our images; but this was not the case.

\subsection{HST}
We obtained UV, optical and IR observations of \wxt\ with
the \textit{Hubble} Space Telescope ({\em HST}). These observations
were obtained on 17-19 Feb 2020, 12-15 Oct 2020 and a final 
epoch on 11 Jan 2021 due to the failure of guidestars during the
optical observations in Oct 2020.\footnote{Observations taken 
in F390W, F475W, F606W and F814W in Oct 2020 were all lost due to a guide star failure. The F606W and F814W observations were repeated in Jan 2021.}

Data were retrieved from the HST archive at MAST\footnote{archive.stsci.edu}, after flat-fielding and bias
correction, and following a correction for the impact
of charge transfer efficiency. The data were subsequently drizzled
to a final pixel scale of 0\farcs025 for UVIS and 0\farcs07 for the IR channel. 

The data clearly show a red source at the location of \wxt\ which
is placed on the complex background of the underlying galaxy (Fig.~\ref{fig:sky_localisation}, right-hand panel). In 
order to estimate the photometry at the time of the first epoch (Feb 2020) we subtract the later data from the earlier images and perform photometry directly on the subtracted images, providing a measurement of the
pure transient light (or a limit thereof). The transient is only detected in the IR, with
non-detections in all UV and optical filters. The resulting photometry
is shown in Table~\ref{tab:hst}.

\begin{table*}
\centering
\setlength{\tabcolsep}{3pt} 
\caption{Log of HST observations of \wxt. Where \wxt\ was not detected on an image, no flux is reported.}
\label{tab:hst}
\begin{tabular}{llcccc}
\toprule
Date          & Instrument & Filter   & Exposure (s) & Flux ($\mu Jy$) \\
\midrule
2020-02-17.1 	& WFC3/UVIS 	& F390W 	& 750.0 	&       - \\
2020-02-17.1 	& WFC3/UVIS 	& F475W 	& 750.0 	&       - \\
2020-02-19.4 	& WFC3/UVIS 	& F606W 	& 750.0 	&       -0.006 $\pm$ 0.010 \\
2020-02-19.5 	& WFC3/UVIS 	& F814W 	& 750.0 	&       -0.005 $\pm$ 0.030 \\
2020-02-19.5 	& WFC3/IR 	    & F125W 	& 1058.8 	&       0.67 $\pm$ 0.04\\
2020-02-19.5 	& WFC3/IR 	    & F160W 	& 1208.8 	&       1.95 $\pm$ 0.06\\
2020-10-14.6	& WFC3/UVIS 	& F225W 	& 2120.000 	& -      \\
2020-10-14.5	& WFC3/IR 	    & F125W 	& 2396.9 	& -      \\
2020-10-14.5	& WFC3/IR 	    & F160W 	& 2396.9 	& -      \\
2020-10-14.7	& WFC3/UVIS 	& F275W 	& 2120.0 	& -      \\
2021-01-11.6 	& WFC3/UVIS 	& F606W 	& 2072.0 	& -      \\
2021-01-11.6 	& WFC3/UVIS 	& F814W 	& 2072.0 	& -      \\
\bottomrule
\end{tabular}
\end{table*}

\subsection{WISE upper limits}

The Wide-field Infrared Survey Explorer (WISE) observed the site of \wxt\ in the $W1$ and $W2$ bands (3.4 and 4.6 $\upmu$m, respectively) on 8 Jan 2020, 20 days after discovery, as part of the NEOWISE Reactivation \citep[NEOWISE-R;][]{mainzer2014} survey. A NEOWISE-R epoch typically consists of $\sim$12--18 exposures across $\sim$2 days, so we used the IRSA NEOWISE Coadder \citep{masci2009} to construct single coadded images in $W1$ and $W2$. We subtracted these images from templates constructed by coadding exposures from previous epochs, obtained well before the transient was first detected. We did not detect a source in the subtracted images at the position of \wxt\ in either filter. We estimate limiting magnitudes in the coadded images of 17.3 in $W1$ and 16.2 in $W2$ in the Vega system. 

\subsection{Optical spectroscopy}

Optical and NIR spectroscopy of \wxt\ was secured from a number of ground based facilities, and a log of all spectroscopic observations is reported in Table \ref{tab:spec}.

\begin{table*}
\centering
\setlength{\tabcolsep}{3pt} 
\caption{Log of spectroscopic observations of \wxt. Spectra with an asterisk beside them contained no flux from the transient.}
\label{tab:spec}
\begin{tabular}{llcccc}
\toprule 
Date          & Phase (d) & Telescope    & Instrument+Grism      & Wavelength (\AA) & Resolution (\AA) \\
\midrule
2019-12-18.8  & 0.68      & LT           & SPRAT+blue            & 4100--7500   & 18    \\ 
2019-12-18.8  & 0.70      & NOT          & ALFOSC+Gr4            & 4000--9640   &  3     \\ 
2019-12-18.9  & 0.75      & GTC          & OSIRIS+R1000B/R2500I  & 3620--9200   & 11/4  \\ 
2019-12-19.0  & 0.90      & NTT          & EFOSC2+Gr13           & 3650--9240   & 18    \\ 
2019-12-19.1  & 0.92      & VLT          & XShooter+UVB/VIS/NIR  & 3200--20300       & 1/1/3 \\ 
2019-12-19.1  & 0.92      & VLT          & FORS2+300V            & 3380--9630   &    3   \\ 
2019-12-19.9  & 1.76      & GTC          & OSIRIS+R1000B/R2500I  & 3620--10150  & 6/4   \\ 
2019-21-21.9  & 3.70      & LT           & SPRAT+red             & 4020--7990   & 18    \\ 
2019-12-24.9  & 6.51      & GTC          & OSIRIS+R1000B/R2500I  & 3620--9200   & 7/5   \\ 
2019-12-26.8  & 8.43      & GTC          & OSIRIS+R1000R         & 5100--9200   & 8     \\ 
2019-12-27.0* & 8.62      & NTT          & EFOSC2+Gr13           & 3650--9240   & 18    \\ 
2020-01-03.3  & 15.66     & Gemini-N     & GMOS+R400             & 4700--9050   & 6     \\ 
2020-01-13.9* & 25.83     & GTC          & OSIRIS+R1000R         & 5100--9200   & 8    \\ 
2020-01-28.8  & 40.29     & GTC          & MEGARA+LRB            & 4330--5230   & 1.0   \\  
2020-01-28.9  & 40.32     & GTC          & MEGARA+LRR            & 6095--7300   & 1.1   \\  
\midrule
2019-12-19.9     & 1.77   & GTC          & EMIR+YJ/HK            &  8500--24200 &  7/13      \\ 
2019-12-19.31   & 1.21 & Gemini       & GNIRS                 &     8000--25000  & 9      \\
2020-01-09.2*   & 22.1 & Gemini       & GNIRS                 &     8000--25000  & 9      \\
2020-01-19.95* & 32.8  & GTC          & EMIR+YJ/KH                 & 8500--24200 &  7/13       \\ 
\bottomrule
\end{tabular}
\end{table*}

Longslit EFOSC2 spectra were taken with Gr\#13 and a 1\farcs0 wide slit. These data were reduced using the PESSTO pipeline; in brief, spectra were overscan and bias subtracted, before being divided by a normalised flat field. Cosmic rays were cleaned using an implementation of the {\sc lacosmic} algorithm \citep{vanDokkum2001}, before one dimensional spectra were optimally extracted, and wavelength calibrated against an arc spectrum taken with the same configuration. A small wavelength shift was then applied to the dispersion solution in order to account for flexure, and bring the wavelengths of the detected sky emission lines into agreement with the expected values. Spectra were flux calibrated using a response function derived from observations of spectrophotometric standard stars, and corrected for second order contamination. Finally, telluric absorptions were removed from the spectrum using a model matched to the strong telluric A and B bands.

GTC+OSIRIS spectra were reduced using standard {\sc iraf} tasks: overscan and bias-subtraction, and flatfielding using a normalised lamp flat. Cosmic rays were identified using the {\sc lacosmic} package, before spectra were optimally extracted. Arc lamp exposures were used to determine the wavelength calibration, while observations of spectrophotometric stars were used to flux calibrate the spectra. Our final GTC spectrum (taken on 13 Jan 2020) contains no flux from the transient.

The NOT+ALFOSC spectrum was reduced using the {\sc alfoscgui} tool\footnote{\url{https://sngroup.oapd.inaf.it/foscgui.html}}, which provides a GUI wrapper to standard {\sc iraf} tasks. Reductions were performed using similar steps as for the EFOSC2 data. 

The GMOS spectrum was obtained using the GMOS-N instrument with the R400 grating and a 1\farcs0 wide slit. The spectrum was reduced using the Gemini \textsc{iraf} package.

The LT+SPRAT spectra were pipeline reduced  using a modified version of the FRODOSpec pipeline \citep{Barnsley12}.\footnote{Described at \url{https://telescope.livjm.ac.uk/TelInst/Inst/SPRAT/}.} Bias, dark and flat field calibrations are applied, before source extraction, sky subtraction and wavelength calibration are performed. The spectra were then flux calibrated within {\sc iraf} using a sensitivity curve derived from a spectrophotometric standard.

The X-shooter data were reduced following \citet{Selsing2019a}. In brief, we used first the tool  {\sc astroscrappy}\footnote{\url{https://github.com/astropy/astroscrappy}}, which is based on the cosmic-ray removal algorithm {\sc lacosmic} \citep{vanDokkum2001}, to remove cosmic-ray hits. Afterwards, spectra were processed using the X-shooter pipeline v3.3.5 and the ESO workflow engine ESOReflex \citep{Goldoni2006a, Modigliani2010a}. The UVB and VIS-arm data were reduced in stare mode to boost the S/N by a factor of $\sqrt{2}$ compared to the standard nodding mode reduction. The individual rectified, wavelength- and flux calibrated two-dimensional spectra files were co-added using tools developed by J.~Selsing.\footnote{\url{https://github.com/jselsing/XSGRB_reduction_scripts}} The NIR spectra were taken using a K-band blocking filter, that increased the S/N in H-band at the expense of reduced wavelength coverage. The NIR data were reduced in nodding mode to ensure a satisfactory subtraction of the night sky-lines. After that, we extracted the one-dimensional spectra of each arm in a statistically optimal way using tools by Selsing. Finally, all spectra were moved to vacuum, and corrected for barycentric motion. The spectra of the individual arms were stitched together by averaging the overlap regions.

The VLT+FORS2 spectrum was processed, extracted and wavelength-calibrated by
using the {\sc pypeit} pipeline \citep{pypeit:joss_pub, pypeit:zenodo}, before {\sc iraf} tasks were used for flux calibration and telluric correction. The sensitivity curve used for flux calibration was derived from all FORS2+300V standard star observations during P104, excluding any star with $\rm{B} - \rm{V} < 0~mag$
for wavelengths $> 6000$\,\AA\ to minimise second-order contamination.

\subsection{NIR spectroscopy}

GNIRS is an echelle spectrograph mounted at the 8.1\,m Gemini North telescope on Maunakea, covering the wavelength range $0.8 - 2.5 \, \mu$m. The spectra were taken in cross-dispersed mode, using the 0\farcs675 wide slit in combination with the 32 l mm$^{-1}$ grating. The resolving power of this setup is $R \sim 1800$, corresponding to $\Delta v = 160$\kms\ at the central wavelength of $1.65\,\mu$m. We obtained six sets of ABBA sequences with 300s exposure times for each frame, resulting in a total integration time of 7200\,s. The slit was positioned along the parallactic angle.
The GNIRS data were reduced using the python-based \textsc{PypeIt} data reduction pipeline \citep{pypeit:joss_pub, pypeit:zenodo}. Raw images have been treated with the cosmic-ray algorithm \textsc{lacosmic} \citep{vanDokkum2001} before being processed by the pipeline. The wavelength calibration is performed using the OH night sky lines visible in the stacked science exposures. Flux calibration was accomplished using observations of the A0V telluric standard HIP 14719 with the same setup as \wxt. 

The EMIR instrument mounted at the Nasmyth ``A'' focus of the GTC was also used to obtain spectroscopic data of \wxt\ on two occasions: 19 Dec 2019 and 19 Jan 2020. In both cases observations were obtained using the YJ and HK grisms for a total exposure time of 1440~s and 1920~sec, respectively. The  A-B nodding pattern customary for NIR (spectroscopic) observations was used, where the exposure time for individual exposures was 120~s in all cases. The A-B throw was 4\arcsec\ for the 19 Dec 2019 observation with the slit at parallactic angle, whereas the A-B throw was 10\arcsec\ with a fixed slit position angle of 5.16$^\circ$ on 19 Jan 2020. For the latter observation we used a bright blind offset star located south of the position of \wxt. An 0\farcs8 slit was used in all observations. The nominal wavelength coverage of the YJ and HK grisms runs from 0.85--1.35 and 1.45--2.42 $\upmu$m, respectively with an approximate resolution of 740 at the centre of the wavelength coverage for each grism given the used slit width.

The GTC/EMIR spectroscopic data were reduced using a GTC pipeline written in python, RedEmIR, with the aim of eliminating the contribution of the sky background in NIR using consecutive A-B pairs of spectra. They were subsequently flat-fielded, calibrated in wavelength, and the different A-B pairs were then co-added to obtain the final spectrum in the K band.
Telluric correction is needed in NIR, and a version of Xtellcor (\citealt{2003PASP..115..389V}) was used for this. The software was improved and tailored to the atmospheric conditions of the La Palma observatory (\citealt{2009ApJ...694.1379R}). The spectrum was then divided by the HIP~10559 A0 star spectrum to remove telluric contamination.

\subsection{IFU spectroscopy}
\label{sect:IFU}

Optical integral-field spectrograph observations of the host galaxy of \wxt\ were carried out on the night of 28 to 29 Jan 2020 (program ID GTC1-18ITP\_0052, PI: P.~Jonker) under dark, spectroscopic sky conditions and a seeing of $\sim$0\farcs8 using MEGARA \citep{gildepaz+18} at the 10.4 m GTC. Two low-resolution (LR) Volume-Phased Holographic (VPH) grisms were used: VPH480-LR, which covers the 4330--5230 \AA\ wavelength range with a spectral dispersion of 0.207 \AA~pix$^{-1}$ and an effective resolution of R$\sim$5000, and VPH675-LR, which covers the 6095--7300 \AA\ wavelength range  with a spectral dispersion of 0.287 \AA\ pix$^{-1}$ and an effective resolution of R$\sim$5900. For each of the VPHs, 3$\times$900 s exposures were taken in order to minimise the impact of cosmic-rays on the data. The field of view of MEGARA is 12\farcs5$\times$11\farcs3.
The MEGARA data were reduced using the {\it megaradrp} v0.11 pipeline \citep{pascual+19, pascual+20}. The pipeline uses several python-based recipes to perform bias subtraction, fibre-tracing,  flat-field correction, wavelength calibration, spectra extraction and sky subtraction. Flux calibration was also performed using observations of the spectrophotometric standard star HR\,3454 obtained with the same instrument settings. The final product is a row-stacked spectrum that is converted into a datacube of 0\farcs2 square spaxel on spatial dimensions using the {\it megararss2cube} task of the {\it megara-tools} suite v0.1.1 \citep{gildepaz+20}. 

\subsection{Radio observations}
\label{sect:radio}

We report here radio observations with the Atacama Large Millimeter/submillimeter Array (ALMA -- \citealt{Wootten2009}) and the enhanced Multi Element Remotely Linked Interferometer Network (e-MERLIN, \citealt{eMERLIN_Garrington2004}). No source was detected at the position of \wxt\ in these observations. Additional radio observations of the source with the \textit{Karl G.~Jansky} Very Large Array have been reported by \cite{Chastain2019GCN}: these were conducted on 20 Dec 2019 and also produced no detection, with $3\sigma$ flux density upper limits of $<0.174$~mJy at 6 GHz and $0.030$~mJy at 22 GHz.

\subsubsection{ALMA}

ALMA observations of \wxt\ were taken on 5 Mar 2020, in Band 3 (113.4 GHz).
The observations were  centred at
01$^h$55$^m$41$^s$.941, Dec +31$^\circ$25$^\prime$04$^{\prime\prime}$.550
and taken in Time Division Mode (4 spectral windows with 128 channels 15.6 MHz wide),
we tuned the band to cover the CO(1-0) transition (rest frame frequency 115.271\,GHz) at a $\rm{z}=0.036$.
The data were calibrated and imaged using manual scripts to quickly access the possible detection of the transient source,
using the software CASA \citep{CASA_website} version 5.6.1.
A continuum image was produced using the Multi Frequency Synthesis technique. The rms noise achieved is 13$\mu$Jy/beam, and a
resolution of 1\farcs66 $\times$~0\farcs9. At the position of the transient no continuum emission is detected: the $3 \sigma$ upper limit on the source detection is $39\mu$Jy/beam. Instead, a source is clearly detected with a S/N larger than 10 offset from the pointing centre (at RA=01:55:41.37, DEC=31:25:04.93). This position is consistent with the location of the centre of the host galaxy. A channel image of the CO(1-0) line has been also obtained from the data, at a spectral resolution of 43\kms.
The host galaxy's rotation disk is clearly detected.

\subsubsection{e-Merlin}

Observations of \wxt\ with the e-MERLIN were carried out in the C band (5 GHz) on 21 and 22 Jul 2020 (C1 and C2 epochs hereafter), and in the L band (1.4 GHz) on 17 and 23 of Sept 2020 (L1 and L2 epochs hereafter).

The C1 and C2 epochs both started around 22:00 UT and lasted 8 and 14 hours respectively. The observations were pointed at
RA 01$^h$55$^m$41$^s$.94, Dec +31$^\circ$25$^\prime$04$^{\prime\prime}$.4 and phase
referenced to the flat-spectrum radio source J0159+3106 (RA 01$^h$59$^m$24$^s$.2542, Dec
+31$^\circ$06$^\prime$47$^{\prime\prime}$.83200, \citealt{Healey2007}). The frequency setup
consisted of four spectral windows spanning the frequency range 4816.5-5328.5 MHz in full polarisation.  Five stations were available (Jodrell Bank's Mark2 was missing) and one spectral window was flagged for Knockin and Defford because of a correlator issue.  Calibration was carried out with the e-MERLIN \texttt{CASA} pipeline and checked interactively.  The data from the two observations were combined in a single dataset and imaged in \texttt{CASA}.
The restoring beam is 44 mas $\times$ 35 mas with a position angle of $25^\circ$ and the image r.m.s.\ noise is 14 $\mu$Jy\,beam$^{-1}$. No pixel brighter than $3\sigma$ is found within $1^{\prime\prime}$ from the position of the target: this implies, assuming a point source, a 3$\sigma$ flux density upper limit $F_\nu(5\,\mathrm{GHz})<42\,\mathrm{\mu Jy}$ for \wxt.  A highly significant ($>7\sigma$) source is detected at $\sim7.4^{\prime\prime}$ west of the target and pointing position.  A circular Gaussian fit to the image plane returns the following parameters for the 
component (nominal uncertainties returned by CASA's ImageFitter): (RA,
Dec)=(01:55:41.36006$\pm0.00012$, +31:25:05.0938$\pm0.0018$); ($b_{maj},
b_{min}, b_{p.a.}$)=(35 mas, 31 mas, 28$^\circ$), $S_\mathrm{peak}=114\pm13
\mu$Jy.  Given the nominal fit uncertainty, the component size and its flux
density, the source is likely unresolved and the uncertainty on its
position is $\sim5$ mas.

The L1 epoch started at 18:15 UT and lasted
16 hours, while L2 started at 22:30 UT and lasted 11 hours.  The frequency setup consisted of eight spectral windows spanning the 1254.65-1766.65 MHz frequency range in full polarisation.  All six e-MERLIN stations participated and no major problems occurred during the observations. Calibration was carried out in \texttt{CASA} with the e-MERLIN \texttt{CASA} pipeline and checked interactively. The restoring beam is 160 mas $\times$ 130 mas with a position angle of $25^\circ$ and the image r.m.s.\ noise is 23 $\mu$Jy\,beam$^{-1}$.  Close to the position of the target, the r.m.s.\ noise is slightly lower, $\sim 21 \mu$Jy\,beam$^{-1}$, with no pixel brighter than $3\sigma$ within $1^{\prime\prime}$ from the pointing
position: again, assuming a point source, a 3$\sigma$ flux density upper limit $F_\nu(1.5\,\mathrm{GHz})<63\,\mathrm{\mu Jy}$ can be set for \wxt. A highly significant ($>7\sigma$) source is detected at the same position as in the C-band observations.  A circular Gaussian fit to the image
plane returns the following parameters for the component: (RA,
Dec)=(01:55:41.3611$\pm0.0004$, +31:25:05.098$\pm0.011$); ($b_{maj},
b_{min}, b_{p.a.}$)=(178 mas, 110 mas, 17.4$^\circ$),
$S_\mathrm{int}=180\pm20 \mu$Jy.  Given the nominal fit uncertainty, the component size and its flux density, the source is likely unresolved and the uncertainty on its position is 15 mas in R.A. and 25 mas in declination.

The comparison of the C- and L-band data (average frequency of 5.07 and 1.5 GHz, respectively) indicate a flat, slightly decreasing spectrum, which can be described by $S_\nu\propto \nu^{-\alpha}$ with $\alpha\sim 0.38$.

\subsection{X-ray observations}

\subsubsection{{\it Swift} XRT}

The site of \wxt\ was observed for 2.6 ks on 18 Dec 2019 (i.e. at maximum light) using the X-Ray Telescope (XRT) on board the {\it Neil Gehrels Swift Observatory}. Using the online {\it Swift}-XRT data products generator to analyse these data\footnote{\url{swift.ac.uk/user_objects/index.php}}, no source was detected at the position of \wxt, and so we follow \cite{Evans20} and calculate a 3$\sigma$ upper limit to the XRT count rate of 5.61$\times 10^{-3}$ ct~sec$^{-1}$. Using the online {\sc WebPIMMS} tool\footnote{\url{heasarc.gsfc.nasa.gov\/cgi-bin\/Tools\/w3pimms\/w3pimms.pl}}, and taking the Galactic column density of HI from \citet{Dickey90} to be $5.06\times10^{20}$ cm$^{-2}$ and a photon index of 2, we calculate an upper limit to the luminosity of \wxt\ in an energy range 0.3-10 keV to be $<6.6\times10^{41}$ erg~s$^{-1}$. As expected given the distance to \wxt, this limit is not particularly constraining, and only the x-ray brightest SNe with strong circumstellar interaction would be detected \citep{Dwarkadas12}.

\subsubsection{\textit{Chandra}}
\label{sec:chandra}

\begin{figure}[h]
\includegraphics[width=\columnwidth]{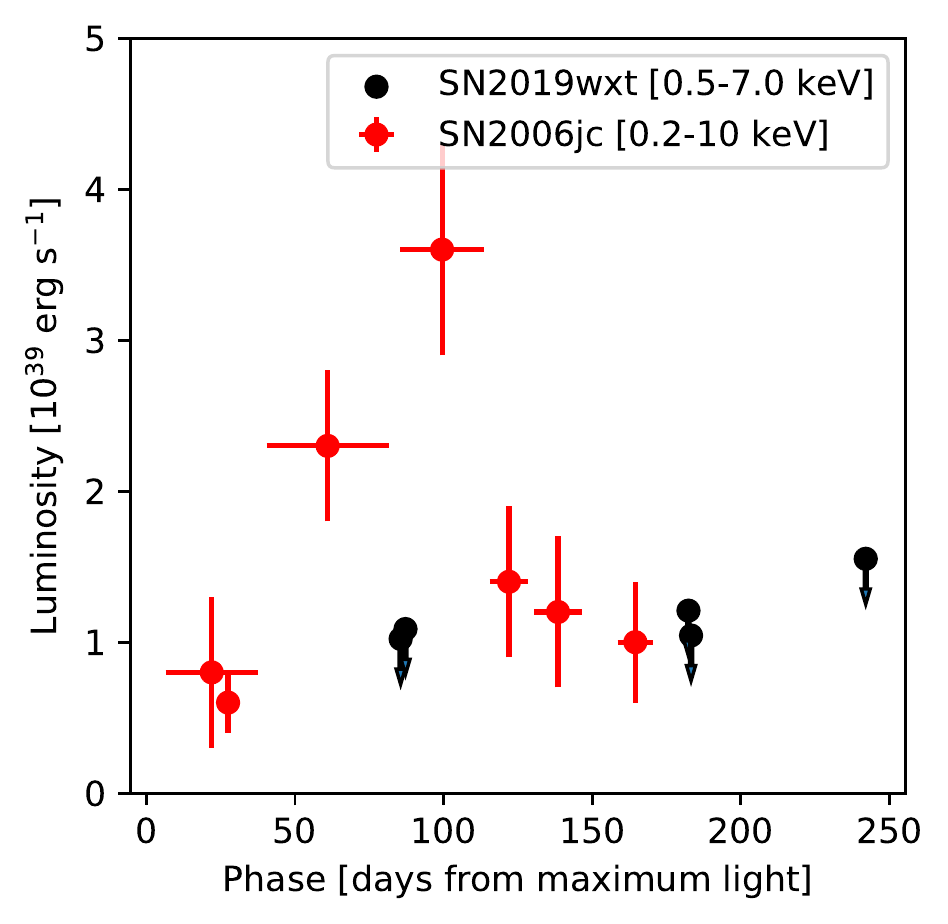}
\caption{\textit{Chandra}+ACIS upper limits to the x-ray luminosity of \wxt, compared to the X-ray lightcurve of the Type Ibn SN\,2006jc. The {\it Swift}+XRT limit $<6.6\times10^{41}$ erg~s$^{-1}$ at 0~d is more than two orders of magnitude shallower than the data plotted here, and hence not plotted.}
\label{fig:chandra}
\end{figure}

A series of deep observations totaling 157.2 ks were taken of the site of \wxt\ using the \textit{Chandra} x-ray telescope + ACIS-S between 12 Mar and 16 Aug 2020. We downloaded these data from the \textit{Chandra} archive, however examination of the ACIS Level 2 imaging event files revealed no source present at the position of \wxt\ (this was also found to be the case by \citealp{Shivkumar22} in their analysis of these Chandra data). We used the CIAO {\sc srcflux} tool to calculate 1$\sigma$ upper limits to the flux in each observation. We calculate these limits in a broad energy range from 0.5 to 7.0 keV, assuming a power law SED for with a photon index of 2, and taking the Galactic column density of HI from \citet{Dickey90}. Finally, we convert our absorption corrected fluxes to luminosities (using the host luminosity distance \hostDL~Mpc), and plot these in Fig. \ref{fig:chandra}.

While no x-ray counterpart was detected for \wxt, we can place constraining upper limits of $\lesssim 10^{39}$~erg~s$^{-1}$ on its luminosity. There are no x-ray detections of ultra-stripped SNe in the literature, and so we compare to the x-ray lightcurve of the Type Ibn SN~2006jc \citep{Immler08} in Fig. \ref{fig:chandra}. A SN of comparable x-ray luminosity to SN~2006jc would have been detected in the \textit{Chandra} data, although as SN~2006jc was brighter due to circumstellar interaction this comparison is somewhat contrived.

\section{Simple bolometric supernova model}\label{sect:simple_SN_model}
The photometric evolution of \wxt\ can be broadly described with the following simple SN model. We modelled the ejecta as a homologously expanding shell of mass $M_\mathrm{ej}$, grey opacity $\kappa$ (in the UVOIR wavelength range), velocity $v_\mathrm{ej}$ and width $\Delta R = R = v_\mathrm{ej}t$. The shell is irradiated by a centrally located radioactive source of mass $M_\mathrm{Ni}$, initially ($t=0$) composed entirely of $^{56}$Ni, whose luminosity $L_\mathrm{\gamma}(t)$ is assumed for simplicity to be entirely emitted in the form of gamma-rays and to follow the time evolution given by \citet{Nadyozhin1994}. The opacity of the ejecta to gamma-rays was assumed to be $\kappa_\gamma=0.03\,\mathrm{cm^2\,g^{-1}}$ \citep{Colgate1980}, leading to a gamma-ray optical depth $\tau_\gamma = \kappa_\gamma \rho\, \Delta R = \kappa_\gamma M_\mathrm{ej}/4\pi v_\mathrm{ej}^2 t^2$. Based on this, we assumed the ejecta to be heated by gamma-ray energy deposition at a total rate $L_\mathrm{heat}(t)=L_\gamma(t)f_\mathrm{nesc}(t)$, where $f_\mathrm{nesc}(t)=1-\exp(-\tau_\gamma(t))$ is the non-escaping fraction of the gamma-ray luminosity. The evolution of the ejecta internal energy $E_\mathrm{int}(t)$ and of its emitted luminosity $L_\mathrm{e}(t)$ was then computed by solving numerically the differential equation \citep{Kasen2010}
\begin{equation}
 \frac{1}{t}\frac{\mathrm{d}}{\mathrm{d} t}\left(E_\mathrm{int}t\right) = L_\mathrm{heat}(t)-L_\mathrm{e}(t),
\end{equation}
where the emitted luminosity was approximated from the diffusion equation,
\begin{equation}
L_\mathrm{e}= \frac{4\pi v_\mathrm{ej} c}{\kappa M_\mathrm{ej}} E_\mathrm{int}t.
\end{equation}
The photosphere was assumed to simply track the shell expansion, $R_\mathrm{ph}=v_\mathrm{ej}t$, and the effective temperature was computed from the Stephan-Boltzmann law, $T_\mathrm{eff}=(L/4\pi\sigma_\mathrm{SB} R_\mathrm{ph}^2)^{1/4}$, where $\sigma_\mathrm{SB}$ is the Stephan-Boltzmann constant. The assumption of photon diffusion in computing the emitted luminosity formally breaks down at the time when the UVOIR opical depth of the shell falls below 1, namely $t_\mathrm{trans}=\sqrt{\kappa M_\mathrm{ej}/4\pi v_\mathrm{ej}^2}$.

\begin{figure*}[h]
    \centering
    \includegraphics[width=\textwidth]{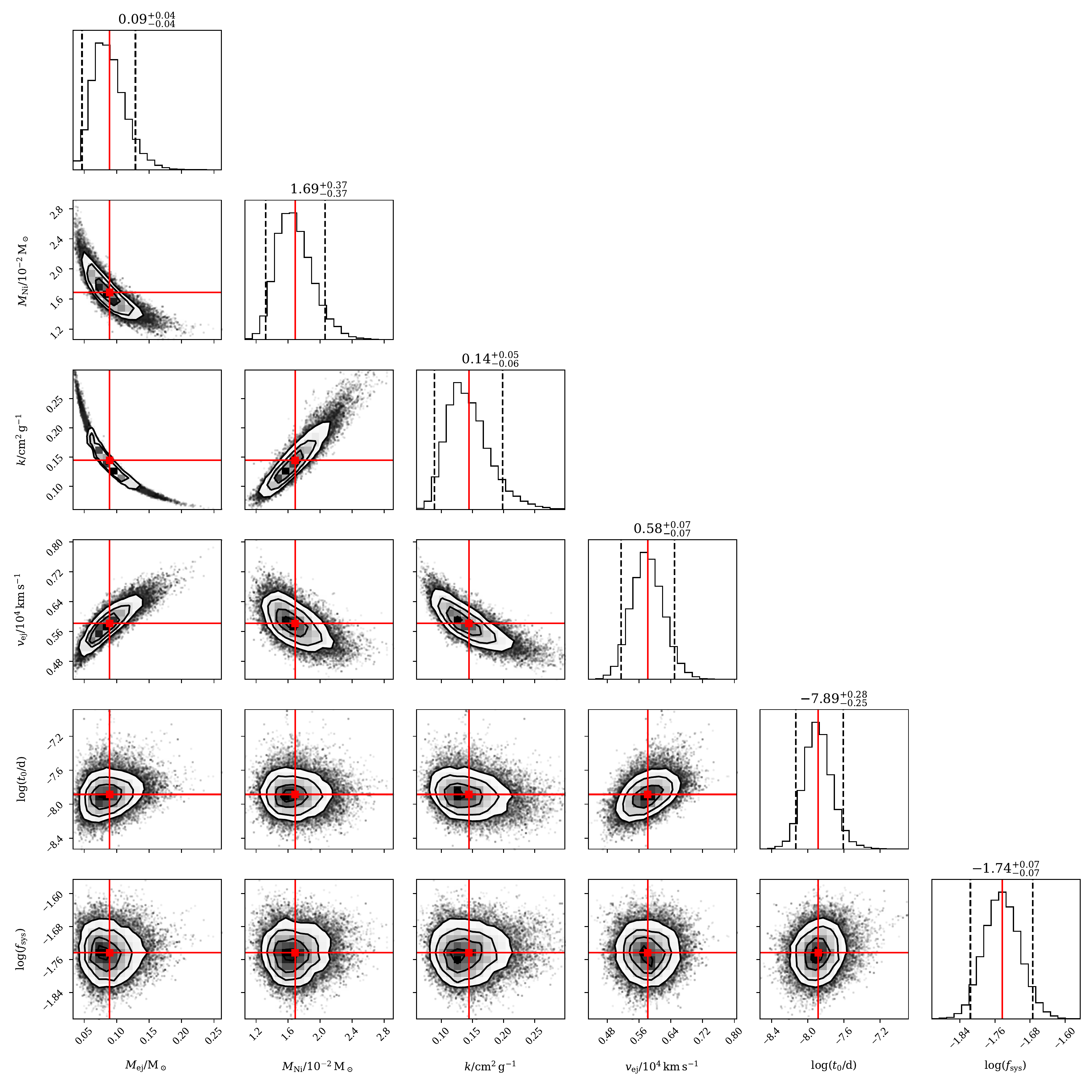}
    \caption{Corner plot showing marginalised one-dimensional and 2-dimensional posterior probability densities on the SN model parameters obtained by fitting the \wxt\ photometric dataset. The red lines and square markers show the estimated position of the maximum \textit{a posteriori}. Dashed lines in the plots on the diagonal bracket 90\% credible ranges. Contours in the two-dimensional plots show credible regions at the 68\%, 95\% and 99.7\% credible level, while black dots are random samples from the posterior, qualitatively showing the behaviour outside the contours. The physical meaning of the parameters is described in the text.} 
    \label{fig:SNfit_corner} 
\end{figure*}

A corner plot demonstrating the results of fitting the above model to the \wxt\ photometric dataset is shown in Figure \ref{fig:SNfit_corner}.

\section{Two-component SED modelling}\label{sec:BB+dust_sed_modelling}

\begin{figure*}[h]
    \centering
    \includegraphics[width=\textwidth]{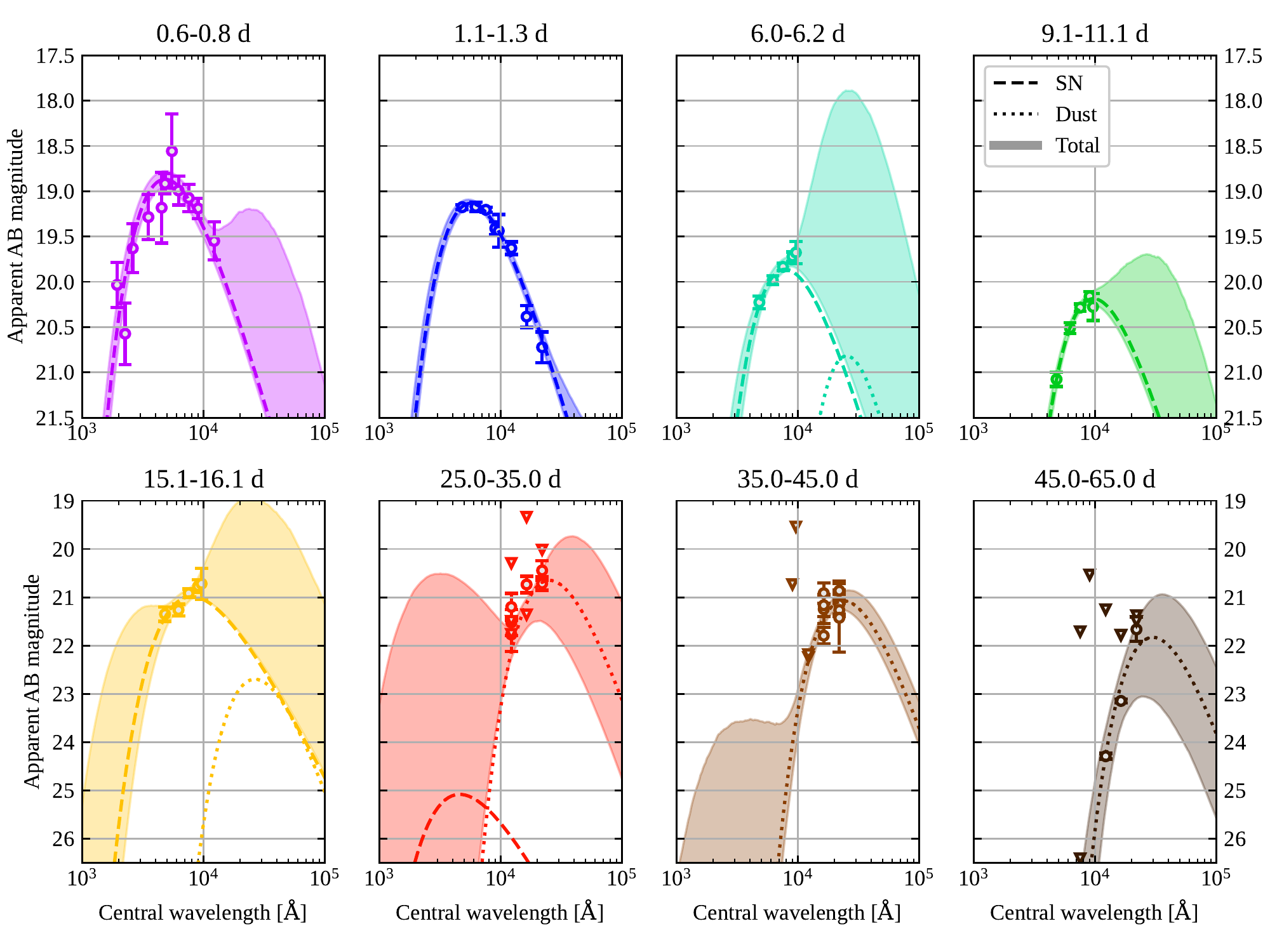}
    \caption{Spectral energy distributions fitted with a blackbody + dust model. 
    Each panel shows a SED of SN2019wxt constructed by considering photometric measurements within a time window (annotated above each panel, in days post $i$-band maximum). The formally best-fitting blackbody (`SN') and modified blackbody (`Dust') components are shown by dashed and dotted lines respectively, while the filled regions span the 5th to the 95th percentile of the model (SN+Dust) magnitudes corresponding to the posterior samples at each wavelength. The fits are performed adopting log-uniform priors on the SN luminosity in the $10^{35}-10^{43}$ erg/s range and on the dust mass in the $10^{-6}-10^{-3}$ M$_\odot$ range, and uniform priors on the SN temperature in the $4000-20000$ K range and on the dust temperature in the $100-2500$ K range.}
    \label{fig:SEDs_BB+dust}
\end{figure*}

\begin{figure}[h]
    \centering
    \includegraphics[width=\columnwidth]{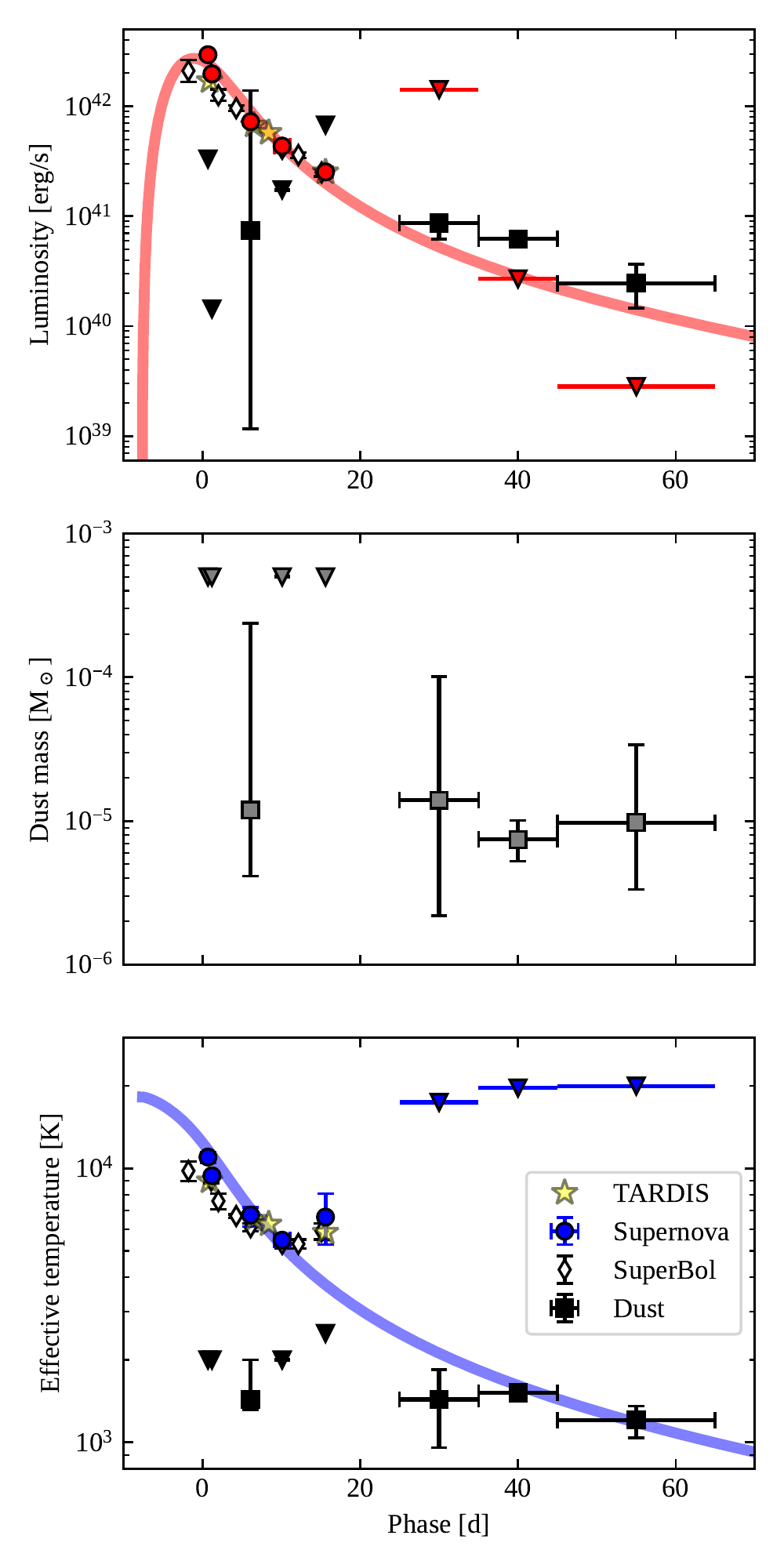}
    \caption{Evolution of SN luminosity, dust mass and effective temperatures from SED fitting. Red circles represent the blackbody (i.e.\ SN) while the black squares represent the modified blackbody (i.e.\ dust) luminosity (upper panel) and effective temperature. Downward-pointing triangles represent 3-sigma upper limits. Error bars show 90\% credible intervals.}
    \label{fig:BB+dust_evolution}
\end{figure}

Here we show the results of modelling the binned SEDs with the blackbody+dust model described in section \ref{sect:dust}. Figure \ref{fig:SEDs_BB+dust} shows the SEDs with the credible regions spanned by the two-component model, while Figure \ref{fig:BB+dust_evolution} shows the evolution of the estimated SN and dust parameters, along with the best-fitting simple SN model (Appendix \ref{sect:simple_SN_model}) obtained taking all the photometric points at $t\geq 20\mathrm{d}$ as upper limits. This yields $M_\mathrm{ej}=0.05\,\mathrm{M_\odot}$, $M_\mathrm{Ni} = 1.96\times\mathrm{10^{-2}\,M_\odot}$, $k = 0.2\,\mathrm{cm^2\,g^{-1}}$,
$v_\mathrm{ej}= 6.1\times \mathrm{10^3\,km\,s^{-1}}$ and $t_\mathrm{0} = -7.7\, \mathrm{d}$. 
 
\begin{table*}
\caption{Blackbody and dust SED fitting results. }
\centering
\renewcommand{\arraystretch}{1.3}
\begin{tabular}{cccccc}
\toprule
Phase & $L_\mathrm{SN}$ & $T_\mathrm{SN}$ & $M_\mathrm{dust}$ & $L_\mathrm{dust}$ & $T_\mathrm{dust}$\\%

[d] & [10$^{41}$ erg/s] & [10$^3$ K] & [10$^{-5}$ M$_\odot$] & [10$^{41}$ erg/s] & [10$^3$ K]\\
\midrule
0.6-0.8 & ${29.5}_{-2.1}^{+1.7}$ & ${11.0}_{-0.5}^{+0.4}$ & $<{49.53}$ & $<{5.15}$ & $<{2.0}$ \\
1.1-1.3 & ${19.7}_{-0.8}^{+1.1}$ & ${9.4}_{-0.3}^{+0.3}$ & $<{49.50}$ & $<{0.10}$ & $<{2.0}$ \\
6.0-6.2 & ${7.2}_{-0.3}^{+0.4}$ & ${6.8}_{-0.8}^{+0.4}$ & ${1.2}_{-0.9}^{+40.0}$ & ${0.8}_{-0.8}^{+12.4}$ & ${1.4}_{-0.1}^{+0.6}$ \\
9.1-11.1 & ${4.3}_{-0.1}^{+0.1}$ & ${5.5}_{-0.2}^{+0.2}$ & $<{49.95}$ & $<{1.82}$ & $<{2.0}$ \\
15.1-16.1 & ${2.5}_{-0.3}^{+0.3}$ & ${6.6}_{-1.4}^{+1.4}$ &  $<{49.9}$ & $<{6.75}$ & $<{2.5}$ \\
25.0-35.0 & $<{19.60}$ & $<{20.0}$ & ${1.2}_{-1.0}^{+8.3}$ & ${0.9}_{-0.3}^{+0.1}$ & ${1.5}_{-0.5}^{+0.4}$ \\
35.0-45.0 & $<{4.78}$ & $<{20.0}$ & ${0.7}_{-0.2}^{+0.2}$ & ${0.6}_{-0.0}^{+0.0}$ & ${1.5}_{-0.1}^{+0.1}$ \\
45.0-65.0 & $<{0.03}$ & $<{20.0}$ & ${0.9}_{-0.6}^{+2.4}$ & ${0.2}_{-0.1}^{+0.1}$ & ${1.2}_{-0.2}^{+0.1}$ \\
\bottomrule 
\end{tabular}
\label{tab:BB+Dust_fitting_result_table}
\end{table*}

\onecolumn

\section{Photometric data tables}

\begin{table*}
\centering
\setlength{\tabcolsep}{3pt} 
\caption{{\it Swift} UVOT photometry of \wxt\ as measured in stacked images in each filter. All magnitudes are in AB system, the upper limit is a $3\sigma$ upper limit.}
\label{tab:phot_uvot}
\begin{tabular}{lllcccccc}
\toprule
Date & MJD       & Phase (d) & $uvw2$    & $uvm2$  & $uvw1$  & $U$   & $B$     & $V$     \\
\midrule
	2017-08-30  & 57995.62  & $-810.48$        &     -    &    -   &    -   & $>$20.76  &   -    &   -    \\
	2019-12-18 & 58835.88  & 0.75        & $20.44^{+0.25}_{-0.20}$ & $21.03^{+0.34}_{-0.26}$ & $19.96^{+0.27}_{-0.22}$ & $19.53^{+0.25}_{-0.21}$ & $19.38^{+0.39}_{-0.29}$ & $18.71^{+0.41}_{-0.30}$ \\    
\bottomrule
\end{tabular}
\end{table*}

\begin{longtable}{ccccccc}
\caption{JHK photometry by autophot - host subtracted - calibrated to 2MASS (Vega magnitudes).
HST observations are in the F125W (J) and F160W (H) bands and are expressed as AB magnitudes.}\label{tab:IR}\\

\toprule
 Date & MJD & Phase & $J$ & $H$ & $K$ & Instrument \\
\midrule
\endhead
\midrule
\multicolumn{7}{r}{{Continued on next page}} \\
\midrule
\endfoot

\bottomrule
\endlastfoot
 2019-12-18 & 58835.80 & 0.68 & 18.68 (0.21) & - & - & TNG \\
 2019-12-19 & 58836.04 & 0.91 & 18.78 (0.14) & $>16.74$ & $>17.04$ & GROND \\
 2019-12-19 & 58836.27 & 1.13 & 18.76 (0.07) & 19.02 (0.12) & 18.89 (0.17) & UKIRT \\
 2019-12-27 & 58844.20 & 8.79 & 19.84 (0.11) & 19.79 (0.14) & 19.20 (0.17) & UKIRT \\
 2019-12-30 & 58847.23 & 11.71 & 19.46 (0.10) & 19.13 (0.13) & 18.64 (0.16) & UKIRT \\
 2020-01-04 & 58852.24 & 16.55 & 20.12 (0.19) & 19.63 (0.24) & - & UKIRT \\
 2020-01-17 & 58865.82 & 29.66 & - & $>17.97$ & - & WHT \\
 2020-01-18 & 58866.33 & 30.15 & 20.33 (0.28) & - & - & UKIRT \\
 2020-01-18 & 58866.93 & 30.73 & $>19.42$ & - & 18.61 (0.20) & WHT \\
 2020-01-19 & 58867.24 & 31.03 & 20.65 (0.27) & - & - & UKIRT \\
 2020-01-19 & 58867.81 & 31.58 & - & 19.37 (0.17) & 18.88 (0.14) & WHT \\
 2020-01-20 & 58868.23 & 31.99 & 20.90 (0.35) & - & - & UKIRT \\
 2020-01-20 & 58868.83 & 32.56 & $>20.62$ & $>19.99$ & $>18.18$ & GTC \\
 2020-01-21 & 58869.20 & 32.92 & $>20.89$ & - & - & UKIRT \\
 2020-01-22 & 58870.25 & 33.94 & $>21.37$ & - & - & UKIRT \\
 2020-01-23 & 58871.25 & 34.90 & - & 19.55 (0.21) & - & UKIRT \\
 2020-01-24 & 58872.24 & 35.86 & - & 19.88 (0.30) & - & UKIRT \\
 2020-01-25 & 58873.24 & 36.82 & - & 19.80 (0.23) & - & UKIRT \\
 2020-01-26 & 58874.23 & 37.78 & - & - & 19.03 (0.20) & UKIRT \\
 2020-01-27 & 58875.20 & 38.71 & - & - & 19.27 (0.25) & UKIRT \\
 2020-01-27 & 58875.95 & 39.44 & - & - & 19.30 (0.25) & GTC \\
 2020-01-28 & 58876.20 & 39.68 & - & - & 19.31 (0.20) & UKIRT \\
 2020-01-29 & 58877.22 & 40.66 & - & - & 19.43 (0.14) & UKIRT \\
 2020-01-29 & 58877.89 & 41.31 & $>21.31$ & - & 19.59 (0.71) & GTC \\
 2020-01-30 & 58878.26 & 41.67 & - & 20.43 (0.17) & - & UKIRT \\
 2020-02-05 & 58884.84 & 48.02 & $>20.38$ & - & 19.83 (0.25) & GTC \\
 2020-02-08 & 58887.82 & 50.90 & - & $>20.42$ & $>19.54$ & GTC \\
 2020-02-18 & 58897.82 & 60.55 & template & $>20.41$ & $>19.66$ & GTC \\
 2020-02-19 & 58898.54 & 61.25 & 24.33 (0.06) & 23.17 (0.04) & - & HST \\
 2020-02-26 & 58905.86 & 68.32 & - & - & template & GTC \\
 2020-02-28 & 58907.85 & 70.24 & - & template & - & GTC \\
\end{longtable}

\newpage
\small
\begin{longtable}{ccccccccccccc}
\caption{Optical photometry calibrated to  Pan-STARRS1 3$\pi$ reference stars. The Pan-STARRS measurements were made with PhotPipe; the NOT and GROND measurements were made with {\sc autophot}. The first three points are non-detections in PS1 and the limits are 3$\sigma$. For HST observations, calibration was carried out using standard HST zeropoints in the $F606W$ and $F814W$ bands, comparable, but not identical to $r$ and $i$. 
}\label{tab:optical}\\
\toprule
 Date & MJD & Phase & $g$ & $r$ & $i$ & $z$ & $w$ & $y$ & Instrument \\
\midrule
\endhead
\midrule
\multicolumn{10}{r}{{Continued on next page}} \\
\midrule
\endfoot

\bottomrule
\endlastfoot
2019-12-11 & 58828.31 & $-6.56$ & -            & -            & -      & $>22.0$ & - & - & PS1 \\ 
2019-12-12 & 58829.34 & $-5.56$ & -            & -            & -      & $>22.2$ & - & - & PS1 \\ 
2019-12-13 & 58830.37 & $-4.57$ & -            & -            & -      & $>21.0$ & - & - & PS1 \\ 
2019-12-15 & 58832.31 & $-2.69$ & -            & -            & $>19.6$      & - & - & - & PS1 \\ 
2019-12-16 & 58833.32 & $-1.72$ & -            & -            & 19.36 (0.12) & - & - & - & PS1 \\
2019-12-18 & 58835.86 & 0.73    & 19.10 (0.11) & 19.12 (0.16) & 19.17 (0.15) & 19.26 (0.11) & - & - & NOT \\
2019-12-19 & 58836.04 & 0.91    & 19.16 (0.08) & 19.25 (0.06) & 19.31 (0.05) & 19.37 (0.06) & - & - & GROND \\
2019-12-19 & 58836.09 & 0.96    & 19.32 (0.04) & 19.25 (0.03) & 19.29 (0.06) & 19.36 (0.06) & - & - & NOT \\
2019-12-19 & 58836.21 & 1.07    & 19.36 (0.03) & 19.30 (0.05) & 19.30 (0.03) & 19.48 (0.07) & - & 19.50 (0.18) & PS1 \\
2019-12-19 & 58836.43 & 1.28    & 19.42 (0.09) & 19.32 (0.07) & 19.32 (0.07) & 19.42 (0.11) & - & 19.36 (0.22) & PS1 \\
2019-12-20 & 58837.11 & 1.94    & 19.86 (0.02) & 19.43 (0.02) & 19.62 (0.01) & 19.58 (0.03) & - & - & PS1 \\
2019-12-22 & 58839.38 & 4.13    & -            & 19.71 (0.33) & 19.64 (0.32) & - & - & - & PS1 \\
2019-12-24 & 58841.21 & 5.90    & 20.41 (0.07) & 20.11 (0.05) & 19.93 (0.04) & 19.79 (0.05) & - & 19.74 (0.12) & PS1 \\
2019-12-28 & 58845.21 & 9.76    & 21.26 (0.08) & 20.64 (0.06) & 20.38 (0.04) & 20.23 (0.05) & - & 20.34 (0.15) & PS1 \\
2019-12-30 & 58847.26 & 11.74   & -            & 20.89 (0.11) & 20.62 (0.08) & 20.30 (0.07) & - & 20.11 (0.18) & PS1 \\
2019-12-31 & 58848.28 & 12.72   & -            & -            & - & - & 21.06 (0.16) & - & PS1 \\
2020-01-02 & 58850.21 & 14.59   & 21.53 (0.15) & 21.39 (0.12) & 21.01 (0.09) & 20.84 (0.13) & - & 20.78 (0.32) & PS1 \\
2020-01-30 & 58878.20 & 41.61   & -            & -            & - & $>20.8$  & - & $>19.6$  & PS1 \\
2020-02-05 & 58884.25 & 47.45   & -            & -            & $>21.8$ & $>20.6$ & - & -  & PS1 \\
2020-02-19 & 58898.45 & 61.16   & -            & $>$27.7      & $>26.5$ & - & - & - & HST\\
\end{longtable}

\end{appendix}

\end{document}